\documentclass[aps,prb,twocolumn,superscriptaddress,floatfix,longbibliography,10pt]{revtex4-2}

\usepackage{amsmath,amssymb} 
\usepackage{bm} 
\usepackage{graphicx} 
\usepackage{comment} 
\usepackage{textcomp} 

\usepackage{enumitem}
\setlist{noitemsep,leftmargin=*,topsep=0pt,parsep=0pt}

\usepackage{xcolor} 
\definecolor{lightgray}{gray}{0.6}
\definecolor{medgray}{gray}{0.4}

\usepackage{hyperref}
\hypersetup{
colorlinks=true,
urlcolor= blue,
citecolor=blue,
linkcolor= blue,
}

\newif\ifptitle
\newif\ifpnumber
\newcounter{para}

\ptitletrue  
\pnumbertrue  

\makeatletter
\newcommand\xleftrightarrow[2][]{%
  \ext@arrow 9999{\longleftrightarrowfill@}{#1}{#2}}
\newcommand\longleftrightarrowfill@{%
  \arrowfill@\leftarrow\relbar\rightarrow}
\makeatother

\newcommand{\mytitle}{Twisted Superfields}

\numberwithin{equation}{section}

\begin{document}

\title{\mytitle}

\author{Scott Chapman}
\email[]{schapman@chapman.edu}
\affiliation{Institute for Quantum Studies, Chapman University, Orange, CA  92866, USA}

\date{\today}

\begin{abstract}
A model is presented that could lead to an interesting extension of the Standard Model.  Like a supersymmetric gauge theory, the model is holomorphic and invariant to local superspace gauge transformations.  However, the model is not invariant to superspace translations, so it is not supersymmetric.  It is proposed that this combination allows the model to have many of the attractive features of supersymmetric theories, while at the same time predicting fewer particles that have not yet been seen experimentally.  For example, the ``superpartners'' of the gauge bosons in the model are quarks. The model is able to generate the symmetries and particles of the Standard Model, but with some significant differences that have observable consequences.  These consequences provide possible explanations for a number of 3-7 sigma deviations from Standard Model calculations that have been found in recent experiments.     
\end{abstract}

\maketitle
\section*{\label{sec:Start}Introduction}

Supersymmetric gauge theories have many attractive properties.  For example, they provide a natural mechanism for cancellation of quadratic divergences and a resolution of the Hierarchy Problem.  They do this by being (i) holomorphic, (ii) invariant to local gauge transformations in superspace, and (iii) invariant to global translations in superspace.  A difficulty with supersymmetric theories, however, is that they predict that for every particle that has been observed, there is another partner particle that has not yet been seen.  As experiments probe higher and higher energies, the fact that no partner particle has been found becomes more problematic.  Early on, supersymmetry practitioners asked whether some of the existing observed particles could actually be supersymmetric partners with each other.  The HLS theorem \cite{HLS} mostly rules out this possibility with some minor exceptions (like the Higgs boson being a slepton \cite{slepton-Higgs}).  But even those exceptions are generally not accepted for other reasons.

The model presented in this paper incorporates the first two features of supersymmetry listed above, but not the third.  Since the model is not invariant to superspace translations, it is not supersymmetric.  Consequently, there is no a priori guarantee that quadratic divergences cancel.  That being said, many quadratic divergences are cancelled for supersymmetric gauge theories primarily due to the fact that the theories are holomorphic and invariant to superspace gauge transformations.  Also, local superspace gauge transformations turn scalar bosons into fermions, and fermions do not have quadratic divergences.  Divergences may cancel in this theory for similar reasons.  To that point, for nonsupersymmetric gauge theories similar to the one presented in this paper, it has been shown explicitly that quadratic divergences cancel to at least the two-loop order \cite{Twisted-Superfields}. 

The superspace gauge transformations of the model are built on the group U(3)$\times$U(3).  The field content of the model includes constructions built in N=1 superspace that are not N=1 superfields but are nonetheless called ``twisted superfields'' by way of analogy. Specifically, the model includes a real ``twisted superfield'', an adjoint-representation chiral ``twisted superfield'', and three flavors of fundamental and anti-fundamental chiral ``twisted superfields''.  

 An advantage to this theory \textit{not} being supersymmetric is that it can allow existing observed particles to be ``superpartners'' with each other (in the sense that superspace gauge transformations change them into each other).  For example, in this model the partners of the gauge bosons (within the real ``twisted superfield'') are quarks.  Also, due to the fact that the theory is holomorphic and supergauge invariant, it is argued that some of the nonperturbative phenomena of supersymmetric gauge theories may apply to the present theory.

After presenting the theory in the first two sections (and the appendix), the third section shows how the theory can reproduce the existing forces and particles of the Standard Model, as well as the observed masses and mixing of neutrinos.  The fourth section shows how gauge anomalies cancel and the theory's coupling constants converge at a unification scale.  The fifth section of this paper shows how the model has the correct structure to reproduce many of the anomalies presented in \cite{Crivellin_2024} and other published papers, where experimental results differ from Standard Model predictions by 3-7$\sigma$. 

In particular, the paper predicts the existence of a seventh quark, three heavy lepton families, a light $Z'$ boson, a second Higgs boson, right-handed weak currents, and many additional scalars.  Current experiments do not rule out these predictions; in fact data are better reproduced if they exist.

Many of the ideas of this paper were originally published by the author in \cite{alternative}.  However, this paper has heavily revised the structure of the theory, the parameter values, and the mapping to experimental results.  This paper replaces that original paper.

\section{U(3) x U(3) Symmetries and Fields}

The theory is constructed in N=1 superspace, extending four-dimensional spacetime by including four additional anticommuting coordinates $\theta _{\alpha } ,\bar{\theta }_{\dot{\alpha }} $.  For reviews of superspace, see \cite{rargurio,SUSY-Martin,SUSY-Haber,SUSY-Bertolini,Binetruy,SUSY-Gates,SQCD-Argyres}; the notational conventions of \cite{rargurio} are used throughout.  The gauge group of the model is U(3)$\times$U(3), and it is described using 6$\times$6 matrices with the gauge fields in the 3$\times$3 diagonal blocks.  The model includes a construction called a real ``twisted superfield'' that has the following attributes: field components in the 3$\times$3 diagonal blocks have an even number of $\theta _{\alpha } ,\bar{\theta }_{\dot{\alpha }} $ factors, while those in the 3$\times$3 off-diagonal blocks have an odd number of $\theta _{\alpha } ,\bar{\theta }_{\dot{\alpha }} $ factors.  An adjoint-representation chiral ``twisted superfield'' has this same structure, and fundamental chiral ``twisted superfields'' have structures consistent with those.

For normal superfields, translations in superspace transform a bosonic component of the superfield into a fermionic one, and vice versa.  For the constructions used here to be superfields, every translation in superspace would have to be accompanied by a gauge rotation.  But the HLS theorem has proven that this kind of rotation is not allowed for N=1 supersymmetric theories \cite{HLS}.  Therefore, the constructions used to develop the model are not N=1 superfields, but the term is used to provide an analogy with supersymmetric theories.

The real ``twisted superfield'' $V=V^{\dag } $ is defined by:
\begin{widetext}
\begin{equation} \label{1.1} 
V=\left(\begin{array}{cc} {C_{1} +N_{1} \theta ^{2} +\bar{\theta }^{2} N_{1}^{\dag } -\bar{\theta }\bar{\sigma }^{\mu } A_{1\mu } \theta +{\tfrac{1}{2}} \bar{\theta }^{2} d_{1} \theta ^{2} } & {\eta \theta +\bar{\theta }\tilde{\eta }^{\dag } +i\bar{\theta }\tilde{\lambda }^{\dag } \theta ^{2} -i\bar{\theta }^{2} \lambda \theta } \\ {\tilde{\eta }\theta +\bar{\theta }\eta ^{\dag } +i\bar{\theta }\lambda ^{\dag } \theta ^{2} -i\bar{\theta }^{2} \tilde{\lambda }\theta } & {C_{2} +N_{2} \theta ^{2} +\bar{\theta }^{2} N_{2}^{\dag } -\bar{\theta }\bar{\sigma }^{\mu } A_{2\mu } \theta +{\tfrac{1}{2}} \bar{\theta }^{2} d_{2} \theta ^{2} } \end{array}\right),  
\end{equation} 
\end{widetext}
where each component field above is a U(3) matrix function of spacetime coordinates $x^{\mu } $.  For example, $A_{1\mu } =A_{1\mu }^{A} \left(x\right)t^{A} $, where $t^{A} $ are 3$\times$3 U(3) matrices normalized by ${\rm tr}\left(t^{A} t^{B} \right)={\tfrac{1}{2}} \delta ^{AB} $.  Lower case letters are used to denote SU(3) adjoint indices $a,b\in \left\{1,2...8\right\}$.  Upper case letters are used to denote U(3)  adjoint indices $A,B\in \left\{0,1,2...8\right\}$ that include the Abelian matrix $t^{0} ={\tfrac{1}{\sqrt{6} }} {\rm diag}\left(1,1,1\right)$.  The $\theta _{\alpha } $ are 2-component anti-commuting Grassman coordinates, and $\bar{\theta }_{\dot{\alpha }} $ are their Hermitian conjugates.  As a result of their $\theta _{\alpha } ,\bar{\theta }_{\dot{\alpha }} $ factors, the fields in the diagonal blocks of $V$ are bosons, while the fields in the off-diagonal blocks are fermions.  

 As mentioned previously, a theory built using the above real twisted superfield is not supersymmetric, since the fermion fields in the superfield are in a different representation of the U(3)$\times$U(3) group than the boson fields.  Despite not being supersymmetric, the real twisted superfield is assumed to transform as follows under a local ``twisted supergauge transformation'':

\begin{equation} \label{1.2}
e^{V} \to e^{i\Lambda ^{\dag } } e^{V} e^{-i\Lambda }.  
\end{equation}

\noindent In the above expression, 
\begin{equation} \label{1.3} 
\Lambda =\left(\begin{array}{cc} {\alpha _{1} \left(y\right)+\theta ^{2} n_{1} \left(y\right)} & {\theta \xi _{1} \left(y\right)} \\ {\theta \xi _{2} \left(y\right)} & {\alpha _{2} \left(y\right)+\theta ^{2} n_{2} \left(y\right)} \end{array}\right) 
\end{equation} 
is a chiral ``twisted superfield'' whose component fields are U(3) matrix functions (e.g. $\alpha _{1} =\alpha _{1}^{A} t^{A} $) of $y^{\mu }=x^{\mu } +i\theta\sigma ^{\mu }\bar{\theta }$.  The twisted supergauge transformation of eq \eqref{1.2} maintains the boson-fermion structure of the real twisted superfield as well as its group structure.  To the latter point, if the group was SU(3)$\times$SU(3) instead of U(3)$\times$U(3), the supergauge transformation would not be consistent, since a general supergauge transformation would generate terms in each block proportional to $t^{0} $.  On the other hand, a U(3)$\times$U(3) twisted supergauge transformation is consistent. 

 Like normal real superfields, the real twisted superfield supports conjugate representations.  To see this, it is helpful to follow the presentation of \cite{SUSY-Martin} and re-express an infinitesimal twisted supergauge transformation as 
\begin{equation} \label{1.4} 
\begin{aligned} 
V & \to V+i\Lambda ^{\dag } -i\Lambda -{\tfrac{1}{2}} i\left[V,\left(\Lambda ^{\dag } +\Lambda \right)\right] \\
& +i\sum _{k=1}^{\infty }\frac{B_{2k} }{\left(2k\right)!} \left[V,\left[V,...\left[V,\left(\Lambda ^{\dag } -\Lambda \right)\right]...\right]\right] ,
\end{aligned}   
\end{equation} 
where $B_{2k} $ are Bernoulli numbers.  Both $V$ and $\Lambda $ can be expanded in terms of component fields multiplied by U(6) matrices $T^{X} $, where the index X runs over the 36 adjoint indices of U(6).  As with any unitary group, the same structure functions $f^{XYZ} $ satisfy both $\left[T^{X} ,T^{Y} \right]=if^{XYZ} T^{Z} $ and $\left[-T^{XT} ,-T^{YT} \right]=-if^{XYZ} T^{ZT} $.  Since products of matrices in eq \eqref{1.4} only enter by way of commutators, a conjugate representation is available by replacing each $T^{X} $ in eq \eqref{1.4} with its negative transpose $\left(-T^{X} \right)^{T} $.  In other words, a twisted real superfield that transforms by eq \eqref{1.2} also transforms as follows:
\begin{equation} \label{1.5}
e^{-V} \to e^{i\Lambda } e^{-V} e^{-i\Lambda ^{\dag } } .  
\end{equation}

 Despite the fact that \eqref{1.1} and \eqref{1.3} are not N=1 superfields and \eqref{1.2} is not a normal supergauge transformation, the word ``twisted'' will be dropped for brevity in much of the rest of the paper.

 A consequence of eq \eqref{1.4} is the fact that one component of the real superfield has a supergauge transformation independent of the other components.  Taking the trace of eq \eqref{1.4}, one finds
\begin{equation} \label{1.6}
\begin{aligned} 
& {\tfrac{1}{\sqrt{3} }} {\rm Tr}\left(V\right)=V_{+}^{0} \to V_{+}^{0} +i\Lambda _{+}^{0\dag } -i\Lambda _{+}^{0} \\
& \Lambda _{+}^{0} ={\tfrac{1}{\sqrt{3} }} {\rm Tr}\left(\Lambda \right), 
\end{aligned}          
\end{equation} 
where ${\rm Tr}$ is the 6$\times$6 trace.  The reason that this supergauge transformation is independent is because all of the commutators in eq \eqref{1.4} are proportional to some 6$\times$6 traceless matrix, so none of them can contribute to eq \eqref{1.6}.  Since by definition, the fermions of the real superfield are all in off-diagonal blocks, the field $V_{+}^{0} $ does not include any fermions, only bosons.  Inside of $V$, the field $V_{+}^{0} $ is multiplied the 6$\times$6 matrix $T_{+}^{0} $ defined via
\begin{equation} \label{GrindEQ__1_7_}
\begin{aligned} 
& T_{\pm }^{A} ={\tfrac{1}{\sqrt{2} }} \left(T_{2}^{A} \pm T_{1}^{A} \right) \\ 
& T_{1}^{A} =\left(\begin{array}{cc} {t^{A} } & {0} \\ {0} & {0} \end{array}\right)\,\,\,\,  T_{2}^{A} =\left(\begin{array}{cc} {0} & {0} \\ {0} & {t^{A} } \end{array}\right).
\end{aligned}     
\end{equation} 
This definition provides another way of saying that $T_{+}^{0} $ is ${\tfrac{1}{2\sqrt{3} }} $ of the 6$\times$6 unit matrix.

 As is often done in superspace gauge theories, the real superfield will be rescaled to explicitly show the coupling constant.  In this case, the following rescaling is performed:
\begin{equation} \label{GrindEQ__1_8_}
\begin{aligned}
& V\to 2gV'+2g_{+} V_{+}^{0} T_{+}^{0} \\  
& V'=V-V_{+}^{0} T_{+}^{0} .   
\end{aligned}       
\end{equation} 
Since $V_{+}^{0} $ has its own, independent supergauge transformation, it also has its own coupling constant.  

Now that the gauge transformation properties of the real superfield have been identified, gauge invariant action terms can be defined.  Just as with normal superfields, the following chiral twisted superfields can be defined:
\begin{equation} \label{GrindEQ__1_9_} 
\begin{aligned}
W'_{\alpha } & =-{\tfrac{1}{8g}} i\bar{D}_{}^{2} \left(e^{-2gV'} D_{\alpha } e^{2gV'} \right) \\
W_{\alpha }^{0} & =-{\tfrac{1}{4}} i\bar{D}_{}^{2} D_{\alpha } V_{+}^{0} ,  
\end{aligned}   
\end{equation} 
where $D_{\alpha } =\partial _{\alpha } +i\sigma _{\alpha \dot{\alpha }}^{\mu } \bar{\theta }_{}^{\dot{\alpha }} \partial _{\mu } $.  Under a supergauge transformation, these fields transform as follows:
\begin{equation} \label{GrindEQ__1_10_} 
\begin{aligned}
W'_{\alpha } & \to e^{i\Lambda} W'_{\alpha } e^{-i\Lambda}  \\
W_{\alpha }^{0} & \to W_{\alpha }^{0} .  
\end{aligned}    
\end{equation} 
As a result, the following terms in the action are supergauge invariant:
\begin{equation} \label{GrindEQ__1_11_} 
\begin{aligned}
S_{V} =&-{\tfrac{1}{2}}\int d^{4} xd^{2} \theta  \left(1+4m_{\lambda } \theta ^{2} \right){\rm Tr}\left(W'^{\alpha } W'_{\alpha } \right) \\
& -{\tfrac{1}{4}}\int d^{4} xd^{2} \theta _{} W_{}^{0\alpha} W_{\alpha }^{0} +h.c.,  
\end{aligned}
\end{equation} 
where $h.c.$ stands for Hermitian conjugate and $m_{\lambda } $ is a ``gaugino mass''.

The action may also include the following gauge-invariant Fayet Iliopoulos term:
\begin{equation} \label{GrindEQ__1_12_} 
S_{\xi} =\tfrac{1}{\sqrt{3}} \xi _{+} \int d^{4} xd^{2} \theta d^{2} \bar{\theta }_{}g_{+} V_{+}^{0}.
\end{equation} 

In addition to the real superfield, the theory includes the following chiral twisted superfield in an adjoint representation of twisted U(3)$\times$U(3):  
\begin{equation} \label{GrindEQ__1_16_} 
\Phi =\left(\begin{array}{cc} {\varphi _{1} \left(y\right)+\theta ^{2} f_{1} \left(y\right)} & {\sqrt{2}\theta \chi \left(y\right)} \\ {\sqrt{2} \theta \tilde{\chi }\left(y\right)} & {\varphi _{2} \left(y\right)+\theta ^{2} f_{2} \left(y\right)} \end{array}\right) 
\end{equation} 
The adjoint superfield transforms as follows:
\begin{equation} \label{GrindEQ__1_17_} 
\Phi \to e^{i\Lambda} \Phi e^{-i\Lambda} .          
\end{equation} 

The following action terms involving this field are supergauge invariant:
\begin{equation} \label{GrindEQ__1_18_} 
\begin{aligned}
S_{\Phi} & =2\int d^{4} xd^{2} \theta d^{2} \bar{\theta }_{} \times \\
& \times{\rm Tr}\Big(\Phi^{\dag } e^{2gV'} \Phi e^{-2gV'} (1-\theta^2\bar{\theta}^2\sum_m m^2_{\Phi m}\sqrt{\tfrac{3}{2}}T_m^0)\Big)\\ 
& -2\int d^{4} xd^{2} \theta  {\rm Tr}\left({\tfrac{1}{2}} m_{\Phi } \Phi ^{2} +{\tfrac{1}{3}} \Gamma _{\Phi } \Phi _{}^{3} \right) +h.c. 
\end{aligned}  
\end{equation} 
The $m^2_{\Phi m}$ terms are scalar mass terms.  Despite the explicit group matrices $T_m^0$, the terms are supergauge invariant since the factor of $\theta^2\bar{\theta}^2$ limits gauge transformations to ones that remain within the same 3$\times$3 diagonal block.  Just like the gaugino mass term (and like analogous soft supersymmetry breaking terms), the scalar mass terms break superspace translation invariance (which is not imposed in this model anyway), but do not break superspace gauge invariance.  

In addition to adjoint-representation chiral fields, the theory also includes three flavors of 6-vector chiral twisted superfields in the fundamental and anti-fundamental representations of twisted U(3)$\times$U(3): 
\begin{equation} \label{GrindEQ__1_13_} 
\begin{aligned} 
& Q_{1F}^{} =\left(\begin{array}{c} {\phi _{1F}^{} +\theta ^{2} f_{1F}^{} } \\ {\sqrt{2} \theta \psi _{2F}^{} } \end{array}\right)\\
& Q_{2F}^{} =\left(\begin{array}{c} {\sqrt{2} \theta \psi _{1F}^{} } \\ {\phi _{2F}^{} +\theta ^{2} f_{2F}^{} } \end{array}\right) \\
& \tilde{Q}_{1F}^{} =\left(\begin{array}{cc} {\tilde{\phi }_{1F}^{} +\theta ^{2} \tilde{f}_{1F}^{} ,} & {\sqrt{2} \theta \tilde{\psi }_{2F}^{} } \end{array}\right) \\
& \tilde{Q}_{2F}^{} =\left(\begin{array}{cc} {\sqrt{2} \theta \tilde{\psi }_{1F}^{} ,} & {\tilde{\phi }_{2F}^{} +\theta ^{2} \tilde{f}_{2F}^{} } \end{array}\right)
\end{aligned}
\end{equation}
where $F\in \left\{1,2,3\right\}$ is a flavor index, and each component field is a chiral 3-vector (or covector).  The supergauge transformation for each of these superfields depends upon their flavor in the following way:
\begin{equation} \label{GrindEQ__1_14_} 
\begin{aligned}
& \tilde{Q}_{2F}^{} \to \tilde{Q}_{2F}^{} e^{-i\Lambda}, \hspace{.5cm}  Q_{2F} \to e^{i\Lambda} Q_{2F},
\hspace{.5cm}F=2,3\\
& \tilde{Q}_{21}^{} \to \tilde{Q}_{21}^{} e^{-i(\Lambda - \Lambda _{+}^{0} T_{+}^{0})}, \hspace{.5cm}
Q_{21} \to e^{i(\Lambda-\Lambda _{+}^{0} T_{+}^{0})} Q_{21} \\
& \tilde{Q}_{1F}^{} \to \tilde{Q}_{1F}^{} e^{-i(\Lambda - \Lambda _{+}^{0} T_{+}^{0})}, \hspace{.5cm}
Q_{1F} \to e^{i(\Lambda-\Lambda _{+}^{0} T_{+}^{0})} Q_{1F}\,.
\end{aligned}           
\end{equation} 
As a result, the following terms in the action are supergauge invariant:
\begin{equation} \label{GrindEQ__1_15_} 
\begin{aligned}
S_{Q } &= \sum _{mF}\int d^{4} xd^{2} \theta d^{2} \bar{\theta }\times \\
& \hspace{0.4cm}\times\Big(\left(1-m^2_{mF}\theta^2\bar{\theta}^2\right)
Q_{mF}^{\dag } e^{2\left(gV'+q_{mF} g_{+} V_{+}^{0} T_{+}^{0} \right)} Q_{mF} \\
& \hspace{0.5cm}+\left(1-\tilde{m}^2_{mF}\theta^2\bar{\theta}^2\right)\tilde{Q}_{mF} e^{-2\left(gV'+q_{mF} g_{+} V_{+}^{0} T_{+}^{0} \right)} \tilde{Q}_{mF}^{\dag } \Big)\\
 &- \sum_{[mFF']} \int d^{4} xd^{2} \theta \times \\
&\hspace{0.4cm}\times\left( \tilde{Q}_{mF} \left(m_{mFF'}+\sqrt{2}\Gamma _{mFF'}\Phi\right)Q_{mF'}\right)\\
&+h.c.   \\
& q_{1F} =q_{21} =-2, \,\,\,\,q_{22} =q_{23} =1\,,
\end{aligned}  
\end{equation}  
where $m\in \left\{1,2\right\}$ and $[mFF']$ means to only sum over combinations where $\tilde{Q}_{mF}$ and $Q_{mF'}$ have the same $q_{mF}$ charge (and corresponding gauge transformations from eq \eqref{GrindEQ__1_14_}).  The $m^2_{mF}$ and $\tilde{m}^2_{mF}$ terms generate additional mass terms for the fundamental and conjugate scalars.

The theory presented above is free of gauge anomalies.  There is a simple reason: for every fermion in the theory, there is another fermion in a conjugate representation with opposite Abelian charges.  Since the theory is a gauge theory and is free of gauge anomalies, the theory is renormalizable.

Since all of the action terms presented above are invariant to twisted supergauge transformations, it is possible to restrict the real superfield to a Wess-Zumino-like gauge.  In that gauge, the real superfield takes the form:
\begin{equation} \label{GrindEQ__1_19_} 
V=\left(\begin{array}{cc} {-\bar{\theta }\bar{\sigma }^{\mu } A_{1\mu } \theta +{\tfrac{1}{2}} \bar{\theta }^{2} d_{1} \theta ^{2} } & {i\bar{\theta }\tilde{\lambda }^{\dag } \theta ^{2} -i\bar{\theta }^{2} \lambda \theta } \\ {i\bar{\theta }\lambda ^{\dag } \theta ^{2} -i\bar{\theta }^{2} \tilde{\lambda }\theta } & {-\bar{\theta }\bar{\sigma }^{\mu } A_{2\mu } \theta +{\tfrac{1}{2}} \bar{\theta }^{2} d_{2} \theta ^{2} } \end{array}\right) .     
\end{equation} 
In \cite{Twisted-Superfields}, it was shown that a Wess-Zumino-like gauge is accessible for a theory with this kind of twisted supergauge invariance.  After imposition of this Wess-Zumino gauge, the residual gauge invariance is just local spacetime gauge invariance.  In the following, the fermions $\lambda $ and $\tilde{\lambda }$ will be referred to as ``gauginos'' despite the fact that they are in the $(3,\tilde{3})$ and $(\tilde{3},3)$ representations of the gauge group, rather than the adjoint representation.

Although the Abelian field $V_+^0$ has its own, independent gauge transformation, the second Abelian field $V_-^0$ (group structure $T_-^0$) does not decouple from the gauge transformations described above.  That is why the same designation $g$ was used for both the $V_-^0$ and nonAbelian couplings above. Nonetheless, in the Wess-Zumino gauge, since $V_-^0$ is an Abelian field, it can accommodate different charges multiplying its coupling constant when acting on different chiral fields.  This flexibility is used to make the charges of $V_-^0$ the same as those for $V_+^0$ defined in eq \eqref{GrindEQ__1_15_}.  In other words, the following replacement is made in that equation: 
\begin{equation}\label{Ab-charge}
q_{mF} g_{+} V_{+}^{0} T_{+}^{0}\to q_{mF} g_{+} V_{+}^{0} T_{+}^{0}+(q_{mF}-1) g V_{-}^{0} T_{-}^{0}\,.
\end{equation}
This change only affects fundamental and conjugate representation fields with with $mF=21$ or $1F$;  it has no effect on fields in the real or adjoint twisted superfields.  For interactions with gauge fields, this modification is equivalent to treating the $q_{mF}=-2$ fundamental or conjugate fields in each SU(3) sector as if they were in the antisymmetric representation of SU(3) for that sector. 

The classical theory has 50 parameters that can be adjusted classically: 2 coupling constants, 2 Abelian charges, 30 masses, 15 superpotential couplings, and a Fayet-Iliopoulos term.  In the unification section of the paper, it is argued that both gauge couplings may be the same at the unification scale ($g_+=g$).  It is also assumed that the following 7 parameters are zero classically: 
\begin{equation} \label{GrindEQ__1_22_}
m_{23F}=m_{223}=\Gamma_{23F}=m_{23}=\tilde{m}_{23}=0
\end{equation} 
Many of the remaining parameters may also be zero classically but acquire values via quantum corrections.

\section{ Dynamical Symmetry Breaking}

This section identifies a minimum of the scalar potential that breaks the gauge symmetry in stages from SU(3)$\times$SU(3)$\times$U(1)$\times$U(1) down to SU(3)$\times$U(1) and labels the fermions in the model based on their Standard Model symmetries.  

The scalar potential for this model can be expressed in terms of its auxiliary fields and scalar mass terms:
\begin{equation} \label{GrindEQ__2_1_}
\begin{aligned} 
\mathcal{V}&={\tfrac{1}{2}} \left(d_{+}^{0} \right)^{2} +{\tfrac{1}{2}} \left(d_{-}^{0} \right)^{2} +{\tfrac{1}{2}} \sum _{m,a}\left(d_{m}^{a} \right)^{2}  \\
&+\sum _{mF}\left(f_{mF}^{\dag } f_{mF} +\tilde{f}_{mF} \tilde{f}_{mF}^{\dag } \right) +2{\rm tr}\left(f_{1}^{\dag } f_{1}\right)+2{\rm tr}\left(f_{2}^{\dag } f_{2} \right) \\
&+\sum _{mF}\left(m^2_{mF}\phi^\dag_{mF}\phi_{mF}+\tilde{m}^2_{mF}\tilde{\phi}^\dag_{mF}\tilde{\phi}_{mF}\right) \\
&+m^2_{\Phi 1}{\rm tr}(\varphi_1^\dag \varphi_1)+m^2_{\Phi 2}{\rm tr}(\varphi_2^\dag \varphi_2),
\end{aligned}
\end{equation}
where $d_{\pm }^{0} ={\tfrac{1}{\sqrt{2} }} \left(d_{1}^{0} \pm d_{2}^{0} \right)$, and lower-case ${\rm tr}$ defines a 3$\times$3 trace.  By their equations of motion, the auxiliary fields are equal to linear or quadratic functions of the scalar fields.  For example, the equations of motion for the d terms of $\mathcal{V}$ result in:
\begin{equation} \label{GrindEQ__2_2_} 
\begin{aligned}
-d_{+}^{0} &={\frac{g_{+}}{\sqrt{12} }} \left(\xi _{+} +\sum _{mF}q_{mF} \left(\phi _{mF}^{\dag }  \phi _{mF} -\tilde{\phi }_{mF} \tilde{\phi }_{mF}^{\dag } \right) \right) \\
-d_{-}^{0} &=-{\frac{g}{\sqrt{12} }}\sum _{mF}q_{mF}\left(-1\right)^{m} \left(\phi _{mF}^{\dag } \phi _{mF}^{} -\tilde{\phi }_{mF}^{} \tilde{\phi }_{mF}^{\dag } \right) \\
-d_{m}^{a} &=2g{\rm tr}\left(t^{a} \left[\varphi _{m} ,\varphi _{m}^{\dag } \right]\right) \\
&+g\sum _{F}\left(\phi _{mF}^{\dag } t^{a} \phi _{mF} -\tilde{\phi }_{mF} t^{a} \tilde{\phi }_{mF}^{\dag } \right)\,. 
\end{aligned}
\end{equation} 
The Abelian charges $q_{mF}$ for $V_\pm^{0}$ (including $d_\pm^{0}$) were defined in eq \eqref{GrindEQ__1_15_}.  Just as for the $d$ auxiliary fields, the equations of motion can also be used to derive expressions for the $f$ auxiliary fields in terms of scalar fields.   

It is assumed that the masses $m_{1F}$ and $\tilde{m}_{1F}$ are large compared to $\xi_+$.  In that case, the minimum of the scalar potential is achieved when the fundamental and conjugate scalars with an $m=1$ subscript have no vacuum expection value (vev):
\begin{equation}\label{1vev}
\left\langle \tilde{\phi }_{1F} \right\rangle =\left\langle \phi _{1F} \right\rangle =0,
\end{equation} 
where $\left\langle \phi _{mF} \right\rangle $ denotes the vev of $\phi _{mF} $.

Due to eq \eqref{GrindEQ__1_22_} along with an assumption that $m_{21}<\xi _{+}$, the $m=2$ fundamental and conjugate scalars may acquire vevs.  Following precedent from Supersymmetric QCD (SQCD) \cite{intriligator-seiberg,SQCD-Argyres}, the vevs in the $m=2$ sector are assumed to take the following form:
\begin{equation} \label{GrindEQ__2_3_} 
\begin{aligned}
\left\langle \tilde{\phi }_{2F}^{} \right\rangle ^{n} &=i\delta _{F}^{n} \bar{\tilde{\phi }}_{2F} \\
\left\langle \phi _{2F} \right\rangle _{n} &=-i\delta _{Fn} \bar{\phi }_{2F} ,  
\end{aligned}    
\end{equation} 
where an overbar on a component of a scalar field (e.g. $\bar{\phi }_{2F}^{} $ above) is used to denote the magnitude (real, positive) of the vev of that component.  In eq \eqref{GrindEQ__2_3_}, the index $n$ represents the SU(3) index of the 3-vectors $\tilde{\phi }_{2F}^{} $ and $\phi _{2F}^{} $.  For example, writing out the SU(3) ``color'' components: $\left\langle \tilde{\phi }_{23} \right\rangle =i\left(0,0,\bar{\tilde{\phi }}_{23} \right)$.  In other words, the vevs $\left\langle \tilde{\phi }_{2F}^{} \right\rangle ^{n} $ and $\left\langle \phi _{2F} \right\rangle _{n} $ form 3$\times$3 diagonal matrices in their flavor-``color'' indices.  The word ``color'' is being used here in order to make a connection with SQCD techniques, but in this model after symmetry breaking, the 3 ``color'' indices of the $m=2$ scalars will actually correspond to 2 isospin doublet indices and 1 singlet index.  The phases of the vevs are chosen to simplify fermion mass matrices in the next section.  

Unlike $m_{21}$, it is assumed that the scalar masses $m_{22}$, $\tilde{m}_{21}$ and $\tilde{m}_{22}$ are large compared to $\xi_+$.  To accommodate a nontrivial minimum in the presence of these masses, the following is assumed classically:
\begin{equation}\label{vevs1}
\bar{\phi}_{22}=\bar{\tilde{\phi}}_{21}=\bar{\tilde{\phi}}_{22}=0 \hspace{0.5cm}\textrm{classically}.
\end{equation}
In the appendix, it is argued that small vevs are generated for these fields quantum mechanically.

The following adjoint vevs are considered for the classical theory:
\begin{equation}\label{GrindEQ__2_4_} 
\begin{aligned}
\left\langle\varphi_1\right\rangle &=0\hspace{0.5cm}\textrm{classically} \\
\left\langle \varphi _{2} \right\rangle &=-i\frac{ \bar{\varphi }_{2}}{\sqrt{2} } \left(\begin{array}{ccc} {0} & {0} & {0} \\ {0} & {0} & {1} \\ {0} & {0} & {0} \end{array}\right) \\ \\
\,. 
\end{aligned}      
\end{equation}
In the next section, in light of the quantum vacuum moduli space, a different assumption will be made for $\left\langle \varphi _{1} \right\rangle $, leading to ${\rm tr}\left\langle\varphi_1^2\right\rangle\ne 0$.

Assuming $m_{211}$ and $m_{222}$ are small classically, the vev of the classical scalar potential is equal to
\begin{equation} \label{GrindEQ__2_6_} 
\begin{aligned}
2\left\langle \mathcal{V}\right\rangle &={\tfrac{1}{12}} g_{+}^{2} \left(\xi _{+} -2\bar{\phi }_{21}^{2} +\Delta\bar{\phi}_3^2 \right)^{2} \\
&+{\tfrac{1}{6}} g^{2} \left(2\bar{\phi }_{21}^{2}  - \Delta\bar{\phi}_3^2\right)^{2} 
+\tfrac{1}{4} g^{2} \left(\bar{\varphi}_2^2-\Delta\bar{\phi}_3^2\right)^{2} \\
&+2m_{21}^2\bar{\phi }_{21}^{2}+\tilde{m}_{\Phi}^2\bar{\varphi}_2^2 \\
\Delta\bar{\phi}_3^2&=\bar{\phi }_{23}^{2} -\bar{\tilde{\phi }}_{23}^{2} 
\hspace{1cm}
\tilde{m}^2_\Phi =m^2_\Phi+m^2_{\Phi 2}\,.  
\end{aligned} 
\end{equation} 

It can be seen that if $\bar{\phi}_{23}^2=\bar{\tilde{\phi }}_{23}^{2}$ but all other vevs, masses and $\xi _{+}$ vanish, then an absolute minimum of the potential is attained: $\left\langle \mathcal{V}\right\rangle=0$.  In this model, it is assumed that $\bar{\phi}_{23}$ and $\bar{\tilde{\phi }}_{23}$ are at a unification scale of $\sim10^{14}$ GeV, but all other vevs, masses and $\xi _{+}$ in eq \eqref{GrindEQ__2_6_} are at the electroweak scale ($\sim10^{2}$ GeV) or smaller.  In that case, deviations from the absolute minimum value $\left\langle \mathcal{V}\right\rangle=0$ only at appear at the electroweak scale.

Well below the unification scale, the superpotential is more complicated since the couplings for different groups run differently.  Nonetheless, by adjusting $\xi _{+}$, $\Delta\bar{\phi}_3^2$ and certain mass relations, classical minima can still be found where the electroweak scale is primarily defined by $\bar{\phi}_{21}$, and a smaller scale is defined by $\bar{\varphi}_2$.

In particular, a classical minimum exists with vevs at the following scales:
\begin{widetext}
\begin{equation} \label{GrindEQ__2_10_} 
\renewcommand{\arraystretch}{1.5} 
\begin{array}{|r|c|l|} \hline
\rm{vev} & \textrm{scale} & \mbox{Symmetry Breaking}\\ \hline
g\bar{\phi}_{23} & \sim 10^{14}\, \rm{GeV}  & \mbox{Unification: SU(3)$\times$SU(3)$\times$U(1)$\times$U(1)$\to$SU(3)$\times$SU(2)$\times$U(1)$\times$U(1)}\\
g_2\bar{\phi }_{21} & \sim 10^{2}\, \rm{GeV}  & \mbox{Electroweak: SU(3)$\times$SU(2)$\times$U(1)$\times$U(1)$\to$SU(3)$\times$U(1)$\times$U(1)}\\
g_2\bar{\varphi }_{2} & \sim 10^{0}\, \rm{GeV}  & \mbox{Bottom quark acquires mass (no gauge symmetry breaking)}\\
g_{Z^\prime}\bar{\varphi }_{2} & \sim 10^{-2}\, \rm{GeV}  & \mbox{$Z'$ acquires mass: SU(3)$\times$U(1)$\times$U(1)$\to$SU(3)$\times$U(1)}\\
\hline
\end{array} 
\end{equation}
\end{widetext}
where $g_2$ is the Weak coupling and $g_{Z^\prime}$ is the $Z'$ coupling that is driven to a very small value by an effective anomaly, as discussed in section 4.  Since $\Delta\bar{\phi}_3^2$ is at the electroweak scale, but $g\bar{\phi}_{23}$ is at the unification scale, the following vev must also be at the unification scale:
\begin{equation} \label{GrindEQ__2_11_}
g\bar{\tilde{\phi}}_{23}\sim 10^{14}  \rm{GeV}.
\end{equation}

In Supersymmetric QCD (SQCD), the classical vacuum does not determine actual values of vevs, but just differences like $\bar{\phi }_{23}^{2} -\bar{\tilde{\phi }}_{23}^{2}$ in eq \eqref{GrindEQ__2_6_}.   But for fewer flavors than colors, the SQCD quantum vacuum causes the vevs involved in those differences to get very large.  In fact, the quantum vacuum drives them to infinity, so that for fewer flavors than colors, SQCD does not have a vacuum \cite{intriligator-seiberg,SQCD-Argyres}.  In the appendix, it is pointed out that similar forces are at work in this theory.  However, they can be counterbalanced by other quantum effects or even a very small mass $m_{233}$.  If $m_{233}^2\bar{\phi }_{23}^{2}\lesssim \xi _{+}^2$, such a mass still allows large vevs for $\bar{\phi }_{23}$ and $\bar{\tilde{\phi }}_{23}$.  But it provides a counterbalance, so that the vevs $\bar{\phi}_{23}$ and $\bar{\tilde{\phi }}_{23}$ do not become infinite.  This is justification for why these two fields have very large unification scale vevs.

Section 4 of this paper determines the numerical value of the unification scale by starting at electroweak energies and running the SU(2) and SU(3) coupling constants up to the scale where they become the same.  In that section, it is argued that the U(1) coupling $g_{+} $ may also unify with the nonAbelian couplings at that same scale. 

A difference from the Standard Model in the above symmetry breaking is that there is an extra U(1) field (the $Z'$) that acquires a mass well below the electroweak scale.  This will be discussed in more detail later in this section and in section 5.  

 In the above symmetry breaking, the SU(3) gluons of the Standard Model come from the $A^a_{1\mu }$ gauge bosons, while the SU(2) weak fields come from the $A^a_{2\mu}$ gauge bosons.  The U(1) fields in the model are a mixture of $A^0_{1\mu }$ and $A^A_{2\mu }$ gauge bosons.  The progression of the U(1) fields through the various stages of symmetry breaking is discussed in detail below.  

Via the Brout-Englert-Higgs mechanism, the scalar vevs $\bar{\phi }_{23}$ and $\bar{\tilde{\phi }}_{23} $ impart unification-scale masses to $A_{2\mu }^{4} ,A_{2\mu }^{5} ,A_{2\mu }^{6}$ and $A_{2\mu }^{7} $ as well as to one diagonal gauge boson.  Consequently, the gauge symmetry is broken down to SU(3)$\times$SU(2)$\times$U(1)$\times$U(1).  To see the group structure of the remaining massless diagonal gauge fields, it is helpful to use the notation of eq \eqref{GrindEQ__1_7_} and re-expand the gauge fields into the linear combinations below:
\begin{widetext}
\begin{equation} \label{GrindEQ__2_12_} 
\left(\begin{array}{c} {A_{\mu }^{Y} } \\ {A_{\mu }^{Y'} } \\ {A_{\mu }^{U} } \end{array}\right)=\left(\begin{array}{ccc} {1} & {0} & {0} \\ {0} & {\cos \phi _{U} } & {\sin \phi _{U} } \\ {0} & {-\sin \phi _{U} } & {\cos \phi _{U} } \end{array}\right)\left(\begin{array}{ccc} {\cos \theta _{U} } & {0} & {\sin \theta _{U} } \\ {0} & {1} & {0} \\ {-\sin \theta _{U} } & {0} & {\cos \theta _{U} } \end{array}\right)\left(\begin{array}{c} {A_{+\mu }^{0} } \\ {A_{-\mu }^{0} } \\ {A_{2\mu }^{8} } \end{array}\right),    
\end{equation} 
where $A_{\mu }^{U} $ acquires a unification-scale mass, but $A_{\mu }^{Y} $ and $A_{\mu }^{Y'} $ remain massless.  In order to achieve the relation
\begin{equation} \label{GrindEQ__2_13_} 
q_{mF} g_{+} A_{+\mu }^{0} T_{+}^{0} + q_{mF}g A_{-\mu }^{0} T_{-}^{0} +gA_{2\mu }^{8} T_{2}^{8} =g_{Y} A_{\mu }^{Y} T_{mF}^{Y} +g_{Y'} A_{\mu }^{Y'} T_{mF}^{Y'} +g_{U} A_{\mu }^{U} T_{mF}^{U} ,    
\end{equation} 
appearing in the action (eq \eqref{GrindEQ__1_15_}), the coupling constants and group matrices must satisfy:
\begin{equation} \label{GrindEQ__2_14_} 
\left(\begin{array}{c} {g_{Y} T_{mF}^{Y} } \\ {g_{Y'} T_{mF}^{Y'} } \\ {g_{U} T_{mF}^{U} } \end{array}\right)=\left(\begin{array}{ccc} {1} & {0} & {0} \\ {0} & {\cos \phi _{U} } & {\sin \phi _{U} } \\ {0} & {-\sin \phi _{U} } & {\cos \phi _{U} } \end{array}\right)\left(\begin{array}{ccc} {\cos \theta _{U} } & {0} & {\sin \theta _{U} } \\ {0} & {1} & {0} \\ {-\sin \theta _{U} } & {0} & {\cos \theta _{U} } \end{array}\right)\left(\begin{array}{c} {q_{mF} g_{+} T_{+}^{0} } \\  {q_{mF} g T_{-}^{0} } \\ {gT_{2}^{8} } \end{array}\right).    
\end{equation} 
\end{widetext}
This is just a generalization of a Weinberg angle rotation.  A more complete generalization could involve a third angle mixing the two massless fields, but that is not needed here.  The reason that the group matrices on the left have an $mF$ dependence is because eq \eqref{GrindEQ__2_13_} involves the Abelian charges $q_{mF}$.  

The angles $\theta _{U} $ and $\phi _{U} $ in eq \eqref{GrindEQ__2_14_} are chosen so that $T^{Y}_{23} $ and $T^{Y'}_{23} $ have zeros in their sixth diagonal slot, so that they get no mass contribution from $\bar{\phi }_{23} $  or $\bar{\tilde{\phi }}_{23} $.  Specifically, the angles are given by:
\begin{equation} \label{GrindEQ__2_15_} 
\begin{aligned}
\tan \theta _{U} &=g_{+}/2g = \tfrac{1}{2}\\ 
\tan \phi _{U} &=-\tfrac{1}{2} \cos \theta _{U} = -\tfrac{1}{\sqrt{5}}. 
\end{aligned}        
\end{equation} 
where the second equalities above assume that $g_+=g$ at the unification scale.  In that case $g_{Y}=g_{Y^\prime}=g$, and the group matrices take the forms:
\begin{equation} \label{GrindEQ__2_16_} 
\begin{aligned}
T_{23}^{Y}=T_{22}^{Y} &={\tfrac{1}{2}} \sqrt{{\tfrac{3}{5}} } {\rm diag}\left({\tfrac{2}{3}} ,{\tfrac{2}{3}} ,{\tfrac{2}{3}} ,1,1,0\right)\\ 
T_{21}^{Y}=T_{1F}^{Y} &=-{\tfrac{1}{2}} \sqrt{{\tfrac{3}{5}} } {\rm diag}\left({\tfrac{4}{3}} ,{\tfrac{4}{3}} ,{\tfrac{4}{3}} ,1,1,2\right)
\end{aligned}     
\end{equation} 
\begin{equation} \label{GrindEQ__2_16b_} 
\begin{aligned}
T_{23}^{Y^\prime}=T_{22}^{Y^\prime} &=\tfrac{1}{\sqrt{10}} {\rm diag}\left(1,1,1,-1,-1,0\right)\\ 
T_{21}^{Y^\prime} = T_{1F}^{Y^\prime} &=\tfrac{1}{\sqrt{10}} {\rm diag}\left(-2,-2,-2,1,1,2\right)
\end{aligned}     
\end{equation} 
When acting on gaugino or adjoint fields, $T_{22}^{Y^\prime}$ should be used, and either version of $T_{mF}^{Y}$ can be used (since $T_{+}^{0}$ commutes with gauginos and adjoint fields). 

It will be seen below that the $T_{mF}^{Y} $ matrices have the correct form for their gauge boson $A_{\mu }^{Y} $ to be identified as the U(1) weak hypercharge field of the Standard Model with $-{\tfrac{1}{2}} \sqrt{{\tfrac{3}{5}} } g_{Y} $ identified as the weak hypercharge coupling.  The $A_{\mu }^{Y'} $ gauge field with its coupling $g_{Y'} $ is a second U(1) gauge boson in this model that remains massless at the unification scale.  Due to eq \eqref{GrindEQ__2_12_}, there is no mixing between the $A_{\mu }^{Y} $ and $A_{\mu }^{Y'} $ gauge fields.  Below the unification scale, the couplings for the SU(3), SU(2), $Y$ and $Y'$ groups run differently, so they are denoted by $g_{3} $, $g_{2} $, $g_{Y} $ and $g_{Y'} $.    

In this model according to eq \eqref{GrindEQ__2_10_}, electroweak symmetry is primarily broken by $\bar{\phi }_{21} $.  This vev gives masses to the $W$ and $Z$ bosons, leaving only the SU(3) gluons, the photon and the $Z'$ boson massless.  To see the structure of the diagonal fields, one may again make a Weinberg-angle rotation:
\begin{widetext}
\begin{equation} \label{GrindEQ__2_17_} 
\left(\begin{array}{c} {eT_{mF}^{\gamma } } \\ {g_{Z'} T_{mF}^{Z'} } \\ {g_{Z} T_{mF}^{Z} } \end{array}\right)=\left(\begin{array}{ccc} {1} & {0} & {0} \\ {0} & {\cos \phi _{Z} } & {\sin \phi _{Z} } \\ {0} & {-\sin \phi _{Z} } & {\cos \phi _{Z} } \end{array}\right)\left(\begin{array}{ccc} {\cos \theta _{Z} } & {0} & {\sin \theta _{Z} } \\ {0} & {1} & {0} \\ {-\sin \theta _{Z} } & {0} & {\cos \theta _{Z} } \end{array}\right)\left(\begin{array}{c} {g_{Y} T_{mF}^{Y} } \\ {g_{Y'} T_{mF}^{Y'} } \\ {g_{2} T_{2}^{3} } \end{array}\right).    
\end{equation} 
\end{widetext}
The angle $\theta _{Z} $ is chosen to make $(T_{21}^{\gamma })_{44}=(T_{22}^{\gamma })_{55}=0 $, so that the photon gets no mass from $\bar{\phi }_{21} $, $\bar{\varphi}_2$, or any of the quantum-generated scalar vevs discussed in section 3.

The angle $\phi _{Z} $ is chosen to make $(T_{21}^{Z'})_{44}=0$, so that it gets no mass from $\bar{\phi }_{21} $.  The resulting angles are given by:
\begin{equation} \label{GrindEQ__2_18_} 
\begin{aligned}
\tan \theta _{Z} &={\sqrt{{\tfrac{3}{5}} } g_{Y} / g_{2} } \\ 
\tan \phi _{Z} &=-\tfrac{2}{\sqrt{10}}\cos\theta_{Z}g_{Y'} / g_{2} \,. 
\end{aligned}       
\end{equation} 
In section 4, it will be argued that $g_{Y'}$ will be driven to a very small value from an effective anomaly.  As a result, the angle $\phi _{Z} $ is very small, and $\theta _{Z} $ is very close to the Weinberg angle $\theta _{W} $ of the Standard Model.

The photon group structure is given by
\begin{equation} \label{GrindEQ__2_20_} 
\begin{aligned}
eT_{22}^{\gamma }=eT_{23}^{\gamma } &=-e_{} {\rm diag}\left({\tfrac{1}{3}} ,{\tfrac{1}{3}} ,{\tfrac{1}{3}} ,1,0,0\right)\\
eT_{21}^{\gamma }=eT_{1F}^{\gamma } &=e_{} {\rm diag}\left({\tfrac{2}{3}} ,{\tfrac{2}{3}} ,{\tfrac{2}{3}} ,0,1,1\right)\,,
\end{aligned}       
\end{equation}
where the following coupling constant normalization is used:
\begin{equation} \label{GrindEQ__2_19_} 
e=-g_{2} \sin \theta _{Z}  \,.
\end{equation} 
For the case where  $|\phi _{Z}|\ll 1,$ the $Z$ boson group structure is approximately:
\begin{equation}\label{Zstructure}
\begin{aligned}
g_ZT_{22}^{Z} &\simeq -(g_2/\cos\theta_W) {\rm diag}\left({\tfrac{1}{3}x} ,{\tfrac{1}{3}x} ,{\tfrac{1}{3}x} ,-\tfrac{1}{2}+x,\tfrac{1}{2},0\right)\\
g_ZT_{1F}^{Z} &\simeq (g_2/\cos\theta_W) {\rm diag}\left({\tfrac{2}{3}x} ,{\tfrac{2}{3}x} ,{\tfrac{2}{3}x} ,\tfrac{1}{2},-\tfrac{1}{2}+x,x\right) \\
x&= \sin^2\theta_Z \simeq \sin^2\theta_W\,.
\end{aligned}
\end{equation}
The group structure of the $Z'$ boson is shown in eq \eqref{Zp-group} of section 5.

 Now that the weak hypercharge and electric charge have been established, it is possible to map the fermions in this model to fermions of the Standard Model.  Based on their SU(3)$\times$SU(2)$\times$U$($1$)_Y$ interactions, the fermions defined in eqs \eqref{GrindEQ__1_13_}, \eqref{GrindEQ__1_16_}, and \eqref{GrindEQ__1_19_} can be labelled:
\begin{widetext}
\begin{equation}\label{GrindEQ__2_21_} 
\begin{aligned}
&\tilde{\lambda }=\frac{1}{\sqrt{2} } \left(\begin{array}{ccc} {\tilde{u}_{W1}^{G} } & {\tilde{u}_{W2}^{G} } & {\tilde{u}_{W3}^{G} } \\ {\tilde{d}_{W1}^{G} } & {\tilde{d}_{W2}^{G} } & {\tilde{d}_{W3}^{G} } \\ {\tilde{d}_{1}^{G} } & {\tilde{d}_{2}^{G} } & {\tilde{d}_{3}^{G} } \end{array}\right) \,\,\,\, 
\lambda =\frac{1}{\sqrt{2} } \left(\begin{array}{ccc} {u_{W1}^{G} } & {d_{W1}^{G} } & {d_{1}^{G} } \\ {u_{W2}^{G} } & {d_{W2}^{G} } & {d_{2}^{G} } \\ {u_{W3}^{G} } & {d_{W3}^{G} } & {d_{3}^{G} } \end{array}\right)\\ 
&\tilde{\chi }=\frac{1}{\sqrt{2} } \left(\begin{array}{ccc} {\tilde{u}_{W1}^{A} } & {\tilde{u}_{W2}^{A} } & {\tilde{u}_{W3}^{A} } \\ {\tilde{d}_{W1}^{A} } & {\tilde{d}_{W2}^{A} } & {\tilde{d}_{W3}^{A} } \\ {\tilde{d}_{1}^{A} } & {\tilde{d}_{2}^{A} } & {\tilde{d}_{3}^{A} } \end{array}\right)  \,\,\,\,
\chi =\frac{1}{\sqrt{2} } \left(\begin{array}{ccc} {u_{W1}^{A} } & {d_{W1}^{A} } & {d_{1}^{A} } \\ {u_{W2}^{A} } & {d_{W2}^{A} } & {d_{2}^{A} } \\ {u_{W3}^{A} } & {d_{W3}^{A} } & {d_{3}^{A} } \end{array}\right)\\
&\psi _{11} =\left(\begin{array}{c} {u_{1}^{\left(1\right)} } \\ {u_{2}^{\left(1\right)} } \\ {u_{3}^{\left(1\right)} } \end{array}\right) \,\psi _{12} =\left(\begin{array}{c} {d_{1}^{\left(2\right)} } \\ {d_{2}^{\left(2\right)} } \\ {d_{3}^{\left(2\right)} } \end{array}\right) \,\psi _{13} =\left(\begin{array}{c} {d_{1}^{\left(3\right)} } \\ {d_{2}^{\left(3\right)} } \\ {d_{3}^{\left(3\right)} } \end{array}\right) 
\,\psi _{21} =\left(\begin{array}{c} {\nu_{W}^{\left(1\right)} } \\ {e_{W}^{\left(1\right)} } \\ {e^{\left(1\right)} } \end{array}\right) \,\psi _{22} =\left(\begin{array}{c} {\nu_{W}^{\left(2\right)} } \\ {e_{W}^{\left(2\right)} } \\ {e^{\left(2\right)} } \end{array}\right) \,\psi _{23} =\left(\begin{array}{c} {\nu_{W}^{\left(3\right)} } \\ {e_{W}^{\left(3\right)} } \\ {e^{\left(3\right)} } \end{array}\right)\\
&\tilde{\psi }_{11}^{T} =\left(\begin{array}{c} {\tilde{u}_{1}^{\left(1\right)} } \\ {\tilde{u}_{2}^{\left(1\right)} } \\ {\tilde{u}_{3}^{\left(1\right)} } \end{array}\right) \,\tilde{\psi }_{12}^{T} =\left(\begin{array}{c} {\tilde{d}_{1}^{\left(2\right)} } \\ {\tilde{d}_{2}^{\left(2\right)} } \\ {\tilde{d}_{3}^{\left(2\right)} } \end{array}\right) \,\tilde{\psi }_{13}^{T} =\left(\begin{array}{c} {\tilde{d}_{1}^{\left(3\right)} } \\ {\tilde{d}_{2}^{\left(3\right)} } \\ {\tilde{d}_{3}^{\left(3\right)} } \end{array}\right) \,\tilde{\psi }_{21}^{T} =\left(\begin{array}{c} {\tilde{\nu}_{W}^{\left(1\right)} } \\ {\tilde{e}_{W}^{\left(1\right)} } \\ {\tilde{e}^{\left(1\right)} } \end{array}\right) \,\tilde{\psi }_{22}^{T} =\left(\begin{array}{c} {\tilde{\nu}_{W}^{\left(2\right)} } \\ {\tilde{e}_{W}^{\left(2\right)} } \\ {\tilde{e}^{\left(2\right)} } \end{array}\right) \,\tilde{\psi }_{23}^{T} =\left(\begin{array}{c} {\tilde{\nu}_{W}^{\left(3\right)} } \\ {\tilde{e}_{W}^{\left(3\right)} } \\ {\tilde{e}^{\left(3\right)} } \end{array}\right).   
\end{aligned}
\end{equation}
\end{widetext} 
In the above labelling, lower numerical indices are fundamental-representation indices for the unbroken SU(3) group (the strong interaction).  Fermions with a ``$W$'' index interact with the $W$ boson (as members of an isodoublet). Based on the magnitude of their electric charges, up-type quarks, down-type quarks, charged leptons, and neutrinos are labelled with $u,d,e,\nu $.  All of the fermion fields are 2-component Weyl fermions with a lower, undotted spin index.  In the convention of \cite{rargurio} (which is also the Wess/Bagger \& Bilal convention of \cite{bilal}), a Weyl fermion with a lower undotted index corresponds to a right-chiral fermion that vanishes when acted on by $1-\gamma_5$ (see appendix A of \cite{2component}).    

In that convention, the $u $ and $d $ fields are right-chiral fermions with electric charges of ${\tfrac{2}{3}} $ and $-{\tfrac{1}{3}} $, respectively, so they are mapped to right-chiral quarks. The fields $\tilde{u}$ and $\tilde{d}$ are right-chiral fermions with electric charges of $-{\tfrac{2}{3}} $ and ${\tfrac{1}{3}} $, respectively, so they are mapped to Hermitian conjugates of left-chiral quarks.   One way that this model differs from the Standard Model is that some of the right-chiral quarks have a ``$W$'' index so they interact with the $W$ boson, while some of the left-chiral quarks lack that index so they do not interact with the $W$ boson.  That difference is discussed in the next section.

\section{ Masses and Mixing of Observed Particles}

This section begins by discussing quantum-generated interactions and their effect on the vacuum.  Given certain assumptions about those quantum interactions, it is shown how this model produces the observed spectrum of particle masses.  To validate the assumptions made, detailed quantum calculations would be needed, and those calculations are not performed in this paper.  Instead, a picture is sketched as to what those calculations would need to produce in order to generate measured masses and mixing.

\subsection{Confinement}

For a supersymmetric gauge theory involving a chiral superfield in the adjoint representation, it has been shown that a tree-level mass term $m_{\Phi } $ for the adjoint superfield will lead to quark confinement \cite{seiberg-witten,seiberg-duality,bilal,intriligator-seiberg}.  The duality inherent in these theories permits moving from a description in terms of strongly coupled scalars with color-electric charge to a description in terms of weakly coupled monopoles with color-magnetic charge.  A tree-level $m_\Phi$ mass term can cause the vacuum to settle on one of two configurations where the vev of the trace of the square of the adjoint superfield does not vanish $\left\langle {\rm tr}\left(\varphi _{1}^{2} \right)\right\rangle \ne 0$.  In one of those configurations, color-magnetic monopoles become massless, condense, and cause quark confinement through a dual Meissner effect \cite{seiberg-witten,seiberg-duality,bilal,intriligator-seiberg}.

The theory of this paper is not supersymmetric, since its ``superfields'' are ``twisted''.  Nonetheless, this theory does have an adjoint-representation ``twisted superfield'' with a mass $m_{\Phi } $.  Also, the scalars in this theory are in the same representation as the scalars in the corresponding supersymmetric theory, so the vacuum moduli spaces of the two theories should be similar, particularly below the scale where the $m=1$ fundamental and conjugate scalars of the theory (aka leptoquarks) get large masses (see section 5).  That being the case, it is speculated that quantum effects similar to those in the supersymmetric theory cause the following vev to form:
\begin{equation}\label{GrindEQ__3_1_} 
\left\langle {\rm tr}\left(\varphi _{1}^{2} \right)\right\rangle \ne 0 \,\,\,   \textrm{quantum mechanically}.
\end{equation}
It is further speculated that the similarity with the corresponding supersymmetric moduli space is sufficient so that the scalars in $\varphi _{1} $ form color-magnetic monopoles that become massless, condense and cause quark confinement through the dual Meissner effect.   

The vev $\left\langle\varphi_1\right\rangle$ will also generate quark mass terms of the following form (connecting gaugino and adjoint quarks):
\begin{equation}\label{conf-mass}
-\sqrt{2}g_3i\int d^4x\rm{tr}\left(\tilde{\chi}\left\langle \varphi _{1}^\dag\right\rangle\lambda-\tilde{\lambda}\left\langle \varphi _{1}^\dag\right\rangle\chi\right)+h.c.,
\end{equation}
where it has been assumed that the strong coupling $g_3$ is the appropriate coupling to use for these quark mass terms.  

There is another quantum effect pertinent to this section.  For supersymmetric theories, instantons can generate nonperturbative low-energy effective superpotential terms.  The appendix proposes that a similar effect occurs for this theory.  To accommodate these terms, the following vevs that are zero classically acquire small but nonzero vevs quantum mechanically:
\begin{equation}\label{phi22} 
\bar{\phi}_{22},\bar{\tilde{\phi}}_{21},\bar{\tilde{\phi}}_{22}\ne 0 \textrm{ quantum mechanically}.
\end{equation}
A quantum superpotential term proportional to $1/\bar{\tilde{\phi}}_{21}^2$ (see eq \eqref{Majorana-mass}) sets the scale for masses of the heavy right-handed neutrinos discussed later in this section.

\subsection{Observed boson masses}

From the symmetry breaking defined in eq \eqref{GrindEQ__2_10_}, the mass of the $W$ boson is primarily determined by $\bar{\phi }_{21}^{2}$.  In other words, at tree level:
\begin{equation} \label{GrindEQ__3_2_} 
M_{W}^{2} \simeq {\tfrac{1}{2}} g_{2}^{2} \bar{\phi }_{21}^{2}.   
\end{equation}    
Phenomenologically, the mass of the $W$ boson determines the vacuum expectation value (vev) $\bar{\phi }_{21}$, with the running coupling $g_{2}^{2} $ evaluated at the $W$ boson mass scale.   

 The $Z$ boson mass in this model is also primarily determined by $\bar{\phi }_{21}$.  Due to the form of eq \eqref{GrindEQ__2_17_}, the $Z$ boson mass in this model differs slightly from the Standard Model expression.  The mass of the $Z$ boson is:
\begin{equation} \label{GrindEQ__3_3_} 
M_{Z} \simeq {M_{W} \mathord{\left/ {\vphantom {M_{W}  \left(\cos \phi _{Z} \cos \theta _{Z} \right)}} \right. \kern-\nulldelimiterspace} \left(\cos \phi _{Z} \cos \theta _{Z} \right)} ,         
\end{equation} 
where the angles are defined by coupling constants as in eq \eqref{GrindEQ__2_18_}.  In section 5 when discussing the $Z'$ boson, $\sin \phi _{Z}$ is estimated to be on the order of $10^{-3}$ at the scale of $\sim$17 MeV.  It could be a little larger at the mass of the $Z$ boson, but should still be very small.  For that reason, the following approximation can be used for most purposes in this paper:
\begin{equation} \label{GrindEQ__3_4_} 
\begin{aligned}
\cos \theta _{Z} &\simeq \cos \theta _{W} \\
\cos \phi _{Z} &\simeq 1, 
\end{aligned}           
\end{equation} 
where $\theta _{W} $ is the Weinberg angle.

 In this model, the scalar vev $\bar{\phi }_{21}^{} $ also generates most of the mass of the observed Brout-Englert-Higgs boson, through the d-term part of the scalar potential.  The d-term has two parts: one part from terms like $\left\langle d_2^3\right\rangle d_2^3$, and the other part where each $d_2^A$ in $\tfrac{1}{2}d_2^A d_2^A$ has one vev and one Higgs field.  Those d-term contributions can be found by first re-expanding the d term part of the scalar potential of eq \eqref{GrindEQ__2_1_} using the following basis of diagonal U(3)$\times$U(3) generators:
\begin{equation} \label{GrindEQ__3_5_} 
g_{U}^{} T_{}^{U} ,g_{Y}^{} T_{}^{Y} ,g_{Y'}^{} T_{}^{Y'} ,g_{2}^{} T_{2}^{3} ,g_{3}^{} T_{1}^{8} ,g_{3}^{} T_{1}^{3} .       
\end{equation} 
The expansion of the complex scalar field $\phi _{21}^{} $ into its component fields includes the following:
\begin{equation} \label{GrindEQ__3_6_} 
i\phi _{21}^{} =\left(\bar{\phi }_{21}^{} +{\tfrac{1}{\sqrt{2} }} h_t\right)\left(1,0,0\right)^{T} +...,        
\end{equation} 
where $h_t $ is a real scalar, and $+...$ includes the fields that get ``eaten'' by the $W$ and $Z$ gauge bosons.  

In the basis of eq \eqref{GrindEQ__3_5_}, the vevs of the auxiliary d fields at the minimum of the scalar potential are approximately:  
\begin{equation}\label{dfields}
\begin{aligned}
-g_U\left\langle d_U \right\rangle &\simeq \sqrt{2}m_{21}^2\\
-g_2\left\langle d_2^3 \right\rangle &= \tfrac{1}{2} g_2^2\bar{\phi}^2_{21} \\
-g_Y\left\langle d_Y \right\rangle &=\tfrac{3\sqrt{3}}{2\sqrt{5}} g_Y^2\bar{\phi}^2_{21} \,,
\end{aligned}
\end{equation}
where it has been assumed that $g_+=g$ at the unification scale and that $g_{Y'}$ is negligible due to reasons discussed below.  The approximation has also assumed that $\bar{\phi}^2_{21}\gg \bar{\varphi}_2$.

Assuming $m_{211}$ is small, the mass of the Higgs boson (before radiative corrections) is:
\begin{equation}\label{GrindEQ__3_7_}
\begin{aligned}
M_{H}^{2} &\simeq m_{21}^2-\tfrac{1}{2}g_2\left\langle d_2^3\right\rangle+\tfrac{\sqrt{3}}{2\sqrt{5}}g_Y\left\langle d_Y \right\rangle+\tfrac{1}{3\sqrt{2}}g_U\left\langle d_U \right\rangle \\
&+2\bar{\phi }_{21}^{2} \left(g_{U}^{2} \left(T_{1}^{U} \right)_{44}^{2} +g_{Y}^{2} \left(T_{1}^{Y} \right)_{44}^{2} +g_{2}^{2} \left(T_{2}^{3} \right)_{44}^{2} \right)\\
&\simeq M_{Z}^{2}+(\tfrac{11}{18}-\tfrac{3}{2}\tan^2\theta_W)M_W^2+\tfrac{2}{3}m_{21}^2 \\
&\simeq \left(97\, {\rm GeV}\right)^{2}+\tfrac{2}{3}m_{21}^2,  
\end{aligned}
\end{equation}
where the scalar mass $m_{21}$ is from eq \eqref{GrindEQ__1_15_}.  The third line above comes from the relations in eqs \eqref{GrindEQ__2_18_}  and \eqref{GrindEQ__3_3_} along with the approximation that $g_U\simeq g_2$.  To get the correct mass squared, the mass parameter $m_{21}$ plus radiative corrections contribute $\sim$80 GeV to the Higgs boson mass. 

Just as in the Standard Model, it is assumed that radiative corrections of the Higgs boson are at a scale similar to the electroweak scale, not the unification scale.  In the Standard Model, it is an open question as to how cancellation of unification-scale corrections are able to generate electroweak-scale corrections to the Higgs mass.  By contrast, supersymmetric theories provide a simple reason for such cancellation:  superspace gauge transformations and translations can change a scalar into a fermion, and fermions do not have quadratic divergences.  Even if some part of the gauge symmetry is broken at a very large scale in a SUSY theory, as long as the superpotential remains equal to zero, the remaining superspace gauge invariance continues to ensure cancellation of quadratic divergences in the symmetry-broken theory.  

Although the theory considered here is not invariant to superspace translations, it is invariant to superspace gauge transformations that change scalars into fermions.  Also, the theory's unification-scale symmetry breaking leaves the superpotential almost equal to zero, with variations from zero only showing up at the electroweak scale.  Parallels with supersymmetry then open the door to the possibility that quadratic divergences naturally cancel in this model down to the electroweak scale.  Further work would be required to prove this.

The $h_t$ real scalar field described above has the same interactions with the $W$ boson, $Z$ boson, and top quark as does the Standard Model Higgs boson.  So this model is consistent with measurements of Higgs boson decays and interactions involving these particles, since those measurements are consistent with the Standard Model.

On the other hand, leptons and quarks (other than the top) have Higgs-like interactions with a different scalar $(\varphi_{2})_{23}$, acquiring their masses from its vev $\bar{\varphi}_{2}$.  Nonetheless, as described in section 5, this model predicts results for almost all Higgs measurements that match those of the Standard Model, and therefore also match data.  One area of difference is that this model predicts significantly smaller cross sections than the Standard Model for ttH production followed by $H\to b\bar{b}$ decay.  Current data are not inconsistent with that prediction.

\subsection{Quark masses and mixing}

Keeping in mind the particle designations of this model defined in eq \eqref{GrindEQ__2_21_}, the up-type quarks can be arranged into the following 3x3 mass matrix (and its Hermitian conjugate):
\begin{equation}\label{GrindEQ__3_14_}
\begin{aligned}
&{\rm Rows:}\,  \tilde{u}^{\left(1\right)} ,\tilde{u}_{W}^{A} ,\tilde{u}_{W}^{G} \,\,\, {\rm Columns:}\,u_{W}^{G} ,u_{W}^{A} ,u^{\left(1\right)}\\
&\mathit{M}_U =\left(\begin{array}{ccc} 
{\hat{g}\bar{\tilde{\phi }}_{21} } & {\tilde{\Gamma}_{211}\bar{\phi}_{21}} & { \tilde{m}_{211}} \\ 
{\tilde{M}_{G}^{(1)}-\Delta} & {\tilde{m}_{\Phi}} & {\tilde{\Gamma}_{211}\bar{\tilde{\phi}}_{21}} \\ 
{\tilde{m}_{\lambda }} & {\tilde{M}_{G}^{(1)}+\Delta} & {\hat{g}\bar{\phi }_{21} } 
\end{array}\right),
\end{aligned}
\end{equation}
where the gaugino coupling $\hat{g}$ is discussed below in eq \eqref{hatg}. A tilde is put on tree-level masses and superpotential couplings to show that they include quantum modifications such as those discussed in the appendix. For all of the fermion mass matrices in this paper (including the above matrix) the rows have an upper undotted spin index (using the convention of \cite{rargurio}) while the columns have a lower undotted spin index (e.g. $\tilde{u}^{\left(1\right)\alpha}u_{W\alpha}^{G}$).  Those indices are summed over, and they are suppressed.

The parameter $\tilde{M}_{G}^{(1)}$ is generated nonperturbatively from terms like eq \eqref{adjoint-gauge} in the appendix.  The mass $\Delta$ is from eq \eqref{conf-mass}. 

The first, second, and third generation up-type quarks correspond to the first, second and third rows and columns of the up-type quark matrix.  To a first approximation, the third-generation quark is just the top quark and its mass is approximately: 
\begin{equation}\label{GrindEQ__3_15_}
m_t \simeq\hat{g}\bar{\phi }_{21}\,\,\,\,\textrm{Top quark ass} .
\end{equation}

Recalling the convention from eq \eqref{GrindEQ__2_21_} that fields with a tilde are Hermitian conjugates of left-chiral quarks while those without one are right-chiral quarks, it can be seen from eq \eqref{GrindEQ__3_14_} that for the third-generation quark, only its left-chiral component interacts with the $W$ boson.  That allows this model to be consistent with top-quark polarization measurements by ATLAS \cite{top-quark-polarization}. 

The first- and second-generation quark interactions with the $W$ boson, however, differ from those of the Standard Model.  For the second-generation up-type quark (mostly charm), both its left- and right-chiral components interact with the $W$ boson.  For the first-generation up-type quark (mostly up), only its right-chiral component interacts with the $W$ boson.  This chiral flipping of the first generation is not ruled out by experiment, since the spin of the proton (and other light hadrons) comes primarily from gluons and orbital angular momentum \cite{pspin1}.  As a result, it is not possible to experimentally disentangle the spins of up and down quarks from other hadronic spins and thereby verify that only left-handed up and down quarks interact with the $W$ boson.  This is discussed further in section 5.

In eq \eqref{GrindEQ__3_2_}, it was shown that the $W$ boson mass is approximately determined by the vev $\bar{\phi }_{21} $ and the SU(2) weak coupling constant $g_{2} $.  In this model, the top quark mass is also determined by $\bar{\phi }_{21}$, but multiplied by the gaugino coupling $\hat{g}$ rather than the weak coupling $g_{2} $ (at the unification scale, these are the same).   

To get this model's prediction for the top quark mass, the calculation should be performed to determine how $\hat{g}$ runs in this model as the scale is lowered from the unification scale.  Such a calculation is outside the scope of this paper.  

Instead, a phenomenological approach is used.  Relations for the $W$ boson and top quark masses require 
\begin{equation}\label{hatg}
\hat{g} \simeq \frac{m_t}{\sqrt{2}M_W} g_2 \simeq 1.5 g_2\simeq 0.8 g_3 ,
\end{equation}
where $g_2$ and $g_3$ are evaluated at the $Z$ boson mass scale.  It is presumed that the gaugino coupling $\hat{g}$ of this model takes the above value at the $Z$ boson mass scale.
 
From eq \eqref{GrindEQ__2_21_}, it can be seen that there are twice as many flavors of down-type quarks in this model $(6L\times6R)$ as there are up-type quarks $(3L\times3R)$.  However, the unification scale vevs generate unification-scale masses for $\tilde{d}^{G}d^{\left(3\right)}$ and $\tilde{d}^{\left(3\right)}d^{G}$, so those down-type quarks decouple.  The remaining 4x4 down-type quark mass matrix has the following structure:
\begin{equation} \label{GrindEQ__3_18_}
\begin{aligned} 
&{\rm Rows:}\,\tilde{d}^{(2)}, \tilde{d}_{W}^{A} ,\tilde{d}_{W}^{G} ,\tilde{d}^{A} \,\,\,  {\rm Columns:}\,d_{W}^{G}  ,d^{(2)},d^{A}, d_{W}^{A}  \\
&M_D=
\left(\begin{array}{cccc} 
{\hat{g}\bar{\tilde{\phi}}_{22}} & {\tilde{m}_{222}} & {...} &   {{\tilde{\Gamma}_{222}\bar{\phi}_{22}}}  \\
{...} & {{\tilde{\Gamma}_{222}\bar{\tilde{\phi}}_{22}}} & ... &  {\tilde{m}'_{\Phi} } \\ 
{\tilde{m}'_{\lambda}  } & {\hat{g}\bar{\phi }_{22}} & {\hat{g}\bar{\varphi}_{2}} &  {...} \\  
{...} & {...} & {\tilde{m}''_{\Phi }} &  {\tfrac{1}{\sqrt{2}}|\tilde{\Gamma}_{\Phi}|\bar{\varphi}_{2}}
\end{array}\right),
\end{aligned}
\end{equation}
where only tree-level couplings are shown explicitly.  Tildes again indicate that these couplings can be modified by quantum-generated couplings discussed in the appendix.  Quantum couplings can also be generated for the components without tree-level contributions; those are denoted by ``$...$''.  Primes and double primes on $\tilde{m}'_{\Phi}$ and $\tilde{m}'_{\lambda}$ are used to show that these masses in the down-type matrix do not need to be the same as those in the up-type matrix. 

To a first approximation, the masses of the $d$, $s$, and $b$ quarks as well as an additional $f$ quark are  given by the diagonal elements of eq \eqref{GrindEQ__3_18_}.  For example, $m_b\simeq\hat{g}\bar{\varphi}_{2}$.  Comparison with the top quark mass sets the value for this vev at $\bar{\varphi}_{2}\simeq (m_b/m_t)\bar{\phi }_{21}$.  The $f$ quark mass is also generated by $\bar{\varphi}_{2}$, but with a superpotential coupling.  None of the down-type quarks get their masses from the vev $\bar{\phi }_{21}$ that gives mass to the top quark.

One model for the numerical values of the quark mass matrices (in GeV) is to take them to be the following:
\begin{equation}\label{3x3fit}
\begin{aligned}
&{\rm Rows:}\,  \bar{u}'_L ,\bar{c}'_L ,\bar{t}'_L \,\,\, {\rm Columns:}\,u'_R ,c'_R ,t'_R\\
&\mathit{M}_U =\left(\begin{array}{ccc} 
{0} & {0.0198} & {0.7302 } \\ 
{-0.1309} & {1.2736} & {0} \\ 
{-0.4147} & {4.0973} & {171.95 }  \end{array}\right).
\end{aligned}
\end{equation}
\begin{equation} \label{4x4fit}
\begin{aligned} 
&{\rm Rows:}\, \bar{d}'_L,\bar{s}'_L,\bar{b}'_L,\bar{f}'_L\,\,\,  {\rm Columns:}\,d'_R,s'_R,b'_R,f'_R  \\
&\mathit{M}_D =\left(\begin{array}{cccc} 
{.0021} & {-0.0164} & {-0.0664 } & {-0.0193} \\ 
{0.0303} & {0.0759} & {0.1792} & {-0.0788} \\ 
{-0.0414} & {-0.3295} & {4.1217} & {0.3959} \\ 
{0.2953} & {-0.0532} & {0.5825} & {-2.8053 }   \end{array}\right).
\end{aligned}
\end{equation}
The rows and columns of these matrices are the same gauge eigenstates as in eqs \eqref{GrindEQ__3_14_} and \eqref{GrindEQ__3_18_}; they have just been renamed to reflect their dominant mass eigenstate.  For example,
\begin{equation}\label{sprime} 
\tilde{d}_W^A= \bar{s}'_L= .97 \bar{s}_L + .22 \bar{d}_L  + .04 \bar{b}_L +.03 \bar{f}_L\,,
\end{equation}
where $\bar{s}_L$ is in the mass eigenbasis (see below) and $\bar{q}_L$ denotes the right-handed field $(q_L)^\dag$.

The numerical values of these quark mass matrices were chosen by a fit to quark masses and CKM data, while not being in conflict with $Z$, $Z'$, and exotic hadron data. The two elements in $\mathit{M}_U$ that are proportional to $\bar{\tilde{\phi}}_{21}$ were set to zero to make them compatible with the large neutrino masses discussed in eq \eqref{Majorana-mass} of the appendix.  In section 5, there is a discussion about how the theory is able to reproduce the many experiments that would seemingly rule out the possibility of an additional light quark. As more data is collected, it may be that rotations of these quark matrices that still reproduce quark masses and CKM data may provide a better global fit.  

Just as in the Standard Model, each quark mass matrix can be diagonalized via a unitary matrix $V$ on each side:
\begin{equation} \label{GrindEQ__3_23_}
\begin{aligned}
&V_L^{U} \mathit{M}_U V_R^{U\dag } = {\rm diag}(m_u,m_c,m_t)\\
&V_L^{D} \mathit{M}_D V_R^{D\dag } = {\rm diag}(m_d,m_s,m_b,m_f)\,,
\end{aligned}
\end{equation}
where the subscript $f$ is used to denote the fourth down-type quark. The unitary matrices that diagonalize the quark mass matrices can be found by first multiplying each mass matrix on the left or on the right by its transpose, then finding the eigenvectors of those product matrices.  The allowed flexibility to introduce complex phases was not used for the fits of this paper, so no attempt was made to fit the experimentally measured complex phases of the CKM matrix.  The mass eigenvalues of the matrices in eqs \eqref{3x3fit} and \eqref{4x4fit} are:
\begin{equation}\label{qmasses}
\begin{aligned}
&m_u =0.002,\,\,\, m_c =1.28,\,\,\,m_t =172\\
&m_d =0.005,\,\,\,m_s =0.095,\,\,\,m_b =4.18,\,\,\,m_f =-2.85\,,
\end{aligned}
\end{equation}
where all mass values are in GeV.   

In the Standard Model, the CKM matrix displays the connections that the $W$ boson makes between up-type and down-type quark mass eigenstates.  Since there are 3 quarks of each type in the Standard Model, the CKM matrix is a $3\times 3$ matrix.  In this model, there are 3 up-type and 4 down-type quarks, so the equivalent ``CKM'' matrix is a $3\times 4$ matrix.  Also, in the Standard Model only the left-handed quarks (and hence $V_L^{U,D} $) matrices have connections to the $W$ boson, but in this model, all four of the matrices have connections to the W.

From the diagonalizing matrices, one may construct the following 2 versions of CKM matrices:
\begin{equation} \label{GrindEQ__3_24_} 
\begin{aligned}
V_{\rm{CKM}}^{\pm } &=V_{R}^{U} \left(\begin{array}{cccc} 
{1} & {0} & {0} & {0} \\ 
{0} & {0} & {0} & {1}  \\ 
{0} & {0} & {0} & {0} 
\end{array}\right)V_{R}^{D\dag } \\
&\pm V_{L}^{U} \left(\begin{array}{cccc} 
{0} & {0} & {0} & {0} \\ 
{0} & {1} & {0} & {0} \\ 
{0} & {0} & {1} & {0} \end{array}\right)V_{L}^{D\dag }  
\end{aligned}
\end{equation} 
The placement of the 1's in the above matrices is based on which quark fields in eqs \eqref{GrindEQ__3_14_} and \eqref{GrindEQ__3_18_} have a $W$ subscript (signifying that they interact with the $W$ boson to change an up-type quark to a down-type quark and vice versa).

In this model, a different CKM matrix should be used depending on whether a vector current or axial vector current process is being considered. Specifically:
\begin{equation} \label{GrindEQ__3_25_} 
\begin{aligned} 
\textrm{Vector current decays: }\,&V_{\rm{CKM}}^{+}\\
\textrm{Axial vector current decays: }\,&V_{\rm{CKM}}^{-}.
\end{aligned}
\end{equation}
For CKM measurements involving vector current decays (e.g. an exclusive semi-leptonic decay from one spin-0 meson to a different spin-0 meson), $V_{\rm{CKM}}^{+} $ should be used for comparison to this model.  For CKM measurements involving axial vector current decays (e.g. the purely leptonic decay of a spin-0 pseudo-scalar meson), $V_{\rm{CKM}}^{-} $ should be used.  

Plugging in the unitary matrices $V$ found from eq \eqref{GrindEQ__3_23_} into eq \eqref{GrindEQ__3_24_}, the following CKM matrices are obtained:
\begin{equation} \label{GrindEQ__3_28_} 
\begin{aligned}
&\left|V_{\rm{CKM}}^{+} \right| =\left(\begin{array}{cccc} 
{0.9740} & {0.2231} & {0.0040} & {0.0021}\\ 
{0.2240} & {0.9732} & {0.0394} & {1.0282}\\ 
{0.0079} & {0.0394} & {0.9890} & {0.1183} \end{array}\right) \\
&\left|V_{\rm{CKM}}^{-} \right| =\left(\begin{array}{cccc} 
{0.9749} & {0.2261} & {0.0046} & {0.0032}\\ 
{0.2182} & {0.9747} & {0.0429} & {0.9712}\\ 
{0.0077} & {0.0394} & {0.9891} & {0.1662} \end{array}\right).  
\end{aligned}        
\end{equation} 

Comparing the first three columns of the above matrices to data presented in \cite{CKM}, it can be seen that despite having both left- and right-handed $W$ boson connections for quarks, the model does a good job of reproducing absolute values of most CKM data.  In fact, comparing to data in \cite{CKM-Leptonic}, it can be seen that the model even does a good job of reproducing the perplexing 3$\sigma$ difference seen between vector- and axial-vector-current data for $|V_{us}|$.  As noted in \cite{Crivellin_2024}, such a difference cannot arise in a model where all quark interactions with the $W$ boson are left-handed.

In this model, to be consistent with the $W$-boson interactions, the $Z$-boson interactions can be Flavor Changing Neutral Currents (FCNCs).  The couplings to the $Z$ boson by quark generation are:
\begin{equation} \label{Zuu} 
\renewcommand{\arraystretch}{1.5} 
\begin{array}{|c|c|c|} \hline
\bar{u}_i \gamma^\mu u_i Z_\mu& g_L & g_R\\ \hline
u' & -\tfrac{2}{3}x &\tfrac{1}{2}-\tfrac{2}{3}x\\
c' &\tfrac{1}{2}-\tfrac{2}{3}x  &\tfrac{1}{2}-\tfrac{2}{3}x \\
t' &\tfrac{1}{2}-\tfrac{2}{3}x  &-\tfrac{2}{3}x\\\hline
{\rm SM} & \tfrac{1}{2}-\tfrac{2}{3}x & -\tfrac{2}{3}x\\
\hline
\end{array} 
\end{equation} 
\begin{equation} \label{Zdd} 
\renewcommand{\arraystretch}{1.5} 
\begin{array}{|c|c|c|} \hline
\bar{d}_i \gamma^\mu d_i Z_\mu & g_L & g_R\\ \hline
d' & \tfrac{1}{3}x &-\tfrac{1}{2}+\tfrac{1}{3}x\\
s' &-\tfrac{1}{2}+\tfrac{1}{3}x  &\tfrac{1}{3}x \\
b' &-\tfrac{1}{2}+\tfrac{1}{3}x  &\tfrac{1}{3}x \\
f' & \tfrac{1}{3}x  &-\tfrac{1}{2}+\tfrac{1}{3}x\\\hline
{\rm SM} & -\tfrac{1}{2}+\tfrac{1}{3}x & \tfrac{1}{3}x\\
\hline
\end{array} \,,
\end{equation} 
where $x=\sin^2\theta_W$ and the ``SM'' rows show the Standard Model couplings for up-type and down-type quarks.  The primed quark variables are gauge eigenstates, just as in eqs \eqref{3x3fit} and \eqref{4x4fit}.

Since the couplings of this model differ by generation, $Z$ boson connections mix mass eigenstates.  For example, the Z mixing matrix for left-chiral down-type quarks is:
\begin{equation}\label{ZmixDL}
\begin{aligned}
V_{ZD}^{L}&= V_L^D\left(\begin{array}{cccc} 
{\tfrac{1}{3}x} & {0} & {0} & {0}\\ 
{0} & {-\tfrac{1}{2}+\tfrac{1}{3}x} & {0} & {0} \\ 
{0} & {0} & {-\tfrac{1}{2}+\tfrac{1}{3}x} & {0} \\ 
{0} & {0} & {0} & {\tfrac{1}{3}x}
\end{array}\right)V_L^{D\dag} \\
&=\left(\begin{array}{cccc} 
{-0.0941} & {-0.0743} & {-0.0058} & {-0.0018} \\ 
{-0.0743} & {-0.4056} & {-0.0004} & {-0.0115} \\
{-0.0058} & {-0.0004} & {-0.4159} & {0.0482} \\
{-0.0018} & {-0.0115} & {0.0482} & {-0.0844}
\end{array}\right) ,
\end{aligned}
\end{equation}
where the diagonalizing matrix $V_L^D$ is the same as was used for CKM matrices, and $x=\sin^2\theta_W=0.2315$ was used.  There are also $Z$ mixing matrices for right-chiral down-type quarks as well as for right- and left-chiral up-type quarks. 

Most of the off-diagonal FCNC elements of these matrices are small, but a few are more significant.  The only Z mixing matrix elements with magnitudes greater than 0.003 are $V_{Zs\bar{d}}^A,V_{Zf\bar{b}}^L,V_{Zf\bar{s}}^L,V_{Zb\bar{d}}^L$, and $V_{Zt\bar{c}}^R$. More detailed work would be needed to verify that the $Z$ boson FCNC mixing of this model is fully consistent with experimental data.

\subsection{Neutrinos}

In addition to three left-handed neutrinos $\nu_W^{\dag(F)}$, this model also has three right-handed heavy neutrinos $\tilde{\nu}_W^{(F)}$.  From eq \eqref{Majorana-mass} of the Appendix, quantum effects generate a mass matrix of the form
\begin{equation} \label{heavy-neutrino} 
\tfrac{1}{3}M_\nu^{(2)} \left(\begin{array}{ccc} 
{1} & {1} & {1} \\ 
{1} & {1} & {1} \\
{1} & {1} & {1}
\end{array}\right)\,,
\end{equation} 
where $M_\nu^{(2)}$ is a very large mass, and both the rows and columns are $\tilde{\nu}_W^{(1)},\tilde{\nu}_W^{(2)},\tilde{\nu}_W^{(3)}$.  A flavor basis rotation can diagonalize this matrix to
\begin{equation} \label{heavy-neutrino2} 
\left(\begin{array}{ccc} 
{0} & {0} & {0} \\ 
{0} & {M_\nu^{(2)} } & {0} \\
{0} & {0} & {0}
\end{array}\right)\,.
\end{equation} 
As described in eq \eqref{RHneutrino-scale}, the upper limit for this mass is $\sim$5200 TeV.  

A flavor rotation that accomplishes this diagonalization is given by:
\begin{equation} \label{rotation} 
U_\nu=
\left(\begin{array}{ccc} 
{\tfrac{2}{\sqrt{6}}} & {\tfrac{1}{\sqrt{3}}} & {0} \\ 
{\tfrac{-1}{\sqrt{6}}} & {\tfrac{1}{\sqrt{3}}} & {-\tfrac{1}{\sqrt{2}}} \\
{\tfrac{-1}{\sqrt{6}}} & {\tfrac{1}{\sqrt{3}}} & {\tfrac{1}{\sqrt{2}}}
\end{array}\right)
\left(\begin{array}{ccc} 
{\cos\theta_\nu} & {0} & {\sin\theta_\nu} \\ 
{0} & {1} & {0} \\
{-\sin\theta_\nu} & {0} & {\cos\theta_\nu}
\end{array}\right)
\end{equation}
for any value of $\theta_\nu$.

At a lower scale in an effective theory that excludes the heaviest neutrino, the same quantum mechanism generates the following new mass term in the rotated basis: 
\begin{equation} \label{heavy-neutrino4} 
\left(\begin{array}{ccc} 
{\tfrac{1}{2}M_\nu^{(1)}} & {0} & {\tfrac{1}{2}M_\nu^{(1)}} \\ 
{0} & {M_\nu^{(2)} } & {0} \\
{\tfrac{1}{2}M_\nu^{(1)}} & {0} & {\tfrac{1}{2}M_\nu^{(1)}}
\end{array}\right)\,.
\end{equation} 
Diagonalization of this can be accomplished via a modified choice of $\theta_\nu$. At an even lower scale, the quantum mechanism generates the third heavy neutrino mass $M_\nu^{(3)}$.  The diagonalized heavy neutrino mass matrix is then:
\begin{equation} \label{heavy-neutrino3} 
\left(\begin{array}{ccc} 
{M_\nu^{(1)}} & {0} & {0} \\ 
{0} & {M_\nu^{(2)} } & {0} \\
{0} & {0} & {M_\nu^{(3)}}
\end{array}\right)\,,
\end{equation}
where $M_\nu^{(2)}>M_\nu^{(1)}>M_\nu^{(3)}$.

Since $\theta_\nu$ can be anything, it is valid to make the choice $\sin\theta_\nu\simeq 1/\sqrt{48}$ for the full rotation.  In that case
\begin{equation}
U_\nu\simeq
\left(\begin{array}{ccc} 
{0.81} & {0.58} & {0.12} \\ 
{-0.30} & {0.58} & {-0.76} \\
{-0.51} & {0.58} & {0.64}
\end{array}\right)\,.
\end{equation}
This rotation is similar to the observed Pontecorvo-Maki-Nakagawa-Sakata (PMNS) neutrino mixing matrix, using sin$\theta_{12}=\sqrt{.309}$, sin$\theta_{13}=-\sqrt{.0222}$, sin$\theta_{23}=-\sqrt{.573}$, and $\delta_{CP}=\pi$ \cite{neutrino_review}.  That matrix is:
\begin{equation} \label{PMNS} 
U_{\rm PMNS}=
\left(\begin{array}{ccc} 
{0.82} & {0.55} & {0.15} \\ 
{-0.27} & {0.61} & {-0.75} \\
{-0.50} & {0.58} & {0.65}
\end{array}\right)\,.
\end{equation}
The similarity of $U_\nu$ to $U_{\rm PMNS}$ is discussed below.

From eq \eqref{quant-mass} in the Appendix, quantum effects generate additional Dirac-mass terms that connect right- and left-handed neutrinos as follows:
\begin{equation}\label{neutDiracmass}
\left(\begin{array}{c} {\tilde{\nu}_{W}^{(1)} } \\ {\tilde{\nu}_{W}^{(2)} } \\ {\tilde{\nu}_{W}^{(3)} } \end{array}\right)^T
\tfrac{1}{9}m_\nu^{\prime}\left(\begin{array}{ccc} {-5} & {4} & {4} \\ {4} & {-5} & {4} \\ {4} & {4} & {-5} \end{array}\right)
\left(\begin{array}{c} {\nu_{W}^{(1)} } \\ {\nu_{W}^{(2)} } \\ {\nu_{W}^{(3)} } \end{array}\right)\,.
\end{equation}
To be consistent with the heavy neutrino basis of eq \eqref{heavy-neutrino3}, the $\tilde{\nu}_{W}^{(F)}$ neutrinos of eq \eqref{neutDiracmass} must be rotated to that basis as follows:
\begin{equation}
\tilde{\nu}_{W}^{(F)}= \tilde{\nu}_{W}^{\prime(F^\prime)}(U_\nu^\dag)_{F^\prime F}\,.
\end{equation}
If the left-handed neutrinos experience the same rotation,
\begin{equation}\label{rotateL}
\nu_{W}^{(F)}=(U_\nu)_{FF^\prime}\nu_{W}^{\prime(F^\prime)}\,,
\end{equation}
then the Dirac-mass matrix is diagonalized to
\begin{equation}\label{neutDiracmass2}
\left(\begin{array}{c} {\tilde{\nu}_{W}^{\prime (1)} } \\ {\tilde{\nu}_{W}^{\prime (2)} } \\ {\tilde{\nu}_{W}^{\prime (3)} } \end{array}\right)^T
m_\nu^{\prime}\left(\begin{array}{ccc} {-1} & {0} & {0} \\ {0} & {\tfrac{1}{3}} & {0} \\ {0} & {0} & {-1} \end{array}\right)
\left(\begin{array}{c} {\nu_{W}^{\prime (1)} } \\ {\nu_{W}^{\prime (2)} } \\ {\nu_{W}^{\prime (3)} } \end{array}\right)\,.
\end{equation}

In this diagonal basis, each flavor of neutrino forms its own seesaw to generate a small left-handed neutrino mass.  The magnitudes of these masses are
\begin{equation}\label{neut-masses}
m_\nu^{(1)} = \frac{(m_\nu^{\prime})^2}{M_\nu^{(1)}}\hspace{.5 cm}
m_\nu^{(2)} = \frac{(\tfrac{1}{3} m_\nu^{\prime})^2}{M_\nu^{(2)}}\hspace{.5 cm}
m_\nu^{(3)} = \frac{(m_\nu^{\prime})^2}{M_\nu^{(3)}}\,.
\end{equation}
To reproduce observations, the quantum-generated masses $m_\nu^{\prime}$ and $M_\nu^{(F)}$ would have to have values that produce the correct differences between masses squared of the above left-handed neutrinos \cite{neutrino_review}.  As one example, in a variant of normal neutrino mass ordering where $(m_\nu^{(1)})^2-(m_\nu^{(2)})^2=7.42\times 10^{-5}{\rm eV}^2$, the correct mass differences are generated by $m_\nu^{\prime}\simeq$ 1.0 MeV, $M_\nu^{(2)}\simeq$ 1800 TeV, $M_\nu^{(1)}\simeq$ 120 TeV, and $M_\nu^{(3)}\simeq$ 20 TeV. For another example, the correct differences are also generated by $m_\nu^{\prime}\simeq$ 0.1 MeV, $M_\nu^{(2)}\simeq$ 100 TeV, $M_\nu^{(1)}\simeq$ 1.2 TeV, and $M_\nu^{(3)}\simeq$ 200 GeV.   

From eq \eqref{rotateL}, $U_\nu$ rotates from the left-handed neutrino mass basis to the flavor basis of this paper.  In the next section, it will be argued that the flavor basis of this paper is similar to the charged lepton mass basis.  $U_{\rm PMNS}$ is the experimentally measured rotation from the left-handed neutrino mass basis to the charged lepton mass basis.  Since $U_\nu$ in this model is similar to $U_{\rm PMNS}$, it is suggested that the mechanisms presented here are a good first approximation for the origin of the observed PMNS rotation of the neutrino flavor basis.

\subsection{Charged Leptons}

From eq \eqref{GrindEQ__2_20_}, the last three diagonal components of the photon field when acting on leptons are $e_{} \left(0,1,1\right)$.  As a result, if the left- and right-chiral components of the three flavors of normal light charged leptons ($l$) are mapped to $l_L^{(F)}=e_{W}^{\dag(F)}$ and $l_R^{(F)}=\tilde{e}^{(F)}$, then those leptons have the correct electric charge of $-1$ as well as the correct isodoublet and isosinglet designations.  These light leptons get their masses from seesaw mechanisms together with this model's additional heavy charged leptons discussed below.

This model has three flavors of heavy charged leptons that are called ``Omega'' leptons.  The three flavors of heavy and light leptons ($\Omega$ and $l$) are identified as follows:
\begin{equation}\label{3.9c}
    \begin{array}{ccc}
{\nu_W^{(F)}=\nu_l^{\dag(F)}\hspace{3em}} &
{e_W^{(F)}=l_{L}^{\dag(F)} \hspace{3em}} &
{e^{(F)}= \Omega_L^{\dag(F)}} \\
{\tilde{\nu}_W^{(F)}=\nu_{\Omega}^{(F)} \hspace{3em} }& 
{\tilde{e}_W^{(F)}=\Omega_R^{(F)} \hspace{3em}} &
{\tilde{e}^{(F)}= l_{R}^{(F)},}
    \end{array}
\end{equation}
From this identification, it can be seen that the negatively charged light leptons have the correct behavior that only their left-handed components interact with the $W$ boson, forming isodoublets with their neutrinos.  On the other hand, the opposite is true for negatively charged Omega leptons; their $W$ boson interactions are right-handed.  The light leptons not only have the same interactions with the photon and $W$ boson as in the Standard Model, they also have the same interactions with the $Z$ boson, as can be seen from eq \eqref{Zstructure}. 

This model's charged leptons have a $6\times 6$ mass matrix of the form:
\begin{equation}\label{leptonmass}
\left(\begin{array}{c} {\Omega_R^{(F)} } \\ {l_R^{(F)} } \end{array}\right)^T
\left(\begin{array}{cc} {\bar{\varphi}_2\tilde{\Gamma}_{1FF^\prime}} & {\tilde{m}_{WFF^\prime}^{\prime}}  \\ {\tilde{m}_{SFF^\prime}^{\prime}} & {0}   \end{array}\right)
\left(\begin{array}{c} {\Omega_L^{\dag(F^\prime)} } \\ {l_L^{\dag(F^\prime)} }  \end{array}\right)\,.
\end{equation}
The couplings and masses in the above matrix could have both tree-level and quantum-generated contributions.  As shown in eq \eqref{quant-mass2} of the appendix, the quantum-generated contribution to $\tilde{m}_{WFF^\prime}^{\prime}$ is diagonal in flavor (only nonzero for $F'=F$). To a first approximation, the whole matrix will be assumed to be diagonal in flavor, becoming three separate seesaw matrices for the three flavors of charged leptons.

In this approximation, the masses of the three flavors of heavy and light leptons are given by:
\begin{equation}\label{lept-masses}
m_\Omega^{(F)} \simeq \bar{\varphi}_2\tilde{\Gamma}_{1FF} \hspace{.5cm}
m_l^{(F)} \simeq \frac{\tilde{m}_{WFF}^{\prime}\tilde{m}_{SFF}^{\prime}}{\bar{\varphi}_2\tilde{\Gamma}_{1FF}}\,.
\end{equation}
As discussed in the anomaly part of section 4, the masses of the Omega leptons should be in similar ranges to those of their partner right-handed neutrinos.  If they were equal to the second neutrino mass example above, the Omega lepton masses could be $m_\Omega^{(1)}\simeq$ 100 TeV, $m_\Omega^{(2)}\simeq$ 1.2 TeV, and $m_\Omega^{(3)}\simeq$ 200 GeV.  One requirement is that the Omega lepton masses must be larger than the lower limit of 103 GeV established in \cite{heavy-lepton}.

Two of the lepton gauge eigenstates from eq \eqref{leptonmass} can be expressed in terms of lepton mass eigenstates as follows:
\begin{equation}\label{lepteigenvec}
\begin{aligned}
{\Omega_R^{(F)} } &\simeq \Omega_{mR}^{(F)} + (\tilde{m}_{SFF}^{\prime}/\bar{\varphi}_2\tilde{\Gamma}_{1FF})l_{mR}^{(F)} \\
{\Omega_L^{\dag(F)} } &\simeq \Omega_{mL}^{\dag(F)} - (\tilde{m}_{WFF}^{\prime}/\bar{\varphi}_2\tilde{\Gamma}_{1FF})l_{mL}^{\dag(F)}\,,
\end{aligned}
\end{equation}
where a subscript ``$m$'' denotes a mass eigenstate. 

By defining the propagating part $h_b$ of the scalar $(\varphi_2)_{32}$ as follows
\begin{equation}\label{varphi32}
{\rm Im}((\varphi_2)_{32}) = \bar{\varphi}_2/\sqrt{2} + \tfrac{1}{2} h_b\,,
\end{equation}
it can be seen that eq \eqref{leptonmass} also implies an interaction between the scalar $h_b$ and the lepton states $\Omega_R^{(F)}$ and $\Omega_L^{\dag(F)}$.  Then using eq \eqref{lepteigenvec}, it can be seen that this model predicts $h_b$ decay to the standard light lepton mass eigenstates:
\begin{equation}\label{hbtoleptons}
h_b \to l_{mL}^{\dag(F)}l_{mR}^{(F)} \propto \frac{\tilde{m}_{WFF}^{\prime}\tilde{m}_{SFF}^{\prime}}{\bar{\varphi}_2^2\tilde{\Gamma}_{1FF}}\simeq \frac{m_l^{(F)}}{\bar{\varphi}_2}\simeq \frac{m_l^{(F)}}{m_b}\,,
\end{equation}
where eq \eqref{lept-masses} was used.

\section{Unification and Anomalies}

As mentioned in section 1, at a scale above any symmetry breaking, the Twisted Superfields theory is free of anomalies.  The reason is simple: for every fermion in every representation, there is another fermion in a conjugate representation of each of the SU(3) groups that also has opposite charges for the Abelian fields $A_{+\mu}^0$ and $A_{-\mu}^0$.  Consequently, all of the chiral gauge anomaly triangle diagrams cancel -- just as they do for SQCD. 

Another way to describe anomaly cancellation is the following:  Because of the theory's fundamental-conjugate structure, it is possible to pair up 2-component fermions into Dirac fermions such that all fermion gauge interactions are vector currents (no axial vector currents).  For example, vector-current $W$ boson interactions connect ($c_L$, $c_R$) and ($s_L$, $f_R$) as well as ($t_L$, $u_R$) and ($b_L$, $d_R$).  In a model with only vector currents, there are no chiral gauge anomalies. 

Below certain energies, mass eigenstates of fermions become better descriptions than the above-described Dirac pairings.  When fermions are put into mass eigenstates, the theory has axial-vector currents connecting them.  In other words, axial-vector currents (and gauge anomalies) are not a structural part of the theory; they are just an effect of mass at low energies.   

The unification scale $\Lambda_U$ of this model is defined as the scale where the vevs $\bar{\phi}_{23}$ and $\bar{\tilde{\phi}}_{23}$ break the original SU(3)$\times$SU(3)$\times$U(1)$\times$U(1) symmetry down to SU(3)$\times$SU(2)$\times$U(1)$\times$U(1).  Below the unification scale, the four couplings $g_3$, $g_2$, $g_Y$ and $g_{Y'}$ for these four groups run differently. As mentioned in section 3, this symmetry breaking causes the quark pairs $\tilde{d}^{G}d^{\left(3\right)}$ and $\tilde{d}^{\left(3\right)}d^{G}$ to acquire unification-scale masses. These heavy fermions are in conjugate representations of the remaining SU(3) group, are SU(2) singlets, and have equal and opposite charges for the Abelian $Y$ and $Y'$ fields.  Consequently, the effective theory below the unification scale that ignores these heavy quarks is also free of local gauge anomalies.

Below the unification scale $\Lambda_U$, the next lower scale is defined by the masses $M_{\nu}^{(2)}$ and $m_\Omega^{(1)}$ of the heaviest right-handed neutrino and charged Omega lepton that were discussed in the last section. At a scale $\mu$ between $\Lambda_U$ and $M_{\nu}^{(2)}$, the running coupling constants for the SU(N) groups obey the following equation:
\begin{equation} \label{GrindEQ__4_1_} 
\begin{aligned}
&\frac{4\pi }{g_{N}^{2} (\mu)} =\frac{4\pi }{g_{N}^{2} (\Lambda_U )} -\frac{b_{N} }{2\pi } \ln \left(\frac{\mu}{\Lambda_U} \right)\\ 
&b_{N} =\left(-\frac{11}{3} N+\frac{1}{3} n_{f} +\frac{1}{6} n_{s} +\frac{2}{3} Nn_{fA} +\frac{1}{3} Nn_{sA} \right), 
\end{aligned}     
\end{equation} 
where $n_{f} $ and $n_{s} $, $n_{fA} $ and $n_{sA} $ are the numbers of fundamental fermion and scalar N-tuplets, and the numbers of fermion and scalar adjoint representations, respectively.  In all cases, the fermions are 2-component fermions and the scalars are complex.  

In this model, $n_{fA} =0$ and $n_{sA} =1$ for each SU(N) group.  The particle content to use between $\Lambda_U$ and $M_{\nu}^{(2)}$ is the following:
\begin{enumerate}
\item  Up quarks: 4 W triplets and 2 non-W triplets
\item  Down quarks:  4 W triplets and 4 non-W triplets 
\item  Charged leptons: 6 W and 6 non-W 
\item  Neutrinos: 6 W
\item  SU(3) scalars: 6 triplets 
\item  SU(2) scalars: 6 doublets (2x flavors 1 \& 2, 2x adjoint)
\item  Singlet scalars: 6 (2x flavors 1 \& 2, 2x adjoint)\,.
\end{enumerate}
\begin{equation}\label{GrindEQ__4_2_}
\end{equation}
In this notation, triplet means an SU(3) color triplet, and a ``W'' fermion is part of an isodoublet that interacts with the $W$ boson.  

The scalars fall into these categories for the following reasons:  The $m=2$ flavor 3 triplets get eaten (or made massive) by the unification-scale symmetry breaking.  Of the 9 components of the $m=2$ U(3) adjoint scalars, 3 are an SU(2) adjoint multiplet, 4 form 2 doublets, and the remaining 2 form 2 singlets.

With the above particle content, the beta factors for running of couplings between $\Lambda_U$ and $M_{\nu}^{(2)}$ are:
\begin{equation} \label{GrindEQ__4_3_} 
\begin{aligned}
b_{3} &=\left(-\frac{11}{3} 3+\frac{1}{3} 14+\frac{1}{6} 6+\frac{1}{3} 3\right)=-\frac{13}{3} \\
b_{2} &=\left(-\frac{11}{3} 2+\frac{1}{3} 12+\frac{1}{6} 6+\frac{1}{3} 2\right)=\frac{1}{3} .     
\end{aligned}  
\end{equation} 

The last section presented the assumption that each of the heavy neutrino and Omega lepton masses fall into similar ranges.  Due to this assumption, in an effective theory that excludes the heaviest of each, all anomalous triangle diagrams involving only SU(2) fields cancel.  However, triangle diagrams involving the Abelian $Y$ and/or $Y'$ fields do not cancel; they generate an anomaly in the effective theory \cite{sterling,dhoker}.

This situation is the one that was studied in \cite{anomaly-scale}: An anomaly-free theory exists, but below a certain scale, an effective theory that excludes some physics has an anomaly.  This anomaly gets resolved by gauge bosons acquiring masses.  Turning it around, given an anomalous effective theory with massive gauge bosons, the paper calculated the upper limit for the scale at which new physics must appear. Those arguments can be used in this theory to calculate an upper limit for the heaviest lepton masses.

In an effective theory without the heaviest neutrino and Omega lepton, two ``mixed'' anomalies are generated.  These come from diagrams involving two SU(2) fields and either a $Y$ or a $Y'$.  From eqs \eqref{GrindEQ__2_16_} and \eqref{GrindEQ__2_16b_}, the mixed anomaly traces for these leptons are:
\begin{equation}
\begin{aligned}
{\cal A}_Y&={\rm Tr}\left(T_2^3T_2^3T_{21}^{Y}\right)=\tfrac{1}{4}\sqrt{\tfrac{3}{5}} \\
{\cal A}_{Y^\prime}&={\rm Tr}\left(T_2^3T_2^3T_{21}^{Y^\prime}\right)=\tfrac{1}{2\sqrt{10}}\,.
\end{aligned}
\end{equation}
From eq (4.13) of \cite{anomaly-scale}, the upper limit on the mass of the heaviest leptons is:
\begin{equation}\label{RHneutrino-scale}
\begin{aligned}
M_{\nu}^{(2)}\sim m_\Omega^{(1)}&\lesssim\frac{64\pi^3 M_Z}{|{\cal A}_Yg_Y|g_2^2}\simeq 5200\, \textrm{ TeV} \\
M_{\nu}^{(2)}\sim m_\Omega^{(1)}&\lesssim\frac{64\pi^3 M_{Z^\prime}}{|{\cal A}_{Y^\prime}g_{Y^\prime}|g_2^2}\,,
\end{aligned}
\end{equation}
where coupling constants are evaluated at the $Z$ and $Z'$ boson mass scales, respectively.  Anomalies are also generated by ``non-mixed'' diagrams having only $Y$ and/or $Y'$ gauge fields.  These anomalies lead to larger upper limits, so will be ignored.

If $M_{\nu}^{(2)}$ is equal to the same percentage of both of the mixed upper limits, the following relation holds:
\begin{equation}
g_{Y^\prime}=\sqrt{\tfrac{3}{2}}\frac{M_{Z^\prime}}{M_Z}g_Y\,,
\end{equation} 
where $M_{Z^\prime}$ is the mass of the model's $Z'$ boson.  In section 5, it is suggested that a good candidate for this model's $Z'$ boson is the X17 boson with a mass of 17 MeV.  If that were the case, then it would imply that $g_{Y^\prime}\simeq 3\times 10^{-4}g_Y$, the correct scale to reproduce X17 data.  In this model, it is assumed that in order to satisfy eq \eqref{RHneutrino-scale}, the effective anomaly drives the $Y'$ coupling down to a much smaller value than would be expected from running of the coupling constant in an anomaly-free effective theory.  

In order to estimate the unification scale, the simplifying assumption is made that the model's leptoquarks have masses similar to those of the heaviest leptons.  As a result, between that heavy mass scale ($M_{\nu}^{(2)}$) and the electroweak scale, $b_3$ is reduced to $-16/3$ due to decoupling of the leptoquarks. In that same range, $b_2$ is also reduced due to decoupling of the heaviest leptons and the one very heavy scalar mentioned in eq \eqref{Majorana-mass} of the appendix.  In other words, for the SU(2) coupling, the 18 fermion doublets are reduced to 17 and the 6 scalar doublets are reduced to 5, resulting in $b_2=-1/6$.

The unification scale can be found by starting with the measured values of the SU(2) and SU(3) coupling constants at the scale of $M_{Z} $, using $b_3=-16/3$ and $b_2=-1/6$ to run the couplings up to the scale of $M_{\nu}^{(2)}$, and then using $b_3=-13/3$ and $b_2=1/3$ to run them up further until they have the same value.  Using $M_{\nu}^{(2)}\simeq$100 TeV, the unification scale is:
\begin{equation} \label{GrindEQ__4_4_} 
\Lambda_U \simeq 10^{14} \textrm{ GeV}.          
\end{equation} 
The inverse of the nonAbelian coupling at the unification scale is:
\begin{equation} \label{GrindEQ__4_5_} 
\alpha _{2}^{-1} \left(\Lambda_U \right)=\alpha _{3}^{-1} \left(\Lambda_U \right)\simeq 29,         
\end{equation} 
where $\alpha _{N} ={g_{N}^{2} \mathord{\left/ {\vphantom {g_{N}^{2}  4\pi }} \right. \kern-\nulldelimiterspace} 4\pi } $. 

In this model, the beta factor for the weak hypercharge coupling $g_{Y} $ between $\Lambda_U$ and $M_{\nu}^{(2)}$ is:
\begin{equation} \label{GrindEQ__4_6_}
\begin{aligned}
b_{Y}&=\frac{3}{20} \left(2\left(\left({\tfrac{1}{3}} \right)^{2} 8+\left({\tfrac{2}{3}} \right)^{2} 4+\left({\tfrac{4}{3}} \right)^{2} 2\right)+\left({\tfrac{4}{3}} \right)^{2} 6\right)\\
&+\frac{3}{20} \left({\tfrac{2}{3}} \left(\left(1\right)^{2} 12+\left(2\right)^{2} 6\right)+{\tfrac{1}{3}} \left(\left(1\right)^{2} 12+\left(2\right)^{2} 2\right)\right)\\
b_{Y} &=\frac{121}{15}
\end{aligned} 
\end{equation} 
where the $m=2$ flavor 2 singlet scalars have zero hypercharge. Between $M_{\nu}^{(2)}$ and the electroweak scale, the 6 lepton families are reduced to 5, the leptoquarks are excluded, and one $m=2$ scalar doublet and singlet is also excluded.  So between $M_{\nu}^{(2)}$ and $M_Z$, $b_Y=131/30$. 

Running the $Y$ coupling up to $M_{\nu}^{(2)}$ and then up to $\Lambda_U$, the hypercharge coupling at the unification scale is
\begin{equation} \label{GrindEQ__4_7_} 
\alpha _{Y}^{-1} \left(\Lambda_U \right)\simeq 28.         
\end{equation} 
So in the simplified scenario presented here, the three couplings come close to unifying at the same scale.  If they do indeed unify, then $g_+$ and $g$ take the same value at the unification scale (motivating that simplifying assumption in section 2).  

If a different scale is taken for $M_{\nu}^{(2)}$ and different assumptions are made about various masses (e.g. leptoquarks), there are scenarios where $g_+(\Lambda_U)$ does not have the same value as $g_2(\Lambda_U)=g_3(\Lambda_U)$.  That makes the expressions for $Y'$ and $Z'$ group structure in sections 2 and 5 more complicated, but it does not qualitatively change the arguments made.

\section{ Experimental Implications}

The model proposed in this paper is significantly different than the Standard Model.  To definitively test this model, more detailed calculations would need to be performed that are outside the scope of this paper.  But even in the absence of these calculations, a number of statements can be made about features (e.g. masses, couplings) the model would have to have in order to reproduce experimental data.

This section argues that the model's structure may allow it to reproduce well-established precision experiments while also providing new physics explanations for data that disagree with Standard Model predictions by 3-7$\sigma$, including many of the anomalies described in \cite{Crivellin_2024}.  The topics addressed in this section are:
\\
\\
\begin{enumerate}
\item  A seventh quark
\item  Hadronic cross sections
\item  Precision Z pole measurements
\item  Weak radiative decay of hyperons
\item  Higgs boson data
\item  Additional scalars
\item  SU(3) running and tau lepton decays
\item  Right-handed $Wud$ interaction
\item  Three heavy lepton families
\item  A light $Z'$ boson
\end{enumerate}

\subsection{A Seventh Quark}

This model predicts the existence of a seventh flavor of quark (denoted here by $f$).  Anomaly cancellation in the presence of seven quarks was addressed in section 4 of this paper.  The model does not predict the mass of the quark, and by choosing a large value for the superpotential parameter $\Gamma_\Phi$, it is possible to construct a version of this model where the $f$ quark has a mass larger than that of the top quark.  However, for reasons discussed below, it is proposed that the mass of the $f$ quark is smaller than the mass of the bottom quark.

A natural question is how an additional low-mass quark could have evaded detection so far. A recent paper shows that if there is a fourth down-type quark with a mass of 2.9 GeV and a light scalar that mediates its interactions, then most of the exotic hadrons discovered over the last twenty years fit nicely into the quark model as normal mesons and baryons rather than as 4- or 5-quark hadrons \cite{new-quark}.  In other words, as opposed to evading detection, it is proposed that the additional quark has been observed hundreds of times. 

If so, then it should be possible to make predictions for how to observe new $f$-quark hadrons or new decay modes of already-observed exotic hadrons.  Indeed, many such predictions have been made for LHCb, Belle II, and BESIII that could be checked by re-analyzing data already collected by these collaborations \cite{HADRON2025,LHCbpres2025}. 

In the Additional Scalars subsection, it is suggested that the scalar  proportional to diag$(1,1,1,1,1,0)$ (called $\varphi_5$) may be very light.  This scalar does not interact with any of the gauge bosons or gaugino quarks below the unification scale, but it does interact with quarks in the adjoint twisted superfield.  The interaction with those quarks is governed by the superpotential couplings $\Gamma_\Phi$ and $\Gamma_{1FF}$ from eqs \eqref{GrindEQ__1_18_} and \eqref{GrindEQ__1_15_}.

$\Gamma_\Phi$ generates the following interactions of quarks with $\varphi_5$:
\begin{equation}\label{eigen2}
\begin{aligned}
&\tfrac{1}{\sqrt{10}}\Gamma_\Phi \varphi_5 \left(2\bar{c}'_Lc'_R+2\left(.22\bar{d}_L+.97\bar{s}_L+.04\bar{b}_L+.03\bar{f}_L\right)f'_R \right)\\
&+\tfrac{1}{\sqrt{10}}\Gamma_\Phi \varphi_5\bar{f}'_Lb'_R+...+h.c.\,,
\end{aligned}
\end{equation}
where one conversion of $q'$ gauge eigenstates (see below eq \eqref{4x4fit}) to $q$ mass eigenstates is shown explicitly, and ``...'' in the last line includes much smaller interactions for the other quark combinations.  Eq \eqref{eigen2} shows that $\varphi_5$ quark interactions are primarily with $c\bar{c}$, $f\bar{s}$, $f\bar{d}$, $f\bar{b}$ and their Hermitian conjugates. 

$\Gamma_{1FF}$ connects $\varphi_5$ to charged leptons.  From eqs \eqref{lepteigenvec} and \eqref{hatg}, the strength of the $\varphi_5e^+e^-$ coupling is $g_3(0.8/\sqrt{5})(\tilde{m}'_{S11}/m_b)$, where $g_3$ is the strong coupling constant, $m_b$ is the bottom quark mass, and $\tilde{m}'_{S11}$ is a mass parameter discussed in the Charged Lepton part of section 3.

Because of these couplings, $f$-quark mesons can be produced in $e^+e^-$ collisions via $e^+e^- \to \varphi_5 \to f\bar{s}$ (or $f\bar{d}$) or $e^+e^- \to c\bar{c}\to\varphi_5\to f\bar{s}$. $f$-quark hadrons can also be produced via decay of a $b$ quark.  For example: $\bar{u}b\to \bar{u}c\bar{c}s \to \bar{u}s+f\bar{s}$, where the first process is mediated by a $W$ boson and the second by $\varphi_5$. Alternatively, $\varphi_5$ could directly mediate $b\to f q\bar{q}'$, where $q$ and $q'$ are $u$, $d$, or $s$ quarks coming from the small $...$ couplings in eq \eqref{eigen2}.

The same scalar also mediates the decay of $f$-quark mesons. For example, $\varphi_5$ mediates the decay $f\bar{u}\to d\bar{u}+c\bar{c}$.  Production and decay mechanisms like these are discussed in detail in \cite{new-quark}, enabling this model to reproduce properties of most of the observed exotic hadrons. 

If the proposed additional quark exists, one might expect it to generate predictions for CKM data that no longer agree with experimental data.  The opposite is true.  Not only can this model reproduce standard CKM data, it can also reproduce CKM data that are in tension with the Standard Model.

Currently there are no CKM measurements that disagree with the Standard Model by 5$\sigma$.  There are, however, some 3$\sigma$ hints of disagreement. Some of these are discussed in the section below on right-handed $Wud$ interactions, but one is addressed here:  The $V_{cb}$ CKM element measured via inclusive decays is 2-3$\sigma$ larger than $V_{cb}$ measured via exclusive decays \cite{CKM-Vcb}.   

In this model, flavor changing neutral current (FCNC) interactions mediated by $\varphi_5$ or the $Z$ boson (see eq \eqref{ZmixDL}) enable $b$-hadron to $f$-hadron decays, where the $f$-hadron then decays to other hadrons involving $c\bar{c}$.  These decays contribute to inclusive measurements of $V_{cb}$ but not to the sum of exclusive measurements of all decays allowed by the Standard Model. 

The additional quark may also provide an alternative explanation for the following:  The LHCb collaboration measured from $pp$ collisions the number of events emerging at very forward rapidities that involved a Z boson and a charm jet \cite{Z-charm}.  The number of measured events was significantly larger than the number predicted by Standard-Model event generators that incorporate Parton Distribution Functions (PDFs) that assume that the only charm quarks inside of a proton are those generated perturbatively (by gluons).  On the other hand, the data were well reproduced by PDFs that assume that a proton has some ``intrinsic charm'' \cite{intrinsic_charm}.

Valence quarks carry a much larger percentage of a proton's momentum than do sea quarks.  In this model a $d$ valence quark in a $pp$ collision can undergo a scalar-mediated transformation into an $f$ quark and a $c\bar{c}$ pair.  Such a process would lead to an excess of charm quarks at very forward rapidities (relative to the Standard Model). The effect should be doubled for neutron neutron collisions, so this model predicts that a much larger excess per nucleon of forward-rapidity charm should be seen in heavy ion collisions compared to $pp$ collisions.

There have been dozens of direct searches for an additional down-type quark.  Most of these searches do not rule out a quark of the type proposed here \cite{PDG-Bp-2020}.  In particular, most model-independent searches do not exclude a new quark with a mass smaller than that of the bottom quark.  

One notable exception is the fact that inclusive hadronic cross section data seem at first glance to rule out the possibility of a light additional quark.  That is the topic of the next subsection.

\subsection{Hadronic Cross Sections}

Over the last fifty years, many experiments have measured $R$, the ratio  ($e^+e^-\to$ hadrons)/($e^+e^-\to \mu^+\mu^-$) \cite{PDG_R}.  As the center-of-mass energy $\sqrt{s}$ is increased above 3 GeV, $R$ data show narrow spikes associated with $J/\psi$ and $\psi(2S)$, the first two $J^{PC}=1^{--}$ $c\bar{c}$ meson resonances.  These mesons are not massive enough to decay to $D\bar{D}$, so they must decay by ``OZI-suppressed'' channels. That makes these mesons long-lived, generating the observed narrow spikes in $R$. For $\sqrt{s}$ greater than those spikes, more $1^{--}$ $c\bar{c}$ meson resonances are seen in $R$ data, but those resonances are much wider since they can decay to $D\bar{D}$.

As $\sqrt{s}$ is increased above 9 GeV, a similar pattern is seen for the lowest-lying $1^{--}$ $b\bar{b}$ meson resonances.  The $R$ data show spikes for mesons not massive enough to decay to $B\bar{B}$ followed by wider resonances for mesons that can decay to $B\bar{B}$.

So if a quark existed that had a mass of 2.9 GeV, one might expect to see a similar pattern of spikes in $R$ data between the $c\bar{c}$ and $b\bar{b}$ spikes.  No such spikes are seen.  But in this model, it is proposed that the $f\bar{d}\to c\bar{c}$ scalar interactions experienced by the $f$ quark pull down the mass of the lowest lying $1^1S_0$ meson of $f\bar{d}+d\bar{f}$ so that it is closer to the mass of $\eta_c$, the $1^1S_0$ state of $c\bar{c}$.  Isospin symmetry pulls the mass of $f\bar{u}$ and $u\bar{f}$ down to a similar scale. The result is that the least massive $1^{--}$ $f\bar{f}$ meson has enough mass to decay to a pair of $f\bar{d}+d\bar{f}$ or $f\bar{u}+u\bar{f}$ mesons.  For that reason, in this model one would not expect to see any OZI-suppressed $f\bar{f}$ spikes, just wide bumps in $R$ between the charmonium and upsilon spikes.

From the fit to the exotic hadron spectrum presented in \cite{new-quark}, the $1^3 S_1$ meson of $f\bar{f}$ in this model has a mass of around 6600 MeV, corresponding to the $X(6600)$ observed by CMS \cite{CMS-6600} and ATLAS \cite{ATLAS-6600}.  The $1^1 S_0$ mesons of $f\bar{d}$ and $f\bar{u}$ of this model are the $X^{\pm,0}(3250)$ observed many years ago \cite{3250,PDG-mesons-2022}.  Given these mappings, one would expect to see a wide bump in the $R$ value at around 6600 MeV.  Although there appear to be normalization inconsistencies between various data sets, the Mark I data in this range do indeed show the kind of bump expected by this model (see for example fig 10 of \cite{lowR2}).  The data for slightly larger $\sqrt{s}$ also appear to be able to accommodate the additional $1^{--}$ $f\bar{f}$ resonances predicted by the model, such as the $2^3S_1$ $f\bar{f}$ resonance predicted at 7250 MeV \cite{exotics}.  Incidentally, $e^+e^-\to \mu^+\mu^-$ data also show a 3.5$\sigma$ hint of a resonance at 7250 MeV \cite{narrow_search}. It would be very helpful to have more up-to-date, precision $R$ value measurements in this CM energy range.

The $R$ value also provides insight to the total number of quarks in a different way.  For $\sqrt{s}$ much larger than quark masses but well below the $Z$ boson peak, the Standard Model predicts that 
\begin{equation}
R = 3\sum_F {\cal Q}_F^2 + \textrm{ quantum corrections}\,,
\end{equation}
where ${\cal Q}_F$ is the electric charge of a quark of flavor $F$. Summing over the five quarks in the Standard Model (excluding top) generates $11/3$ for the first term of $R$ (tree-level).  Quantum corrections then bring the total $R$ in this range up to $\sim$3.8.   

The $R$ value in the $20<\sqrt{s}<45$ GeV range has been measured by many different experiments, generating $R\sim 3.8-4.1$. \cite{PDG_R}.  The data can be reproduced by the Standard Model (at the low end of the range), but they could also be reproduced by another model that predicted a higher value. 

This model has another quark with charge $-1/3$, so the tree-level contribution to $R$ becomes 4. Gluon quantum corrections (with a smaller $\alpha_s$ as discussed in the SU(3) coupling subsection) bring the total $R$ up to $\sim$4.1, consistent with the high end of the range of data.  But this model also has additional scalar-field quantum corrections that are not present in the Standard Model and that reduce the $R$ value.   

At lowest order, gluon-mediated perturbative corrections increase $R$ by $\alpha_s/\pi$, where $\alpha_s=g_3^2/4\pi$ and $g_3$ is the strong coupling constant.  That first-order contribution comes from two sources: the square of a gluon-emission diagram and the interference between a tree-level diagram and a gluon-mediated vertex-correction diagram.  The gluon emission contribution is positive (since it is the absolute square of a matrix element), and the interference contribution is negative.  Both terms have infrared logarithmic divergences, but they cancel in the sum, with the remainder being the $\alpha_s/\pi$ positive contribution.

The topology of scalar-mediated perturbative diagrams is the same, but there are three main differences: Scalars have mass, scalar-mediated interactions can change quark flavor (see for example eq \eqref{eigen2}), and the scalars in the $m=1$ octet form a condensate.

To the last point, this model includes an octet of scalars $\varphi_1^a$ that interact with gluons but no other gauge bosons.  It is argued in section 3 that in the vacuum, these scalars become massless, form monopoles and condense, generating confinement via a dual Meissner effect.  It is assumed that in their condensed state, these scalars do not participate significantly in the decay of most hadrons.  On the other hand, above a certain scale (estimated below to be $\sim$4 GeV), there is a phase transition, and these scalars participate in processes such as perturbative corrections to $e^+e^-\to q\bar{q}$.

The interactions of $\varphi_1^a$ with quarks are governed by the supergauge interaction $2g{\rm Tr}\left(\Phi^\dag [V,\Phi]\right)$,
\begin{equation}\label{octetg}
\tfrac{1}{\sqrt{2}}g_3 \varphi_1^{\dag a} \left( \bar{c}'_L t^a u'_R - \bar{t}'_L t^a c'_R + \bar{s}'_L t^a d'_R - \bar{b}'_L t^a f'_R  \right)+h.c.,
\end{equation}
the superpotential coupling:
\begin{equation}\label{octetPhi}
\Gamma_\Phi \varphi_1^a\left(\bar{c}'_Lt^a c'_R+\bar{s}'_Lt^a f'_R +\bar{f}'_Lt^a b'_R\right)+h.c.\,,
\end{equation}
and various quantum-generated couplings, including 
\begin{equation}\label{octetcu}
i\tilde{\Gamma}_{cu} \varphi_1^a\bar{u}'_Lt^a c'_R+h.c.\,,
\end{equation}
where the last was discussed in appendix eq \eqref{charm-decay}.  All of these can contribute to scalar-mediated corrections to the $R$ value, but the focus here initially will be on the supergauge coupling of eq \eqref{octetg}.

Consider the last term in eq \eqref{octetg}; it changes a $b'_L$ to an $f'_R$ and vice versa.  In a scalar-mediated vertex loop diagram, this term adds a negative correction proportional to $\ln^2(s/M_1^2)$ to the $e^+e^-\to b_L\bar{b}_L$ cross section (and also to the $e^+e^-\to f_R\bar{f}_R$ cross section), where $M_1$ is the mass of $\varphi_1$.  This is the negative infrared divergent term discussed above (and in eq \eqref{negR} below).  The positive $\ln^2$ term in the scalar emission diagram does not cancel this, since it produces $e^+e^-\to b_L\bar{f}_R\varphi_1^{a}+h.c.$, an end state with two different quark flavors.  In other words, this model predicts that above the upsilon resonances, the cross section for $e^+e^-\to b\bar{b}$ should be smaller than in the Standard Model.

Recently, an analysis was performed to adjust $\sqrt{s}\sim$11-11.3 GeV BABAR cross section data \cite{Babar2008} for the effects of Initial State Radiation \cite{bbar_China}.  After this adjustment, these data can be compared with the Standard Model calculation of $R_b=\sigma(e^+e^-\to b\bar{b})/\sigma(e^+e^-\to \mu^+\mu^-)$.  It was found that the values of the adjusted BABAR data were 10\% (3$\sigma$) smaller than the Standard-Model predictions.  The authors of \cite{bbar_China} suggested that the BABAR data may have been normalized incorrectly.  Another possibility is that the data show evidence of the scalar-mediated $R_b$ reduction predicted by this model.  Additional $R_b$ data in this range would help distinguish between these possibilities.

It was mentioned above that below a certain scale, this model should have a phase transition where a scalar condensate forms.  $R$ data may indicate the energy scale for this phase transition.  Values of nonperturbative condensates with dimension 2, 4, 6, etc. can be extracted from low-energy $R$ data by considering its moments.  In the Standard Model, there are no gauge-invariant dimension-2 condensates, so one would expect the data to generate zero for its value.  This model, however, has a gauge-invariant dimension-2 condensate: tr$\langle\varphi_1\varphi_1\rangle$.  

In \cite{dim2con}, it was shown that $R$ data imply a nonzero value for the dimension-2 condensate (although the value is consistent with zero within large uncertainties).  The analysis shows that perturbative methods that ignore the condensate should only be used for $\sqrt{s}>M_c\sim$4 GeV. In this paper, it is assumed that the condensate exists, and that an infrared cutoff of $M_c$ should be used for $R$-value diagrams that involve emission of $\varphi_1$ scalars.

But the mass of virtual $\varphi_1$ scalars used in vertex correction diagrams does not have to be exactly the same as the infrared condensate cutoff $M_c$.  For example, suppose the $\varphi_1$ mass $M_1$ was $\sim$3 GeV.  In that case, the logarithmic contributions of the scalar emission and vertex correction diagrams do not exactly cancel for scalar-mediated $R$ value calculations.  At some large $\sqrt{s}$, the sum of these two corrections is:
\begin{equation}\label{negR}
\frac{\delta R}{R_0} = \tfrac{4}{3}\frac{\alpha_s}{2\pi}\left[\tfrac{3}{2}-\ln^2\left(\frac{s}{M_1^2}\right)+\ln^2\left(\frac{s}{M_c^2}\right)+3\ln\left(\frac{M_c^2}{M_1^2}\right)\right]\,.
\end{equation}
At $\sqrt{s}=30$ GeV, $\delta R/R_0$ is $\sim -1.2$ times the positive gluon correction of $\alpha_s/\pi$.  This would reduce this model's prediction of the $R$ value at that energy to a little below 4, further improving agreement with the data.

The above net negative correction gets larger for larger $s$.  For example, the correction from supergauge couplings of $\varphi_1$ becomes $\sim -3$ times the gluon correction at the $Z$ pole ($\sqrt{s}=m_Z$).  But at that higher energy, superpotential couplings play a larger role than supergauge couplings.  This is used in the discussion below about reproducing $Z$ pole data. 

The $\bar{b}'_L \varphi_1^{\dag} f'_R$ term in eq \eqref{octetg} has another interesting consequence.  This model predicts the following to be significant: $e^+e^-\to c\bar{c} \to f\bar{b}$, where the first part is mediated by a photon and the second by a scalar.  Therefore, it may be possible to see $f\bar{b}$ meson resonances in $R$ data.

It indeed appears that the MD1 data for $R$ in the 7.5-10 GeV range would be better fit by including resonances at around 8.1 and 8.9 GeV (see for example fig. 10 of \cite{lowR2}).  These resonances would have the correct masses to be identified with the $1^3S_1$ and $1^3D_1$ mesons of $\tfrac{1}{\sqrt{2}}(f\bar{b}+b\bar{f})$.

\subsection{Precision Z pole measurements}

The experiments at LEP and SLD measured the inclusive hadronic cross section at the $Z$ pole, as well as partial width and asymmetry data for $Z$ decay to leptons, $b\bar{b}$ and $c\bar{c}$ \cite{Z-summary}.  The experimental data agreed to high precision with all Standard-Model predictions.  This subsection presents a scenario whereby this model could also reproduce those data.

One of the main observables at these experiments was the cross section $\sigma^0(e^+e^-\to Z\to {\rm hadrons})$.  In the Standard Model (SM), $\sigma^0$ is proportional the sum of squares of couplings of quarks with the $Z$ boson.  But this model (7Q) has different $Z$ boson couplings as shown below:
\begin{equation} \label{Zqq} 
\renewcommand{\arraystretch}{1.5} 
\begin{array}{|c|c|c|} \hline
{\rm SM} & g_L & g_R\\ \hline
u' & \tfrac{1}{2}-\tfrac{2}{3}x & -\tfrac{2}{3}x\\
d' & -\tfrac{1}{2}+\tfrac{1}{3}x & \tfrac{1}{3}x\\
s' & -\tfrac{1}{2}+\tfrac{1}{3}x & \tfrac{1}{3}x\\
c' & \tfrac{1}{2}-\tfrac{2}{3}x & -\tfrac{2}{3}x\\
b' & -\tfrac{1}{2}+\tfrac{1}{3}x & \tfrac{1}{3}x\\
& & \\
\hline
\end{array} \,\,\,{\rm vs.}\,\,\,
\begin{array}{|c|c|c|} \hline
{\rm 7Q} & g_L & g_R\\ \hline
u' & -\tfrac{2}{3}x & \tfrac{1}{2}-\tfrac{2}{3}x\\
d' & \tfrac{1}{3}x  & -\tfrac{1}{2}+\tfrac{1}{3}x\\
s' & -\tfrac{1}{2}+\tfrac{1}{3}x & \tfrac{1}{3}x\\
c' & \tfrac{1}{2}-\tfrac{2}{3}x & \tfrac{1}{2}-\tfrac{2}{3}x\\
b' & -\tfrac{1}{2}+\tfrac{1}{3}x & \tfrac{1}{3}x\\
f' & \tfrac{1}{3}x  & -\tfrac{1}{2}+\tfrac{1}{3}x\\
\hline
\end{array}\,,
\end{equation} 
where $x=\sin^2\theta_W$ and $q'$ denotes a quark gauge eigenstate as described below eq \eqref{4x4fit}.  At a tree level, this model predicts a 33\% larger $\sigma^0$ than the SM for two reasons: a larger $c_R$ coupling and an additional quark.  

Just like the strong gauge coupling $g_3$, the supergauge couplings of this model run to smaller values at higher energies.  As a result, at $\sqrt{s}=m_Z$ the superpotential couplings of eqs \eqref{octetPhi} and \eqref{octetcu} may be a few times larger than the supergauge couplings of eq \eqref{octetg}.  A detailed calculation of the $R$ value in this model should include all of these couplings as well as ones with other scalars.  Since scalars change quark flavor, such a calculation would also have diagrams where a photon creates $t\bar{t}$, but a scalar-mediated vertex correction changes the outgoing quarks to $u\bar{u}$ or $c\bar{c}$.  In lieu of a detailed calculation, a very simple approximation will be made to argue that this model has a structure that could allow it to reproduce the $Z$ pole measurements.  In this approximation, all scalars other than $\varphi_1$ and all supergauge couplings are ignored.

Within that approximation, the following scenario allows the model to reproduce the data: The $\tilde{\Gamma}_{cu}$ vertex corrections add $\sim$69\% of the $Z\to u'_L\bar{u}'_L$ tree-level coupling to the $Z\to c'_R\bar{c}'_R$ tree-level coupling value to get a corrected $Z\to c'_R\bar{c}'_R$ coupling, and vice versa.  Also, the $\Gamma_\Phi$ vertex correction subtracts $\sim$35\% of the $c'_L$ and $s'_L$ tree-level couplings from the $c_R$ and $f_R$ couplings, respectively, and vice versa. Assuming $\sin^2\theta_W\simeq 0.231$, the net result is that scalar-mediated vertex corrections modify the couplings of eqs \eqref{Zqq}  as follows:
\begin{equation} \label{ZddVA} 
\renewcommand{\arraystretch}{1.5} 
\begin{array}{|c|c|c|} \hline
{\rm 7Q} & g_L & g_R\\ \hline
u' & -.154 & .346\\
d' & .077  &-.423 \\
s' & -.423 & .077\\
c' & .346 & .346\\
b' & -.423  & .077 \\
f' & .077  &-.423 \\
\hline
\end{array}
 \to
\begin{array}{|c|c|c|} \hline
{\rm 7Q} & g_L & g_R\\ \hline
u' & .083 & .346\\
d' & .077  &-.423 \\
s' & -.273 & .077\\
c' & .224 & .118\\
b' & -.423  & .077 \\
f' & .077  &-.273 \\
\hline
\end{array}\,.
\end{equation} 
After these scalar-mediated corrections, the sum of squares of couplings for the six non-top quarks of this model is equal to the sum of squares of the tree-level couplings for the five non-top quarks of the Standard Model.  Corrections like these could allow this model to predict the same hadronic cross section $\sigma^0_{\rm had}$ at the $Z$ pole as the Standard Model -- and therefore agree with the data.

Since the light leptons of this model have the same coupling to the $Z$ boson as the leptons in the Standard Model, the above scenario would also allow this model to reproduce $R_e$, $R_\mu$ and $R_\tau$ at the $Z$ pole (as well as lepton asymmetry measurements). 

In eq \eqref{ZddVA}, there is no scalar-mediated correction to the $Z\to b'\bar{b}'$ diagrams.  Also, from eq \eqref{Zqq}, the $Z$ interactions with $b'_L$ and $b'_R$ in this model are the same as the $Z$ interactions with $b_L$ and $b_R$ in the Standard Model. Since these interactions are the same and $\sigma^0_{\rm had}$ is the same, the $b$-quark partial width ($R_b$) and asymmetry predicted by this model at $\sqrt{s}=m_Z$ are the same as in the Standard Model and therefore reproduce the data.

From eqs \eqref{octetPhi} and \eqref{octetcu}, the $f$ quarks produced in $Z\to f\bar{f}$ subsequently decay mostly by $f_L\to s_R c_L \bar{c}_R$. For this reason, it is assumed that many $Z\to f\bar{f}$ events could have been interpreted experimentally as $Z\to c\bar{c}$ events.  In fact, since the $f\bar{f}$ events can produce 2 $c$ quarks per $f$ quark, the efficiency in asymmetry measurements for detecting $f$ quarks interpreted as $c$ quarks could even exceed the efficiency of directly detecting $c$ quarks.  A scenario fitting the data can be achieved if it is assumed that in partial width measurements, $\sim0.76\varepsilon$ of $f\bar{f}$ events are detected and interpreted as $c\bar{c}$ events, while in asymmetry measurements, $\sim 1.1\varepsilon$ of $f\bar{f}$ events are, where $\varepsilon$ is the $c$-quark detection efficiency.  In that case, the effective charm couplings ($c_{V{\rm eff}}=c_V+.76f_V$ and $c_{A{\rm eff}}=c_A+1.1f_A$) are equal to the Standard-Model $Z$ to charm couplings.

The above simple scenario is not meant to be a proof that this model reproduces the precision data at the $Z$ pole.  It simply demonstrates that this model has the possibility of reproducing the data.  Future work would be needed to determine whether the actual values of quantum-generated couplings and effects of the scalar condensate were able to reproduce these results.

\subsection{Weak Radiative Decay of Hyperons}  

For several decades, there has been a debate about how to reproduce data from weak radiative decays of hyperons \cite{WRHD}.  Hara's theorem states on general grounds that in the limit of flavor SU(3) symmetry, these decays should have very small parity violation \cite{hara}.  The large asymmetry observed in these decays can be interpreted in one of two ways: (a) flavor SU(3) symmetry is significantly broken for weak interactions or (b) Hara's theorem is violated.  Baryon magnetic moments show that flavor SU(3) symmetry provides a good approximation for electromagnetic interactions, so if Hara's theorem is not violated, a reason must be provided for why flavor SU(3) is significantly broken just for weak interactions.  On the other hand, violation of Hara's theorem would require violation of electromagnetic gauge invariance, locality (at the hadron level), or $CP$ symmetry.  

This model assumes Hara's theorem is not violated and provides an alternative explanation for the breaking of flavor SU(3) for Weak interactions but not electromagnetic interactions.  As in the Standard Model, the $d$ and $s$ quarks of this model have the same electric charge, so SU(3) flavor symmetry should be a good approximation for electromagnetic interactions.  Unlike the Standard Model, they have different weak interaction couplings, as can be seen from eq \eqref{Zqq}.  Therefore, even in the limit that the $d$ and $s$ quark masses are the same, flavor SU(3) is significantly broken for hadronic matrix elements involving the Weak interaction.

\subsection{Higgs Boson Data}

In the Standard Model (SM), a single Higgs boson gives mass to all fermions.  In this model, the scalar $(\phi_{21})_1$ gives mass to the top quark, but a different scalar $(\varphi_2)_{23}$ gives mass to the $b$ and $f$ quarks and all leptons.  Consequently, this model predicts that ttH production followed by $H\to b\bar{b}$ decay should have a cross section significantly below the SM expectation.  Interestingly, recent data for ttH then $H\to b\bar{b}$ are 2-3$\sigma$ below the SM expectation \cite{ATLASttHb,CMSttHb}.  This difference from the SM is discussed below.  But first is an explanation for how this 2-Higgs model can reproduce other Higgs data that agree with the Standard Model.  The Scalar subsection below also describes how the charged doublet partner of the second Higgs gets a large quantum-generated mass that allows it to evade the limits on 2-Higgs models imposed by $b\to s\gamma$ data \cite{Hpluslimit}. 

From eq \eqref{GrindEQ__3_6_}, the imaginary part of $(\phi_{21})_1$ has a vacuum expectation value (vev) $\bar{\phi}_{21}$ and a scalar field $h_t$.  The value of $\bar{\phi}_{21}$ is the same as the Higgs vev in the Standard Model.  As a result, the amplitudes for $h_t\to VV$ (meaning $h_t\to W^+W^-$ or $h_t\to ZZ$) are the same as the amplitudes for $H\to VV$ in the Standard Model.  This is because in both models, the amplitudes are proportional to the weak coupling $g_2$ and the same vev. 

From eq \eqref{GrindEQ__3_18_}, the bottom quark mass is $m_b\simeq \hat{g}\bar{\varphi}_2$, where $\hat{g}$ is the gaugino coupling discussed in eq \eqref{hatg}. This can be compared to the top quark mass $m_t\simeq \hat{g}\bar{\phi}_{21}$ from eq \eqref{GrindEQ__3_15_}. The ratio of these vevs is $\bar{\varphi}_2/\bar{\phi}_{21}\simeq m_b/m_t$.  As a result, the amplitude for $h_b\to VV$ (where ${\rm Im}((\varphi_2)_{23}) = \bar{\varphi}_2/\sqrt{2} + \tfrac{1}{2} h_b$) is suppressed by $m_b/m_t$ relative to $h_t\to VV$. Also, form factors for ggH-generated quark loops show that top quarks are at least 15 times more prevalent in these loops than other quarks.

Therefore, this model interprets that the scalar $h_t$ is by far the main contributor to Higgs data for VH, VBF, ggF or ttH production followed by $H\to VV$ or $H\to \gamma\gamma$ decays. $h_b$ makes negligible contributions to these.  Since the vev and couplings of $h_t$ are the same as those of the Standard-Model Higgs, this model's expectations for those processes are the same as the SM ones, allowing this model to reproduce those data. 

The model can also reproduce Higgs data for VH production followed by $H\to b \bar{b}$ decay.  In this model, $h_b$ decays to $b \bar{b}$, but $h_t$ has no classical coupling to $b \bar{b}$ (only the small quantum-generated one discussed below). In the Standard Model, the value of the $H\to b \bar{b}$ vertex is $m_b/m_t$ smaller than the $H\to t \bar{t}$ vertex, since the Yukawa coupling of the former is smaller by that amount.  In this model, $h_b\to b \bar{b}$ and $h_t\to t \bar{t}$ have the same coupling $\hat{g}$. However, in this model, the $VV\to h_b$ vertices are suppressed by $m_b/m_t$ relative to the $VV\to h_t$ vertices.  That means that VH production of $h_b$ followed by $H\to b \bar{b}$ decay of $h_b$ in this model has the same amplitude as the Standard Model, as long as the masses of $h_b$ and $h_t$ are similar.  The data from \cite{ATLAShbmass,CMShbmass} imply that the mass of $h_b$ must be within a few GeV of 125 GeV (the mass of $h_t$).  That being said, if the mass of H from $H\to b\bar{b}$ data was fit independently, a slightly larger mass might provide a better fit.

{\bf ttH then Hbb} This model predicts a much smaller cross section than the SM for ttH production followed by $H\to b \bar{b}$ decay.  In SM ttH production, two gluons create 2 top pairs; a member of each pair fuses to form the Higgs boson, leaving the remaining $t\bar{t}$ as spectators.  In this model, the scalar formed by 2 tops fusing is $h_t$.  Classically, this model has no $h_t \to b \bar{b}$ coupling, but it does have a small quantum-generated coupling described below \eqref{adjoint-mass}) in the appendix.  Consequently, this model predicts a much smaller cross section than the SM for ttH production followed by $H\to b \bar{b}$ decay.  As mentioned above, recent ATLAS and CMS data for this process are 2-3$\sigma$ below the SM expectation \cite{ATLASttHb,CMSttHb}.  

However, the same quantum-generated term generates a much larger expectation for ttH production followed by $H\to c \bar{c}$, $H\to f \bar{b}$, or $H\to f \bar{s}$ decay (+ h.c.).  If an experiment was inclusively looking at all heavy quark decays coming from ttH production and trying to categorize them into either decays to $c\bar{c}$ or to $b\bar{b}$, that experiment might interpret a number of $f \bar{b}$ (or possibly even $f \bar{s}$) decays as $b\bar{b}$ decays.  In that case, the experiment would get a larger value for ttH production followed by $H\to b \bar{b}$ than an exclusive measurement.  That is this model's alternative explanation for why the inclusive CMS measurement in \cite{CMSttHb2} agrees with the SM prediction, while the exclusive CMS measurement of \cite{CMSttHb} shows a cross section $\sim 3\sigma$ below it.

That same quantum term enables the decay $h_t \to c\bar{c}$, where $h_t$ could be produced via any mechanism, including VH and ttH production. It is also possible that some $h_t\to f \bar{s}$ decays could be interpreted as $H \to c\bar{c}$ (since $f$ quickly decays to $c$).  Depending on the size of the quantum-generated term, this model could accommodate $h_t \to c\bar{c}$ being the same magnitude or even larger than the SM expectation.

As further described below eq \eqref{adjoint-mass}, a similar quantum-generated interaction could cause a different scalar $(\varphi_2)_{32}$ to also decay to $c\bar{c}$. The Lagrangian term with two covariant derivatives connects $WW$ and $ZZ$ with a pair of these scalars.  This opens the door to VH production of a pair of $(\varphi_2)_{32}$ scalars, where one is seen decaying to $c\bar{c}$. If this scalar had a mass of around 155 GeV, it could show up as a second bump in $H\to c\bar{c}$ at that mass.  Preliminary CMS data could be interpreted as hinting at a second bump at about that mass \cite{VHcc}.

The same quantum-generated interaction also predicts that $(\varphi_2)_{32}$ could decay at a much smaller rate to $b\bar{b}$.  ATLAS data in \cite{ATLAShbmass} could be interpreted as hinting at a very small second bump at 155 GeV, the same location as the second bump mentioned above in $H\to c\bar{c}$ data.

As discussed further in the Scalar subsection below, since the $(\varphi_2)_{32}$ is created in pairs (associated production), it is also a good candidate to reproduce the various hints of signals at 152 GeV mentioned in \cite{Crivellin_2024}.

Another difference in this model arises in Vector Boson Fusion (VBF) production of a Higgs boson.  In VBF, two high energy quarks each emit a W or each emit a Z; those two vector bosons then fuse, forming a Higgs boson along with the two quark jets.  Due to the quark jets, VBF is a primary contributor to qqH production.  This model has six non-top quark flavors, as compared to five in the SM. Consequently, this model predicts VBF production of both $h_b$ and $h_t$ to exceed SM predictions by as much as $(6/5)^2\sim$1.44. CMS and ATLAS data are not inconsistent with this (see \cite{CMSVBF} fig. 11 and \cite{ATLASHiggs} fig.3).

In the Standard Model, ggF production (gluon fusion) is completely dominated by top quark loops since the bottom-Higgs coupling is suppressed by $m_b/m_t$ relative to the top-Higgs coupling.  No such suppression is present in this model.  However, ggF production of $h_b$ involves both b-quark and f-quark loops, since they both connect to $h_b$.  An interesting possibility supported by the fit in eq \eqref{4x4fit} is that the $h_bb\bar{b}$ and $h_bf\bar{f}$ couplings have opposite signs, generating destructive interference.  The magnitude of the contribution of these loops (relative to SM $t$ loops) is set by the fact that the form factors in ggF production diagrams with b and f loops are about 15-20 times smaller than the form factor in a $t$ loop.  Putting these two effects together could result in ggF production followed by $H\to b\bar{b}$ decay in this model generating fewer events than predicted by the Standard Model.  This is not in conflict with current data (e.g. \cite{CMSVBF}).   

This model also predicts more bbH production followed by $H\to b\bar{b}$ decay than in the Standard Model.  This is because gluons creating 2 $b\bar{b}$ pairs with one fusing to form a scalar is not suppressed by $(m_b/m_t)^2$, like it is in the SM.

From eq \eqref{hbtoleptons}, $h_b$ decays to charged leptons (tau, muon, electron) with an amplitude proportional to $m_l/m_b$, where $m_l$ is the lepton mass.  By this mechanism, VH production followed by $H\to l^+l^-$ has the same expectation for each lepton flavor as in the SM. But the decay rate could also be enhanced by the same quantum-generated terms discussed above that lead to $H\to c\bar{c}$ decay.

\subsection{Additional Scalars}

The model includes many more scalar fields than the ones mentioned in the Higgs data subsection above. The full list for this model includes 6 triplets, 1 octet, and 1 singlet for each of the two original U(3) groups (before the symmetry of the $m=2$ group is broken to SU(2) at the unification scale).  This subsection describes a possible scenario for the masses and interactions of those 54 complex scalars that is motivated by recent experimental data.

The six $m=1$ scalar triplets are leptoquarks since they connect leptons with both gaugino quarks and adjoint-representation quarks.  The model requires the scalar mass parameters $m_{1F}$ and $\tilde{m}_{1F}$ defined in section 1 to be much larger than the electroweak scale, so they are chosen to be larger than the lower limits from leptoquark searches.  If the flavor-1 leptoquarks have masses of around 10 TeV, they could be at least partly responsible for the non-resonant di-electron anomaly discussed in \cite{Crivellin_2024}.

As discussed in section 3, in an $N=2$ Super Yang-Mills (SYM) theory where the adjoint superfield has a tree-level mass, the adjoint scalars form massless color monopoles that condense and cause confinement \cite{seiberg-witten,seiberg-duality,bilal,intriligator-seiberg}.  Due to the similar structure of this theory, it was proposed in section 3 that the octet of $m=1$ SU(3) adjoint scalars of this theory behave the same way.  Above the confinement scale, these scalars are assumed to have masses of a few GeV and to participate in the processes mentioned in the hadronic cross section subsection above and the SU(3) running coupling subsection below.

The remaining scalar in the $m=1$ sector is an adjoint singlet that will be discussed below.

In the $m=2$ sector of the theory, two of the scalar triplets, $\phi_{23}$ and $\tilde{\phi}_{23}$, get unification-scale masses.  Their vevs are responsible for breaking the $m=2$ sector's U(3) gauge symmetry down to SU(2).

In the Higgs subsection above, it was proposed that there are two main scalars that generate the data that are commonly associated with a single Higgs boson.  A vev in the SU(2) doublet within the $\phi_{21}$ triplet gives mass to the top quark and is responsible for more than 99.9\% of the mass of the $W$ and $Z$ bosons.  Of the four real components of the complex scalar doublet, three get eaten by the $W$ and $Z$ bosons, leaving only $h_t$ as a propagating scalar.  The mass of the $h_t$ is 125 GeV.

The vev of one of the $m=2$ adjoint scalars $(\varphi_2)_{23}$ gives mass to $b$ and $f$ quarks, all leptons, and the $Z'$ boson (discussed further in a subsection below).  It is also responsible for a tiny part of the mass of the $W$ and $Z$ bosons.  Of the two real components of this complex scalar, one gets eaten by the $Z'$ boson, leaving only $h_b$ as a propagating scalar. As discussed in the Higgs data subsection, the mass of $h_b$ is taken to be 125-130 GeV. As discussed below, its charged isodoublet partner $(\varphi_2)_{13}$ has a much larger mass that allows it to evade the limits on 2-Higgs models from \cite{Hpluslimit}.

The Higgs data subsection also mentioned the possibility that another neutral scalar $(\varphi_2)_{32}$ has mass of around 155 GeV.  It was proposed to produce secondary bumps in $H\to c\bar{c}$ and $H\to b\bar{b}$ decays. This same scalar could potentially also be responsible for the hints of signals at 152 GeV described in \cite{Crivellin_2024}.

The adjoint scalar $\sqrt{2}\varphi_2^3=(\varphi_2)_{11}-(\varphi_2)_{22}$ is proposed to have a mass of around 95 GeV, a bit larger than the radiative correction contribution to the $h_t$ mass, as discussed in section 3.  If so, it could be responsible for some of the 95 GeV excesses mentioned in \cite{Crivellin_2024}.  This scalar has supergauge couplings to $t'_L\bar{c}'_R$, $c'_L\bar{u}'_R$, $b'_L\bar{f}'_R$, and $s'_L\bar{d}'_R$ and superpotential couplings to $\bar{c}'_L c'_R$ and $\bar{s}'_Lf'_R$ (which has an $f\bar{f}$ component).  Since it connects to quarks, this scalar can be produced by gluon fusion and can decay by $\gamma\gamma$, so it could generate the 95 GeV di-photon excess.  

Via the superpotential coupling $\Gamma_{1FF}$, $\varphi_2^3$ is also connected to $\Omega_R^{(F)}l_L^{\dag (F)}$, where $\Omega^{(F)}$ are the heavy leptons discussed in section 3.  From eq \eqref{lepteigenvec}, this includes a connection to $\tau^+\tau^-$, so this scalar can decay to  $\tau^+\tau^-$.  Since the scalar does not have a vev, it does not decay to $WW$ or $ZZ$, and it cannot be created by vector boson fusion.  In addition, its connection to $b\bar{b}$ is very small, so it would have negligible bbH production.  As a result, this scalar has consistent interactions with the fact that CMS saw an excess in $S\to\tau^+\tau^-$ when $S$ is produced via ggF, but not when it is produced via bbH (see fig. 10  of \cite{CMS95GeV}).

As discussed in section 1, the scalar mass parameters $m_{22}$, $\tilde{m}_{22}$, and $\tilde{m}_{21}$ are assumed to be large. Due to the experimental hints mentioned below, the following values are chosen for these parameters: $\sim$680 GeV, $\sim$950 GeV, and $>$10 TeV. These parameters impart mass-squared terms with those masses to the scalar triplets $\phi_{22}$, $\tilde{\phi}_{22}$ and $\tilde{\phi}_{21}$, respectively. 

In eq \eqref{Majorana-mass} of the appendix, it is shown how quantum interactions generate a much larger mass-squared term for the $(\tilde{\phi}_{21})_1$ component of $\tilde{\phi}_{21}$.  The mass of this component is similar to that of the heaviest right-handed neutrino, estimated in section 3 to fall in the 100-1800 TeV range.  Masses of the other two components of that triplet (the charged components) are governed by $\tilde{m}_{21}>$10 TeV.

As mentioned in the appendix and the beginning of section 3, quantum effects impart a small vev to the scalar triplet $\phi_{22}$, despite the fact that it has a relatively large mass $m_{22}\sim680$ GeV.  Since it has a vev, the neutral component $(\phi_{22})_2$ has supergauge connections with $WW$, $ZZ$.  Consequently, this scalar has the right properties to reproduce the 3$\sigma$ excesses in $\gamma\gamma$ and $ZZ$ events seen at that mass \cite{Crivellin_2024}.  The scalar also has supergauge and superpotential interactions with $b'_L\bar{s}'_R$ and $\bar{d}'_Lf'_R$.

The parameter $m_{22}$ imparts the same mass of $\sim680$ GeV to $(\phi_{22})_3$.  Via a quantum-generated coupling $\bar{\tilde{\phi}}_{22}\varphi_2^3(\varphi_2)_{23}(\phi_{22})_3$ this scalar can decay to the $\sim$95 GeV scalar and the $h_b$ 125-130 GeV scalar.  If $\varphi_2^3\to f\bar{b}$ is interpreted experimentally as a decay to $b\bar{b}$, then $(\phi_{22})_3$ has the right properties to reproduce the 650 GeV di-Higgs excess mentioned in \cite{Crivellin_2024}. 

The charged scalar $(\phi_{22})_1$ also gets a mass contribution from $m_{22}$, but it could get more radiative corrections due to its charge, so it could be more massive, $\sim$800-950 GeV.  Due to the vev of its neutral partner, this scalar has a small connection with $WZ$.  It has much larger supergauge and superpotential couplings with $t'_L \bar{s}'_R$ and $\bar{d}'_L c'_R$, respectively.  As a result of these couplings, $(\phi_{22})_1$ can decay to two jets (a dijet).  This charged scalar has a mass just above the lower limit implied by $b\to s\gamma$ decays \cite{Hpluslimit}.

As mentioned above, a value of $\tilde{m}_{22}\simeq$950 GeV is chosen for the masses of the three members of the $\tilde{\phi}_{22}$ triplet (although the charged component could be somewhat more massive).  The charged scalar $(\tilde{\phi}_{22})_1$ has supergauge and superpotential couplings to $d'_L\bar{u}'_R$ and $\bar{c}'_Ls'_R$.  The neutral scalar $(\tilde{\phi}_{22})_2$ has supergauge and superpotential couplings to $d'_L\bar{d}'_R$ and $\bar{s}'_Ls'_R$.  The neutral scalar $(\tilde{\phi}_{22})_3$ has a superpotential coupling to $\bar{f}'_Ls'_R$.  All of these combinations can produce dijets that could be responsible for the 950 GeV dijet excesses mentioned in \cite{Crivellin_2024}. 

Eq \eqref{adjoint1mass} of the appendix describes how quantum-generated f-terms generate large mass terms for the charged adjoint scalar pairs $(\tilde{\phi}_{21})_2$ and $(\varphi_2)_{21}$, $(\tilde{\phi}_{21})_3$ and $(\varphi_2)_{31}$ and also $(\varphi_2)_{13}$ and $(\phi_{21})_3$, (where $(\varphi_2)_{13}$ is the charged SU(2) doublet partner of the second $h_b$ Higgs boson discussed in the Higgs data section above). For phenomenological reasons discussed below, the mass terms for the first pair are assumed to be $\sim$3.5 TeV, while those for the second two pairs are assumed to be larger ($>$8 TeV).  Since it was assumed above that the tree-level parameter $\tilde{m}_{21}>$10 TeV, that mass-squared term dominates over the $\sim$3.5 TeV term for $(\tilde{\phi}_{21})_2$, so that scalar's net mass is $>$10 TeV.  But the scalar $(\varphi_2)_{21}$ still has a mass of $\sim$3.5 TeV. All six of these charged scalars have masses well above the lower limit implied by $b\to s\gamma$ decays \cite{Hpluslimit}.

For $(\varphi_2)_{21}$, the superpotential coupling $\Gamma_{222}$ enables the following decay: $(\varphi_2)_{21}\to (\tilde{\phi}_{22})_2(\phi_{22})_1$.  Each of the produced scalars can in turn produce dijets, one with mass $\sim$950 GeV, and the other with mass $\sim$800-950 GeV. This could potentially allow the scalar $(\tilde{\phi}_{22})_2$ to reproduce the di-dijet data from CMS and ATLAS that is mentioned in \cite{Crivellin_2024}.

The scalar $(\varphi_2)_{21}$ also has supergauge couplings with  $b'_L\bar{c}'_R$ and $s'_L\bar{u}'_R$ and a superpotential coupling with $\bar{c}'_Lf'_R$.  In the CKM fit of this paper, $\sim$3\% of $c'_R$ is $t_R$, so $(\varphi_2)_{21}$ could contribute to the $b\bar{t}$ excess at $\simeq$3.5 TeV seen in \cite{ATLAStb2023}.

Via quantum-generated couplings, this scalar also connects charged leptons with their light neutrino partners.  In the tau lepton subsection below, it is described how $(\varphi_2)_{21}$ contributes to tau lepton decay and to $B\to D^{(*)}\tau\bar{\nu}$ decay.

The only charged scalar not yet discussed is the adjoint scalar $(\varphi_2)_{12}$. It does not get a large mass via the mechanisms described above, so its mass is assumed to be similar to but a bit heavier than that of its SU(2) triplet partner $\varphi^3$.  It will be assumed to have a mass in the 100-120 GeV range.  It has supergauge and superpotential couplings with $t'_L\bar{f}'_R$, $c'_L\bar{d}'_R$ and $\bar{s}'_Lc_R$.  From the CKM fit of this paper, the $\bar{b}t$ and $\bar{t}s$ parts of the above coupling are very small, much smaller than the CKM connections between these quarks in the Standard Model.  As a result, even with its relatively light mass, this charged scalar's interactions do not conflict with $b\to s\gamma$ data.

$(\varphi_2)_{12}$ also has a large superpotential coupling to heavy right-handed neutrinos via $\bar{\nu}_R^{(F)}l_L^{(F)}$.  Since the mass eigenstates of the three light neutrinos contribute a negligible amount to heavy neutrino gauge eigenstates, $(\varphi_2)_{12}$ only connects to leptons negligibly.  

The model's remaining scalars that have not yet been discussed are $\varphi_1^0$, $\varphi_2^0$ and $\varphi_2^8$.  None of these interact with any gauge bosons below the unification scale.  These three scalars can be re-expressed as $\varphi_4$, $\varphi_5$ and $\varphi_6$ with the following group structures: $T_{4}=\sqrt{\tfrac{3}{20}}(\tfrac{2}{3},\tfrac{2}{3},\tfrac{2}{3},-1,-1,0)$, $T_5=\tfrac{1}{\sqrt{10}}(1,1,1,1,1,0)$ and $T_6=\tfrac{1}{\sqrt{2}}(0,0,0,0,0,1)$.  

The scalar $\varphi_4$ has supergauge interactions with many different gaugino quarks.  For this reason, it is assumed that radiative corrections cause it to have a large mass.  The scalars $\varphi_{5}$ and $\varphi_{6}$ have no interactions with gaugino quarks, so they have no supergauge interactions at all.  For that reason, they may be much lighter than the other adjoint scalars.  In the Seventh Quark subsection above, it was suggested that the $\varphi_{5}$ scalar mediates production and decay of most exotic hadrons.

From the CKM fit of this paper, the $\varphi_{5}$-mediated contributions of $\bar{s}'_Lf'_R$ and $\bar{f}'_Lb'_R$ to $\bar{s}_Lb_R$ offset.  Therefore, the model predicts that scalar-mediated $b\to sc\bar{c}$ decays are much rarer than $W$-mediated $b\to sc\bar{c}$ decays.  This is important for reproducing experimental data.  

$\varphi_{5}$ has a $\sim$7\% connection to $s\bar{s}$, so it could mediate $c\bar{c}\to s\bar{s}$ decays.  One consequence is that it leads to a higher rate for charmonium decays than in the Standard Model; this is discussed at the end of the next subsection.  

The second consequence is that if the scalar has a mass of around 1.7-1.8 GeV, it is possible that it has been observed as the $f_0(1710)$ scalar particle.  More very light $0^{++}$ scalars have been observed experimentally than can be accounted for as mesons in the quark model, so it is thought that one or more of them may not be mesons.  The $f_0(1710)$ has been considered a candidate for a glueball. Glueballs should have a similar rate of decay to pions and gluons, but the $f_0(1710)$ decays almost exclusively to kaons, casting doubt on that interpretation \cite{1710scalar}.  One possibility is that the real part of the $\varphi_{5}$ scalar is the $f_0(1710)$ and the imaginary part is the $X(1835)$.  Another possibility is that $\varphi_{5}$ has a smaller mass, possibly even making it a dark matter candidate. 

Due to its more limited interactions, the scalar $\varphi_{6}$ should have a mass smaller than that of $\varphi_{5}$. Its only connection is to $\bar{f}'_Lb'_R$, but this does allow it to mediate $b\bar{b}\to f\bar{s}$ decays of Upsilon resonances.  This is discussed at the end of the next section.  Due to its limited connections and very low mass, it is the most promising dark matter candidate of this theory.  

Some of the arguments above rely on quantum corrections that have been qualitatively discussed, but not calculated in detail.  Future work would be needed to ensure that the quantum corrections of this model are consistent with the picture painted above.

\subsection{SU(3) running and tau lepton decays}

This section discusses how scalar fields affect (i) running of the SU(3) strong coupling, (ii) decays of tau leptons, and (iii) evidence for more charged-current semi-leptonic decays of B mesons to tau leptons than to muons.

Measurements of the strong coupling constant over a wide range of energies show very good agreement with Standard-Model predictions for the SU(3) beta factor that governs the curvature of the effective coupling at various energies.  The beta factor can be calculated using eq \eqref{GrindEQ__4_1_}.  In the Standard Model, calculation between the bottom and top masses uses 5 4-component quark triplets (10 2-component triplets) and no colored scalars.  This leads to $b_3=-23/3$.  In this model, there is an additional 4-component quark triplet (the $f$ quark) that changes $b_3$ to $-21/3=-7$.  The colored leptoquark scalars are assumed to be very heavy, so they do not affect this calculation.   However, there are also colored adjoint scalars. 

As mentioned in section 3 and the hadronic cross section subsection above, adjoint scalars form a condensate in the vacuum, but perturbative methods involving those scalars are valid above a scale of $\sim$4 GeV.  So above the bottom quark mass scale, $b_3$ for the strong coupling changes to $-6$.  Since $b_3$ in this model is less negative than in the Standard Model, it predicts less curvature in its plot of the running strong coupling.  

Specifically, if the magnitude of this model's $\alpha_s(M_Z)$ is set to match the data, then the model predicts an $\alpha_s(m_\tau)$ that is about 20\% below the value derived from hadronic tau lepton decays assuming only Standard-Model interactions. In the Standard Model, $\sim$20\% of the hadronic tau lepton decay rate contribution comes from a perturbative QCD correction, approximately proportional to $\alpha_s$ (at one loop). Due to its 20\% smaller $\alpha_s(m_\tau)$, this model's pQCD correction only contributes $\sim$16\% to the decay rate.  To be consistent with data, this theory needs another channel that contributes $\sim$2\% to the amplitude of hadronic tau lepton decays.  Squaring the amplitude results in a $\sim$4\% larger decay rate, making up for the 20\% smaller $\alpha_s(m_\tau)$.

The additional needed channel is provided by the adjoint scalar $(\varphi_2)_{21}$ mentioned in the scalar subsection.  $(\varphi_2)_{21}$ has a mass of $\sim$3.5 TeV and has a large superpotential coupling with $\Omega_R^{\dag(F)}\nu_L^{(F)}$, where $\Omega^{(F)}$ are the heavy leptons discussed in section 3.  From eq \eqref{lepteigenvec}, this includes a connection to $\tau^+\nu_\tau$.  $(\varphi_2)_{21}$ also has supergauge coupling to $\bar{s}'_Lu'_R$, so the scalar mediates the decay $\tau^+\to\bar{\nu}_\tau u_R \bar{s}_L$ (and its Hermitian conjugate).  With these interactions and 3.5 TeV mass, if the lepton seesaw parameter $\tilde{m}'_{S33}$ (see \eqref{lepteigenvec}) has a mass in the dozens of GeV range, $(\varphi_2)_{21}$ mediates the needed amount of additional hadronic tau decay.

But that same scalar also has a supergauge coupling to $\bar{b}'_Lc'_R$, so it also has an effect on $B\to D^{(*)}\tau\bar{\nu}$ decays.  The same parameter $\tilde{m}'_{S33}$ increases the rate of these decays by about 20\% relative to decays only being mediated by $W$ bosons.  If the seesaw mass $\tilde{m}'_{S22}$ for muons is much smaller (not surprising due to the much smaller muon mass), then the enhancement to $B\to D^{(*)}\mu\bar{\nu}$ decays is much smaller than for decays to tau leptons.  In that case, this model produces a ratio that is $\sim$20\% larger for decays to tau leptons vs. to muons.  This is the correct amount to reproduce the $\sim3\sigma$ deviation from the Standard Model mentioned in \cite{Crivellin_2024}.

Besides tau decay, other low-energy data points for the running of the strong coupling come from charmonium and bottomium decays.  In the Standard Model, quantum corrections make a percentage of these decays proportional to $\alpha_s$, and this model predicts a $\sim$15\% smaller  $\alpha_s$ in that energy range.  To make up for this, the model must have additional channels for decay of charmonium and bottomium. As discussed in the Scalars subsection, the scalars $\varphi_5$ and $\varphi_6$ provide additional channels that could make up that difference.

\subsection{Right-handed W-ud interaction}

It has long been assumed that the $W$-boson interaction connecting up and down quarks is left-handed, but that assumption is not required to reproduce experimental data.  The handedness of the $Wud$ interaction can be probed by measurements of polarized neutron decay or deep inelastic scattering. Both of these lead to the following conclusion:  If the spin of quarks inside a nucleon is more aligned than anti-aligned with the overall nucleon spin, then the $Wud$ interaction is left-handed.  But if the spin of quarks is more anti-aligned than aligned with nucleon spin, then the $Wud$ interaction is right-handed.

In polarized neutron decay, experiments measure correlation coefficients (conventionally labelled $A$ and $B$) that determine the magnitude and sign of the ratio $\lambda$ of the axial vector to vector coupling constants (see for example eqs 5 and 6 of \cite{Gorchtein_2023}).  Assuming the $Wud$ interaction is left-handed, the data imply a positive value for $\lambda$ ($\sim$1.27).  But those same data are fit equally well if it is assumed that the $Wud$ interaction is right-handed and $\lambda$ is negative ($\sim -$1.27).

If one assumes that the net quark spin is aligned with the nucleon, then a simple nonrelativistic calculation (outlined below) generates a value of $\lambda = +\tfrac{5}{3}$.  But if quark spin is net anti-aligned, that same calculation produces $\lambda = -\tfrac{5}{3}$, compatible with a right-handed $Wud$ interaction.  

Nonrelativistically, the axial current charge in the ``3'' direction of space and of isospin is (see for example Appendix A of \cite{Sasaki_2003}):
\begin{equation}
Q^{33}_A = \sum_i\left(b^\dag_{u\uparrow i}b_{u\uparrow i}-b^\dag_{u\downarrow i}b_{u\downarrow i}-b^\dag_{d\uparrow i}b_{d\uparrow i}+b^\dag_{d\downarrow i}b_{d\downarrow i}\right)\,,
\end{equation}
where $b^\dag$ are quark creation operators and $i$ is a fundamental color index.
The wave function of a proton whose net quark spin is aligned with the proton spin is (see for example eq 11 of \cite{Simonov_2009}):
\begin{equation}\label{pwavefunc}
\begin{aligned}
\sqrt{18}|p_+ \uparrow\rangle&=\left(b^\dag_{u\uparrow 1}b^\dag_{u\downarrow 2}b^\dag_{d\uparrow 3}+5\,{\rm perms}\right)|0\rangle \\
&-2\left(b^\dag_{u\uparrow 1}b^\dag_{u\uparrow 2}b^\dag_{d\downarrow 3}+2\,{\rm perms}\right)|0\rangle\,,
\end{aligned}
\end{equation}
where the ``$+$'' subscript on $p_+$ indicates that quark spins are net aligned with nucleon spin. The well-known nonrelativistic value of $\lambda$ is calculated as follows:
\begin{equation}
\lambda = \langle p_+ \uparrow|Q^{33}_A|p_+ \uparrow\rangle = +\tfrac{5}{3}\textrm{(aligned)}\,.
\end{equation}

The wave function of a proton whose net quark spin is anti-aligned with the proton spin can be found by simply flipping the spin of every quark in eq \eqref{pwavefunc}.  After that is done, one finds
\begin{equation}
\lambda = \langle p_- \uparrow|Q^{33}_A|p_- \uparrow\rangle = -\tfrac{5}{3}\,\textrm{(anti-aligned)}\,.
\end{equation}
This opens the possibility to reproducing neutron decay data with a right-handed $Wud$ interaction as long as quark spins are net anti-aligned with nucleon spin. 

The standard way to reproduce the data from polarized deep inelastic scattering experiments is to assume that the $Wud$ interaction is left-handed and that quark spins are net aligned with nucleon spin.  But those same data can be fit equally well by assuming right-handed $Wud$ interactions and net anti-aligned quark spins.  Either way, the alignment is surprisingly small \cite{protonspincrisis}.  The data show that only about $1/3$ of the magnitude of nucleon spin comes from quark spins; the rest must come from orbital angular momentum (gluon spin is thought to contribute negligibly) \cite{pspin1}.  The value of $1/3$ is very different from the value of $1$ that comes from the naive quark state of eq \eqref{pwavefunc} that has no orbital angular momentum.

Since the data indicate that quarks inside of nucleons have a lot of orbital angular momentum, a simple model to test is the following: Each quark has total angular momentum $J=\tfrac{1}{2}$ (similar to the MIT bag model assumption), but each quark is in a ``P'' state with orbital angular momentum $L=1$. A value of $m_J=+\tfrac{1}{2}$ can come from $m_L=+1$ and $m_S=-\tfrac{1}{2}$ or from $m_L=0$ and $m_S=+\tfrac{1}{2}$.  From Clebsch-Gordon coefficients, the relative amplitudes of these are $\sqrt{2}$ and $1$, respectively.  

The proton wave function in this simple P-state model starts with $|p_+ \uparrow\rangle$ from eq \eqref{pwavefunc} and makes the following substitutions:
\begin{equation}
b^\dag_{q\uparrow i}\to\tfrac{\sqrt{2}}{\sqrt{3}}b^\dag_{q,+\downarrow i}+\tfrac{1}{\sqrt{3}}b^\dag_{q,0\uparrow i}
\hspace{.5cm}
b^\dag_{q\downarrow i}\to\tfrac{\sqrt{2}}{\sqrt{3}}b^\dag_{q,-\uparrow i}+\tfrac{1}{\sqrt{3}}b^\dag_{q,0\downarrow i}\,,
\end{equation}
where the additional $+,-,0$ subscripts denote $m_L$.  The expectation value of spin of this state is $-1/3$ of the total nucleon spin, and it has $\lambda=-5/3$.  This has the correct magnitude to reproduce the deep inelastic scattering data mentioned above, and it is as close to the $\lambda$ value for right-handed currents as the usual model is for left-handed currents.  In both cases, pion cloud and other refinements could generate a $\lambda$ closer to the measurements.

Although the P-state model does a good job phenomenologically of reproducing data, a theoretical problem with it is to come up with some reason why the lowest baryon state would have its quarks in P states rather than S states. 

One speculative theoretical possibility is the following:  At the beginning of section 3, it was proposed that in this model, adjoint scalars form color-magnetic monopoles that become massless, condense, and generate color confinement via a dual Meissner effect. In another model where adjoint scalars form monopoles, it was shown that spherically symmetric zero modes of quarks combine with the monopoles \cite{solitonfermion}.  To isolate propagating quarks, these zero modes must be removed.  If that had to be done in this model, then removal of the zero modes could necessitate removing some or all of the S modes of quarks, leaving a P mode as the lowest-energy quark state inside a nucleon.  

Returning to experimental evidence, one might expect a right-handed $Wud$ interaction to generate predictions for CKM data that do not agree with experimental data.  The opposite is true.  

Currently there are no CKM measurements that disagree with the Standard Model by 5$\sigma$.  There are, however, some 3$\sigma$ hints of disagreement.  For example, measurements of CKM matrix elements lead to a first-row unitarity calculation of $0.9985\pm 0.0005$ \cite{CKM}, which is a 3$\sigma$ variation from the unitary value of 1 predicted by the Standard Model.  In addition, the following 2.9$\sigma$ difference is measured:
\begin{equation}\label{2}
\begin{aligned}
\textrm{Vector Current: }&|V_{us}|=0.2231\pm 0.0006 \\
\textrm{Axial Vector Current: }&|V_{us}|=0.2254\pm 0.0006,
\end{aligned}
\end{equation}
where the first result above is from semi-leptonic kaon decay and the second is from leptonic \cite{CKM-Leptonic}.   

Since the Standard Model has only left-handed $W$ boson interactions, its CKM matrix should be unitary and there should be no difference between vector-current and axial-vector-current CKM matrices.  If, as in this model, there are both left- and right-handed $W$ boson interactions, then there is no requirement for the CKM matrix to be unitary, or for vector-current and axial-vector-current CKM data to be the same.  

The model presented here is able to reproduce all of the standard CKM data and also the above 3$\sigma$ differences.  In fact, the quark mass matrices of eqs \eqref{3x3fit} and \eqref{4x4fit} were chosen to do this, as can be seen from the $V_{us}$ elements generated by those matrices (see eq \eqref{GrindEQ__3_28_}).  If increased precision causes the above discrepancies to exceed 5$\sigma$, it could be construed as evidence for right-handed currents \cite{Crivellin_2024}.

\subsection{Three Heavy Lepton Families}

This model predicts the existence of three additional heavy charged leptons (referred to in this paper as ``Omega leptons'') and their isodoublet partner heavy right-handed neutrinos.  As mentioned in section 3, direct searches have ruled out an additional charged lepton with a mass of less than 103 GeV \cite{heavy-lepton}.  As discussed in section 4, the upper limit for the heaviest charged lepton mass and heaviest neutrino mass is 5200 TeV, but that heaviest mass could be lower.  In section 3, two scenarios are presented for masses of the three lepton families that would allow reproducing observed neutrino masses and mixing.  Scenario 1: 1800 TeV, 120 TeV, and 20 TeV.  Scenario 2: 100 TeV, 1.2 TeV, and 200 GeV.

Due to the large superpotential couplings $\Gamma_{1FF}$, the heavy leptons decay very quickly to the three Standard-Model light lepton families.

\subsection{A Light Z' Boson}

A consequence of the structure of this model is the existence of an additional light U(1) gauge boson -- a $Z'$.  In order to be consistent with precision electroweak experiments, the angle $\phi _{Z} $ (from eqs \eqref{GrindEQ__2_17_} and \eqref{GrindEQ__2_18_}) that determines the mass and coupling of the $Z'$ must be very small.  In section 4, a mechanism was suggested for this very small value.  The calculations in this paper are not detailed enough to predict the mass and coupling of the $Z'$ boson.  Instead, a phenomenological approach is taken; an observed experimental anomaly is studied and mapped to the $Z'$ of this model.  

A group at the Institute for Nuclear Research ATOMKI in Hungary has published evidence consistent with the existence of a neutral boson with a mass of $\sim$17 MeV \cite{X17-2016,X17-2018,X17-2019,X17-2022,X17-2023}.  So far, only ATOMKI and a second group in Hanoi that included some ATOMKI personnel \cite{X17-Hanoi} have detected this ``17 MeV anomaly'' (aka the X17).  That being said, the X17 anomaly has been observed in experiments involving three different nuclei ($^8$Be, $^4$He, $^{12}$C) and at a 7$\sigma$ level of significance.  The analysis below assumes that the $Z'$ boson of this model is the X17 with a mass of $\sim$17 MeV. 

In section 4, it is mentioned that the heavy leptons generate an anomaly for $Z'$.  Below the mass of the top quark, the $Z'$ anomaly is exacerbated when the top quark decouples, since the anomaly is not resolved by the $Z$ boson mass.  Below the top-quark mass scale, it is assumed that the coupling for $Y'$ interactions with $Q_{22}$ fields (with an effective quark anomaly) run down more quickly than those with $Q_{21}$ fields (without one).  This effect is parametrized by a factor $\kappa$ by which the $g_{Y'}$ coupling to $Q_{22}$ is less than its coupling to $Q_{21}$ at the $Z'$ mass scale.

The $T^{Z'}_{22}$ group structure is then derived by the following modification of the Weinberg rotation of eq \eqref{GrindEQ__2_17_}:
\begin{equation}
\begin{aligned}
g_{Z'}T^{Z'}_{22} =& \kappa\cos\phi_Z g_{Y'}T^{Y'}_{22} \\
+&\sin\phi_Z\left(\cos\theta_Z g_2T_2^3-\sin\theta_Z g_Y T^Y_{22}\right)\,.
\end{aligned}
\end{equation}
Using the angles from eq \eqref{GrindEQ__2_18_}, the $Z'$ matrices at the 17 MeV scale take the form:
\begin{widetext}
\begin{equation}\label{Zp-group}
\begin{aligned}
g_{Z'}T_{21}^{Z^\prime}=g_{Z'}T_{1F}^{Z^\prime}&=-{\rm{diag}}
\left(1+\tfrac{2}{3}x,1+\tfrac{2}{3}x,1+\tfrac{2}{3}x,
0,x-1,x-1\right)e\sin\phi_Z/\sin\theta_W\cos\theta_W \\
g_{Z'}T_{22}^{Z^\prime}&={\rm{diag}}
\left(\tfrac{1}{2}\kappa+\tfrac{1}{3}x,\tfrac{1}{2}\kappa+\tfrac{1}{3}x,\tfrac{1}{2}\kappa+\tfrac{1}{3}x,
x-\tfrac{1}{2}-\tfrac{1}{2}\kappa,\tfrac{1}{2}-\tfrac{1}{2}\kappa,0\right)e\sin\phi_Z/\sin\theta_W\cos\theta_W  \,,
\end{aligned}
\end{equation}
\end{widetext}
where $x=\sin^2\theta_W$.  The $Z'$ group structure when interacting with gaugino and adjoint fields is the same as $T_{22}^{Z^\prime}$.

Given those group structures along with the definitions of quarks and leptons earlier in the paper and the quark mass matrices of section 3, $Z'$ couplings to left- and right-handed up and down quarks, electrons and neutrinos are:
\begin{equation} \label{ZpLR} 
\renewcommand{\arraystretch}{1.5} 
\begin{array}{|r|c|c|} \hline
 & g_L & g_R\\ \hline
u &1+\tfrac{2}{3}x  &-\tfrac{1}{2}-\kappa+\tfrac{2}{3}x\\
d & \tfrac{1}{2}\kappa+\tfrac{1}{3}x   & \tfrac{1}{2}\kappa+\tfrac{1}{3}x\\
e & x-1  & x-1\\
\nu_e & 0   & 0 \\
\hline
\end{array}\,\,\,\times \frac{e\sin\phi_Z}{\sin\theta_W\cos\theta_W}\,.
\end{equation}
In order to make the $Z'$ protophobic (see below), the following phenomenological choice is made:
\begin{equation}
\kappa\simeq\tfrac{1}{3}+\tfrac{2}{3}x\,.
\end{equation}

If $\sin\phi_Z$ takes the following value
\begin{equation}\label{ScaleZ}
\frac{e\sin\phi_Z}{\sin\theta_W\cos\theta_W} \simeq 1.5\times 10^{-4}\,,
\end{equation}
then the quark couplings can be translated into Vector (V) and Axial-vector (A) proton, neutron and electron couplings as follows:
\begin{equation} \label{ZpVA} 
\renewcommand{\arraystretch}{1.5} 
\begin{array}{|r|c|c|} \hline
 & V & A\\ \hline
p & C_p\simeq 0  & a_p \simeq 5.4 \\
n & C_n\simeq 1.5   & a_n \simeq -1.1 \\
e & |\epsilon| \simeq 7.6  & 0\\
\hline
\end{array}\,\,\,\times 10^{-4}\,,
\end{equation}
where notation from \cite{barducci} (Appendix C) is being used for protons and neutrons, $x\simeq .238$ was used, and $\epsilon$ is the electron vector coupling divided by $e\simeq 0.3$.  

With these choices, the $Z'$ boson of this model can reproduce $X17$ data while satisfying all applicable constraints.  From fig. 5 of \cite{barducci}, it can be seen that the proton and neutron vector couplings can reproduce the $X17$ $^{12}$C data, while the axial vector couplings can reproduce the $^8$Be and $^4$He data.  The vector coupling is protophobic ($C_p=0$), satisfying the NA48 constraint.  The fact that there are no interactions with electron neutrinos allows the model to satisfy neutrino constraints \cite{X17-neutrino}. The electron interaction strength is above the limit imposed by NA64 but below the level that would contradict electron $g-2$ data \cite{NA64-2020}.  Finally, as described in Appendix F of \cite{barducci}, the very stringent atomic parity constraints on mixed-parity models do not apply when the electron has only a vector interaction (no axial vector), as in this model.

The mass of the $X17$ in this model is generated by the vev $\bar{\varphi}_2$.  From the form of $g_{Z'}T_{22}^{Z^\prime}$ in eq \eqref{Zp-group}, the tree-level mass is:
\begin{equation}
m_{Z^\prime}^2\simeq \tfrac{1}{2}(\kappa-1)^2m_b^2 (1.5\times 10^{-4})^2\,,
\end{equation}
where eq \eqref{ScaleZ} and $\bar{\varphi}_2\simeq m_b$ was used.  The remainder of the assumed 17 MeV mass of the $Z'$ in this model comes from radiative corrections.

\section*{Discussion}

The theory presented in this paper is being proposed as an alternative to the Standard Model. The paper has taken a two-pronged approach:  theoretical and phenomenological.

In sections 1, 2, 4 and the appendix, a model is presented that has a number of attractive theoretical features.  For example, it is holomorphic, invariant to local superspace gauge transformations, supports coupling constant unification, and is similar at the unification scale to a theory that has been shown to be free of quadratic divergences to at least two loops \cite{Twisted-Superfields}.

In sections 3 and 5, detailed experimental data are considered, including data that differ by 3-7$\sigma$ from the Standard Model.  The results that nonperturbative calculations would have to generate in order to reproduce the data are identified.

The theory looks promising on the theoretical side and also on the phenomenological side.  More work needs to be done to tie these two sides together.  Would actual nonperturbative calculations support the parameter values required to reproduce data?

But even without that work, the model provides a couple of interesting explanations and makes a number of predictions.  The model provides explanations for the mechanisms of confinement and neutrino oscillations.  The model predicts three additional charged leptons, a seventh quark (without an eighth), a $Z'$ boson, right-handed quark interactions with the $W$ boson, and dozens of additional scalar particles.  So far, these predictions do not appear to be ruled out by existing data; in fact, they could provide new physics explanations for many of the anomalies discussed in \cite{Crivellin_2024}.

\appendix
\section{Quantum Effective potential}

This appendix proposes possible superpotential terms that may be generated by nonperturbative quantum effects. 

In \cite{seiberg-witten,seiberg-duality,intriligator-seiberg,Nonpert-SQCD-1,Nonpert-SQCD-2,Nonpert-SQCD-Talk}, it is shown how holomorphy and symmetry arguments can be used to determine the exact superpotential terms that get generated nonperturbatively for Supersymmetric QCD (SQCD) at low energies.  That analysis is anchored in the fact that SQCD is an asymptotically free theory with an ultraviolet renormalization group fixed point.

The U(3)xU(3) theory of this paper has Abelian groups, so it not entirely asymptotically free.  However, in this appendix it is implicitly assumed that this theory is an effective theory of a more general asymptotically free theory, applicable above the unification scale. With that assumption in hand, this appendix makes arguments similar to those used for SQCD and proposes general features of nonperturbative superpotential terms.

It is possible that instanton calculations could be employed to determine the exact nonperturbative superpotential terms for this model, but that is outside the scope of this paper.  Instead, allowed functional forms of terms are derived, and then the magnitude of the quantum-generated couplings that would be required for this model to fit experimental data are phenomenologically identified. 
 
The first step in deriving the effective superpotential terms is to specify the beta function for the two SU(3) groups, evaluated for the case where all tree-level masses and superpotential couplings are zero.  The beta function for an SU(N) theory is:
\begin{equation}\label{A1}
\begin{aligned}
&\beta =g\left(\frac{g^{2} }{16\pi ^{2} } \right)b_{N} \\
&b_{N} =\left(-\frac{11}{3} N+\frac{1}{3} n_{f} +\frac{1}{6} n_{s} +\frac{2}{3} Nn_{fA} +\frac{1}{3} Nn_{sA} \right),
\end{aligned}
\end{equation}
where $n_{f} $ and $n_{s} $, $n_{fA} $ and $n_{sA} $ are the numbers of fundamental fermion and scalar N-tuplets, and fermion and scalar adjoint-representation multiplets.  In all cases, the fermions are 2-component Weyl fermions and the scalars are complex.  

In the model presented in this paper, $n_{f} =18$, $n_{s} =6$, $n_{fA} =0$ and $n_{sA} =1$ for each SU(3) group.  The fundamental scalar number comes from 3 flavors of both fundamental and anti-fundamental representations.  The fundamental fermions have those plus another 6 fundamental fermion triplets from the gauginos and another 6 from the adjoint superfield.  Putting that together, one finds:
\begin{equation}\label{A2}
b_{3} =-3.
\end{equation}
Following standard techniques of integrating the one-loop beta function, one finds: 
\begin{equation}\label{A3}
\ln \left({\Lambda _{}^{2} \mathord{\left/ {\vphantom {\Lambda _{}^{2}  \mu ^{2} }} \right. \kern-\nulldelimiterspace} \mu ^{2} } \right)=-{16\pi ^{2} \mathord{\left/ {\vphantom {16\pi ^{2}  \left(g^{2} \left(\mu \right)\left(-b_{3} \right)\right)}} \right. \kern-\nulldelimiterspace} \left(g^{2} \left(\mu \right)\left(-b_{3} \right)\right)},
\end{equation}
where $\mu $ is the scale at which the SU(3) coupling is evaluated and $\Lambda $ is the quantum-generated scale of each SU(3) gauge theory.

The next step is to determine which fermion representations generate an axial anomaly.  In this model, and using notation similar to that of \cite{SQCD-Argyres}, the axial anomaly is proportional to:
\begin{equation}\label{anomaly}
\begin{aligned}
&\sum_f\rm{Tr}_{R_f}\left( T^0_{A_{R}+}\left(\{T^B_{V_{R}+},T^C_{V_{R}+}\}+\{T^B_{V_{R}-},T^C_{V_{R}-}\}\right) \right)\\
+&
\sum_f\rm{Tr}_{R_f} \left(2T^0_{A_{R}-}\{T^B_{V_{R}+},T^C_{V_{R}-}\}\right).
\end{aligned}
\end{equation}
In the above expression, $R$ represents the representation of fermion $f$. A representation's contribution to a local gauge current proportional to 
$T^B_{\pm}$ (as defined in eq \eqref{GrindEQ__1_7_}) is denoted by $T^B_{V_{R}\pm}$.  Similarly, $T^0_{A_{R}\pm}$ denote a representation's contributions to the global axial currents proportional to $T^0_{\pm}$.

Since they are in the $(3,\tilde{3})$ and $(\tilde{3},3)$ representations, the gauginos and the fermions from the twisted adjoint superfield contribute zero to both $T^0_{V_{R}+}$ and $T^0_{A_{R}+}$.  In addition, for the second term of \eqref{anomaly} they generate a factor of $f^{bad}d^{cad}$ which vanishes.  Therefore, those fermions do not contribute to the axial anomaly.  

It is speculated that the Abelian gauge field $T^0_{V_{R}\pm}$ contributions to the axial anomaly are nontopological and can be ignored, although in \cite{SQCD-Argyres}, it is pointed out that this is not always the case.  If they indeed can be ignored, then the remaining contributions to the anomaly for each SU(3) group can be treated independently, and they are identical to those for SQCD with 3 flavors and colors.  

In the SQCD derivation of nonperturbative low-energy superpotential terms, the next step is to consider instanton effects.  Following that approach, the instanton amplitude in Euclidean space is proportional to:
\begin{equation}\label{A4}
\exp \left(-S_{\rm{Inst}} \right)=\exp \left(-{8\pi ^{2} \mathord{\left/ {\vphantom {8\pi ^{2}  g^{2} \left(\mu \right)}} \right. \kern-\nulldelimiterspace} g^{2} \left(\mu \right)} \right)=\left({\Lambda \mathord{\left/ {\vphantom {\Lambda  \mu }} \right. \kern-\nulldelimiterspace} \mu } \right)^{3},
\end{equation}
where the second equality uses eqs \eqref{A2} and \eqref{A3}. For three flavors of fundamental (and anti-fundamental) fields, the instanton amplitude carries an axial U(1) charge of 6.  For eq \eqref{A4} to be consistent, the axial U(1) selection rule implies that $\Lambda $ should carry an axial U(1) charge of 2.

The anomaly-generating axial U(1) charges of the superfields $Q_{mF} $, $\tilde{Q}_{mF} $, and $\Phi $ are 1, 1, and 0, respectively.  Therefore, the following factors are invariant to most of the U(3)$\times$U(3) local gauge symmetry and obey the axial selection rule:
\begin{equation}\label{A5}
\begin{aligned}
&{\rm Tr}\left(\Phi ^{n} \right) \\
&M_{mm'FF'}^{\left(n\right)} =\Lambda^{-1}\tilde{Q}_{mF} \Phi ^{n} Q_{m'F'},
\end{aligned}
\end{equation}
where $n$ is a non-negative integer.  From eq \eqref{GrindEQ__1_14_}, all of the ``meson'' configurations ($M$) above are invariant to nonAbelian supergauge transformations, but some are not invariant to Abelian supergauge transformations dependent on $\Lambda_+^0$. This is addressed below.

It should be noted that $M_{mm'FF'}^{\left(n\right)}$ with $m\ne m'$ is proportional to $\theta $, so terms involving these factors do not contribute to the vev of the scalar potential or its minimization, although they do contribute to lepton masses and neutrino mixing as described below.

In SQCD, the allowed nonperturbatively generated terms also have to respect flavor symmetries.  In this model, that is complicated for quantum-generated mass terms since $Q_{1F}$ and $\tilde{Q}_{1F}$ have a $3\times 3$ flavor symmetry (since $q_{11}=q_{12}=q_{13}$), but $Q_{2F}$ and $\tilde{Q}_{2F}$ only have a $1\oplus 2\times 2$ symmetry (since $q_{21}\ne q_{22}=q_{23}$).  The flavor symmetry will be incorporated as follows: First, one of the $m=1$ flavors will be identified with the $m=2$ flavor $F=1$ and terms involving flavor determinants will be constructed.  Then the other two $m=1$ flavors will be identified with the $m=2$ flavor $F=1$ and analogous terms will be constructed.
  
Identifying the $m=1$ flavor $F=1$ with the $m=2$ flavor $F=1$, the following combination chiral fields and meson operators can be constructed:
\begin{equation}\label{QF}
\begin{aligned}
&Q'_{1F}=Q_{1F}+Q_{2F}\hspace{1cm}\tilde{Q}'_{1F}=\tilde{Q}_{1F}+\tilde{Q}_{2F} \\
&M_{1FF'}^{(n)} =\Lambda^{-1}\tilde{Q}'_{1F} \Phi ^{n} Q'_{1F'}\,.
\end{aligned}
\end{equation}
The subscript ``1'' on $Q'$, $\tilde{Q}'$, and $M$ denotes that the $m=1$ flavor $F=1$ was identified with the $m=2$ flavor $F=1$ in the above definitions.  Next, as mentioned above, the $m=1$ flavors are redefined with $1\to 2\to 3\to 1$ (e.g. $Q'_{21}=Q_{12}+Q_{21}$ or $Q'_{31}=Q_{13}+Q_{21}$). 

Just as flavor determinants are used in the quantum vacuum of SQCD, the following flavor determinant is defined:
\begin{equation}\label{A5b}
\Delta^{(pqr)}_{i} = {\rm det}\left|\begin{array}{ccc} 
{M_{i11}^{(p)}} & {M_{i12}^{(q)}} & {M_{i13}^{(r)}} \\ 
{M_{i21}^{(p)}} & {M_{i22}^{(q)}} & {M_{i23}^{(r)}} \\ 
{M_{i31}^{(p)}} & {M_{i32}^{(q)}} & {M_{i33}^{(r)}}  \end{array}\right|\,.
\end{equation}
As mentioned above, some of the $M$ factors are not invariant to Abelian supergauge transformations dependent on $\Lambda_+^0$.  

Any terms in the determinant that are not completely supergauge invariant are set to zero. That means for example that in the term $M_{111}^{(0)}M_{123}^{(0)}M_{132}^{(0)}$,
\begin{equation}\label{nogo}
M_{111}^{(0)}\tilde{Q}_{12} Q_{23}\tilde{Q}_{23}Q_{22}=0\,,
\end{equation}
since under a supergauge transformation, $\tilde{Q}_{12} Q_{23}\to\tilde{Q}_{12} Q_{23}e^{i\Lambda_+^0/\sqrt{12}}$ and there is nothing else in that term that can cancel the $\Lambda_+^0$ gauge dependence.  On the other hand,
\begin{equation}\label{tricky}
M_{111}^{(0)}\tilde{Q}_{12} Q_{23}\tilde{Q}_{23}Q_{12}\ne 0\,,
\end{equation}
since the $\tilde{Q}_{11} Q_{22}$ and $\tilde{Q}_{22} Q_{11}$ factors have $\Lambda_+^0$ gauge dependencies that cancel each other. 

Following symmetry arguments similar to those used for SQCD, quantum interactions will generate low energy effective superpotential terms with mass dimension 3 and one power of $\Lambda$ in the numerator.  These terms are constructed from the above flavor determinants as well as factors of $\left({\rm Tr}(\Phi ^2) \right)^{s}$ and other gauge invariant factors such as those mentioned later in this Appendix. 

An example of a term meeting the above criteria is the following:
\begin{equation}\label{A8}
O_0=\left({\rm Tr}\left(\Phi ^{2} \right)\right)^2\left(\tfrac{1}{3}\sum_{i}\Delta^{(000)}_{i} \right)^{-1/3} \,.
\end{equation}
The scalar potential can be derived from superpotential terms like the one above by taking Taylor-expansion derivatives to extract ``f terms'' proportional to $\theta^2 f$ and inserting them into the f-term part of eq \eqref{GrindEQ__2_1_}.  The quantum vacuum can be found by minimizing the vev of the scalar potential after including these quantum contributions.

In this process, vevs of scalar potential terms derived from $O_0$ will be proportional to
\begin{equation}\label{vevO0}
\begin{aligned}
\bar{O}_0&=\left\langle O_0\right\rangle = \frac{\Lambda\bar{\varphi}_1^4}{\left(\bar{\phi}_{21}\bar{\tilde{\phi}}_{21}\bar{\phi}_{22}\bar{\tilde{\phi}}_{22}\bar{\phi}_{23}\bar{\tilde{\phi}}_{23}\right)^{1/3}} \\
\bar{\varphi}_1^2&={\rm tr}\left\langle \varphi_1^2\right\rangle\,,
\end{aligned}
\end{equation}
where it was argued at the beginning of section 3 that ${\rm tr}\left\langle \varphi_1^2\right\rangle\ne 0$ in the quantum vacuum.  In section 2, it was noted that if the model includes large tree-level masses $m_{22}$, $\tilde{m}_{21}$ and $\tilde{m}_{22}$, then the vevs $\bar{\phi}_{22}$, $\bar{\tilde{\phi}}_{21}$ and $\bar{\tilde{\phi}}_{22}$ vanish classically.  However, in order to stabilize expressions such as the one above, it is assumed that these vevs acquire small values quantum mechanically.  This is mentioned in section 3.  As shown below, those quantum-generated small vevs lead to very large right-chiral neutrino Majorana mass terms that allow the model to reproduce observed neutrino masses and mixing.

The classical scalar potential only restricts the difference $\bar{\phi}_{23}^2-\bar{\tilde{\phi}}_{23}^2$.  However, it can be seen from eq \eqref{vevO0} that the quantum scalar potential will try to make each of these vevs go to infinity while maintaining the difference.  A mechanism like that is what causes Supersymmetric QCD with fewer flavors than colors to not have a vacuum solution.  But as described in section 2, a nonzero tree-level or quantum-generated superpotential coupling provides a small counterbalancing effect that stops $\bar{\phi}_{23}$ from becoming infinite. 

Once the scalar vevs have been adjusted and a minimum of the quantum potential has been achieved, second-derivative Taylor-expansion terms of the quantum superpotential can also produce fermion and scalar mass terms.  

For example, taking derivatives with respect to $\tilde{Q}'_{i1}$ and $\tilde{Q}'_{j1}$, evaluating that expression at its vev, then multiplying by $\tilde{Q}'_{i1}$ and $\tilde{Q}'_{j1}$ generates the following mass terms
\begin{equation}\label{Majorana-mass}
\frac{4\theta^2\bar{O}_0}{9\bar{\tilde{\phi}}_{21}^2}\left(2(\tilde{f}_{21})_1(\tilde{\phi}_{21})_1-\tfrac{1}{9}\sum_{FF^\prime}
\tilde{\nu}_W^{(F)}\tilde{\nu}_W^{(F^\prime)}\right)\,,
\end{equation}
where $2\theta\tilde{\nu}_W^{(F)}\theta\tilde{\nu}_W^{(F^\prime)}=-\theta^2 \tilde{\nu}_W^{(F)}\tilde{\nu}_W^{(F^\prime)}$ was used.  Since the quantum-generated vev $\bar{\tilde{\phi}}_{21}$ is very small, the second term generates a very large neutrino Majorana mass matrix.  The first term generates the corresponding scalar mass, where the notation $(\tilde{\phi}_{21})_1$ denotes the first SU(3) component of the $\tilde{\phi}_{21}$ scalar triplet.

Taking derivatives of $O_0$ with respect to $\tilde{Q}'_{i1}$ and $Q'_{j1}$ generates the following mass terms:
\begin{equation}\label{quant-mass}
\begin{aligned}
&-\frac{\bar{O}_0}{3\bar{\tilde{\phi}}_{21}\bar{\phi}_{21}}\left(
\tilde{Q}_{21}Q_{21}
-\tfrac{4}{3}(\tilde{Q}_{21})_4(Q_{21})_4
\right) \\
&+\frac{\bar{O}_0\theta^2}{9\bar{\tilde{\phi}}_{21}\bar{\phi}_{21}}
\sum_F\left(\tilde{\nu}_W^{(F)}\nu_W^{(F)}
-\tfrac{4}{9}\sum_{F^\prime}\tilde{\nu}_W^{(F)}\nu_W^{(F^\prime)}\right)
\,,
\end{aligned}
\end{equation}
where  a variant of eq \eqref{tricky} was used.  The second line are neutrino Dirac masses that can form seesaws with the heavy neutrinos.  This is discussed in the Neutrino part of section 3.

Taking derivatives with respect to $\tilde{Q}'_{i2}$ and $Q'_{j2}$ generates more mass terms:
\begin{equation}\label{quant-mass2}
\begin{aligned}
&-\frac{\bar{O}_0}{3\bar{\tilde{\phi}}_{22}\bar{\phi}_{22}}\left(
\tilde{Q}_{22}Q_{22}
-\tfrac{4}{3}(\tilde{Q}_{22})_5(Q_{22})_5
\right) \\
&+\frac{\bar{O}_0\theta^2}{9\bar{\tilde{\phi}}_{22}\bar{\phi}_{22}}
\sum_F\tilde{e}_W^{(F)}e_W^{(F)}
\,.
\end{aligned}
\end{equation}
The masses in the second line form part of a seesaw to generate the observed charged lepton masses.  Due to the applicable variant of eq \eqref{nogo}, this term is diagonal in the charged lepton flavor (unlike neutrino mass terms). 

Derivatives can also be taken with respect to adjoint fields.  Taking derivatives with respect to the adjoint fermions $\chi$ and $\tilde{\chi}$ generates the following quark mass terms:
\begin{equation}\label{adjoint-mass}
\frac{\bar{O}_0\theta^2}{\bar{\varphi}_1^2}\left(
\bar{c}'_L c'_R + \bar{s}'_L f'_R + \bar{f}'_L b'_R\,,
\right)
\end{equation}
where the notation described below eq \eqref{4x4fit} for quark gauge eigenstates is being used.  From the CKM fit to quark mass matrices in eq \eqref{4x4fit}, $f'_L=.99f_L+.14b_L+...$.  So the above quantum generated term has a small $b\bar{b}$ component.

For each of the three cases above, if a third derivative is taken with respect to $Q_{21}$ before setting the expression to its vev, then $h_t$ Higgs interaction terms are generated.  These are proportional to the expressions above but multiplied by $(\phi_{21})_1/\bar{\phi}_{21}\simeq \hat{g}(\phi_{21})_1/m_t$, where eq \eqref{GrindEQ__3_15_} was used.  These terms allow $h_t$ to decay to leptons, to $c\bar{c}$ and to a lesser extent, to $b\bar{b}$. These decay mechanisms are mentioned in the Higgs decay part of section 5. 

If a third derivative with respect to $(\varphi_2)_{32}$ is taken instead, then new scalar interaction terms are generated for the scalar $(\varphi_2)_{32}$.  These terms are proportional to the expressions above but multiplied by $2(\varphi_2)_{32}\bar{\varphi}_2/\bar{\varphi}_1^2$.  So they enable decays of $(\varphi_2)_{32}$ to leptons, $c\bar{c}$ and to a lesser extent, to $b\bar{b}$.

The operator $O_0$ could be generalized by inserting various powers of adjoint fields in the flavor determinant.  One simple example generalization is
\begin{equation}\label{O1}
\Delta^{(000)}_i\to \Delta^{(000)}_i+\sum_n\alpha_n\Delta^{(00n)}_i/\left({\rm Tr}\left(\Phi ^{2} \right)\right)^{n/2}\,,
\end{equation}
where $\alpha_n$ are constants.  Of course this could be generalized further by placing various multiples of adjoint fields in different parts of the determinant. These additional terms not change the vev $\bar{O}_0$, but they do generate additional mass terms.  

For example, with $n=1$, if derivatives are taken with respect to $\tilde{Q}_{21}$ and $(\varphi_2)_{k1}$ then set to its vev, then multiplied by those fields, the following scalar mass terms are generated:
\begin{equation}\label{adjoint1mass}
-\frac{\bar{O}_0\theta^2}{3\bar{\varphi}_1\bar{\tilde{\phi}}_{21}}
\left((\tilde{f}_{21})_k(\varphi_2)_{k1}+(\tilde{\phi}_{21})_k(f_2)_{k1}\right)
\end{equation}
Solving the equations of motion for $(\tilde{f}_{21})_2$ and $(\tilde{f}_{21})_3$, this term generates large masses for $(\varphi_2)_{21}$ and $(\varphi_2)_{31}$.  It does not cause a large mass for $(\varphi_2)_{11}$, since the equation of motion for $(\tilde{f}_{21})_1$ is dominated by the much larger mass term of eq \eqref{Majorana-mass}.  The equations of motion for $(f_2)_{21}$ and $(f_2)_{21}$ generate those same large masses for $(\tilde{\phi}_{21})_2$ and $(\tilde{\phi}_{21})_3$.  

Since these masses are generated at a scale much smaller than the unification scale and they only involve the $m=2$ sector, they could be generated by an effective theory after the $m=2$ SU(3) has been broken down to SU(2). In that case, the mass terms for $(\varphi_2)_{21}$ and $(\tilde{\phi}_{21})_2$ do not need to be the same as those for $(\varphi_2)_{31}$ and $(\tilde{\phi}_{21})_3$.  In section 5, it is estimated that the mass of $(\tilde{\phi}_{21})_2$ is $\sim$3.5 TeV and that of $(\varphi_2)_{31}$ is larger, $>8$ TeV.

There is another consequence of the mass generation happening in the sub-unification SU(2) theory.  Since SU(2) is a self-conjugate group, a transformation generates the additional mass term $(\varphi_2)_{13}(f_{21})_3+(f_2)_{13}(\phi_{21})_3$.  This results in the same large mass ($>8$ TeV) for the scalars $(\varphi_2)_{13}(f_{21})_3$ and $(\phi_{21})_3$. 

By taking $n=2$ in eq \eqref{O1}, the following quantum superpotential term is generated:
\begin{equation}\label{charm-decay}
-i\frac{\sqrt{2}\theta^2\bar{O}_c}{3\bar{\varphi}_1^2\bar{\tilde{\phi}}_{21}}\left(\tilde{u}^{(1)}\varphi_1 u_W^A\right)\,.
\end{equation}
Just as for the Majorana neutrino masses, since $\bar{\tilde{\phi}}_{21}$ is very small, this quantum-generated coupling is large.  This coupling is used in section 5 when describing a way that the model could potentially reproduce $Z$ pole observables.

Another expression that is supergauge invariant is 
\begin{equation}\label{adjoint-gauge}
{\rm Tr}\left(W^{0\alpha} W'_\alpha\Phi \right)
\end{equation}
The above expression has dimension 4 and no axial anomaly charge, so it can replace $\left({\rm Tr}\left(\Phi ^{2} \right)\right)^2$ in $O_0$.  Since from eq \eqref{dfields}, $\left\langle W^{0\alpha}\right\rangle\ne 0$, eq \eqref{adjoint-gauge} generates quark mass terms that mix gaugino quarks with adjoint quarks, such as $\tilde{M}_{G}^{(1)}$ in eq \eqref{GrindEQ__3_14_}.  

Presumably, detailed nonperturbative calculations could be performed to determine exact forms of superpotential terms for this theory.  In the absence of those calculations which would produce the exact parameter values and scalar vevs at the quantum minimum, this paper has phenomenologically determined the values those parameters and vevs would have to take in order to reproduce experimental data.


\begin{thebibliography}{76}%
\makeatletter
\providecommand \@ifxundefined [1]{%
 \@ifx{#1\undefined}
}%
\providecommand \@ifnum [1]{%
 \ifnum #1\expandafter \@firstoftwo
 \else \expandafter \@secondoftwo
 \fi
}%
\providecommand \@ifx [1]{%
 \ifx #1\expandafter \@firstoftwo
 \else \expandafter \@secondoftwo
 \fi
}%
\providecommand \natexlab [1]{#1}%
\providecommand \enquote  [1]{``#1''}%
\providecommand \bibnamefont  [1]{#1}%
\providecommand \bibfnamefont [1]{#1}%
\providecommand \citenamefont [1]{#1}%
\providecommand \href@noop [0]{\@secondoftwo}%
\providecommand \href [0]{\begingroup \@sanitize@url \@href}%
\providecommand \@href[1]{\@@startlink{#1}\@@href}%
\providecommand \@@href[1]{\endgroup#1\@@endlink}%
\providecommand \@sanitize@url [0]{\catcode `\\12\catcode `\$12\catcode
  `\&12\catcode `\#12\catcode `\^12\catcode `\_12\catcode `\%12\relax}%
\providecommand \@@startlink[1]{}%
\providecommand \@@endlink[0]{}%
\providecommand \url  [0]{\begingroup\@sanitize@url \@url }%
\providecommand \@url [1]{\endgroup\@href {#1}{\urlprefix }}%
\providecommand \urlprefix  [0]{URL }%
\providecommand \Eprint [0]{\href }%
\providecommand \doibase [0]{https://doi.org/}%
\providecommand \selectlanguage [0]{\@gobble}%
\providecommand \bibinfo  [0]{\@secondoftwo}%
\providecommand \bibfield  [0]{\@secondoftwo}%
\providecommand \translation [1]{[#1]}%
\providecommand \BibitemOpen [0]{}%
\providecommand \bibitemStop [0]{}%
\providecommand \bibitemNoStop [0]{.\EOS\space}%
\providecommand \EOS [0]{\spacefactor3000\relax}%
\providecommand \BibitemShut  [1]{\csname bibitem#1\endcsname}%
\let\auto@bib@innerbib\@empty
\bibitem [{\citenamefont {Haag}\ \emph {et~al.}(1975)\citenamefont {Haag},
  \citenamefont {Lopuszanski},\ and\ \citenamefont {Sohnius}}]{HLS}%
  \BibitemOpen
  \bibfield  {author} {\bibinfo {author} {\bibfnamefont {R.}~\bibnamefont
  {Haag}}, \bibinfo {author} {\bibfnamefont {J.}~\bibnamefont {Lopuszanski}},\
  and\ \bibinfo {author} {\bibfnamefont {M.}~\bibnamefont {Sohnius}},\
  }\bibfield  {title} {\bibinfo {title} {{All Possible Generators of
  Supersymmetries of the S-Matrix}},\ }\href@noop {} {\bibfield  {journal}
  {\bibinfo  {journal} {Nuclear Physics B}\ }\textbf {\bibinfo {volume} {88}},\
  \bibinfo {pages} {257} (\bibinfo {year} {1975})}\BibitemShut {NoStop}%
\bibitem [{\citenamefont {Biggio}\ \emph {et~al.}(2016)\citenamefont {Biggio},
  \citenamefont {Dror}, \citenamefont {Grossman},\ and\ \citenamefont
  {Ng}}]{slepton-Higgs}%
  \BibitemOpen
  \bibfield  {author} {\bibinfo {author} {\bibfnamefont {C.}~\bibnamefont
  {Biggio}}, \bibinfo {author} {\bibfnamefont {J.}~\bibnamefont {Dror}},
  \bibinfo {author} {\bibfnamefont {Y.}~\bibnamefont {Grossman}},\ and\
  \bibinfo {author} {\bibfnamefont {W.}~\bibnamefont {Ng}},\ }\href@noop {}
  {\bibinfo {title} {{Probing a slepton Higgs on all frontiers}}} (\bibinfo
  {year} {2016}),\ \Eprint {https://arxiv.org/abs/1602.02162} {1602.02162}
  \BibitemShut {NoStop}%
\bibitem [{\citenamefont {Chapman}(2025{\natexlab{a}})}]{Twisted-Superfields}%
  \BibitemOpen
  \bibfield  {author} {\bibinfo {author} {\bibfnamefont {S.}~\bibnamefont
  {Chapman}},\ }\bibfield  {title} {\bibinfo {title} {A new form of soft
  supersymmetry breaking?},\ }\href
  {https://doi.org/10.1007/s40509-024-00355-2} {\bibfield  {journal} {\bibinfo
  {journal} {Quantum Studies: Mathematics and Foundations}\ }\textbf {\bibinfo
  {volume} {12}},\ \bibinfo {pages} {12} (\bibinfo {year}
  {2025}{\natexlab{a}})},\ \Eprint {https://arxiv.org/abs/2104.03898}
  {2104.03898} \BibitemShut {NoStop}%
\bibitem [{\citenamefont {Crivellin}\ and\ \citenamefont
  {Mellado}(2024)}]{Crivellin_2024}%
  \BibitemOpen
  \bibfield  {author} {\bibinfo {author} {\bibfnamefont {A.}~\bibnamefont
  {Crivellin}}\ and\ \bibinfo {author} {\bibfnamefont {B.}~\bibnamefont
  {Mellado}},\ }\bibfield  {title} {\bibinfo {title} {Anomalies in particle
  physics and their implications for physics beyond the standard model},\
  }\href {https://doi.org/10.1038/s42254-024-00703-6} {\bibfield  {journal}
  {\bibinfo  {journal} {Nature Reviews Physics}\ }\textbf {\bibinfo {volume}
  {6}},\ \bibinfo {pages} {294–309} (\bibinfo {year} {2024})},\ \Eprint
  {https://arxiv.org/abs/2309.03870} {2309.03870} \BibitemShut {NoStop}%
\bibitem [{\citenamefont {Chapman}(2022)}]{alternative}%
  \BibitemOpen
  \bibfield  {author} {\bibinfo {author} {\bibfnamefont {S.}~\bibnamefont
  {Chapman}},\ }\bibfield  {title} {\bibinfo {title} {{An Alternative to the
  Standard Model}},\ }\href {https://doi.org/10.1007/s40509-021-00268-4}
  {\bibfield  {journal} {\bibinfo  {journal} {Quant. Studies: Math and
  Foundations}\ }\textbf {\bibinfo {volume} {9}},\ \bibinfo {pages} {235}
  (\bibinfo {year} {2022})}\BibitemShut {NoStop}%
\bibitem [{\citenamefont {Argurio}(2017)}]{rargurio}%
  \BibitemOpen
  \bibfield  {author} {\bibinfo {author} {\bibfnamefont {R.}~\bibnamefont
  {Argurio}},\ }\href {http://homepages.ulb.ac.be/~rargurio/susycourse.pdf}
  {\bibinfo {title} {Phys-f-417 supersymmetry course}} (\bibinfo {year}
  {2017}),\ \bibinfo {note} {{Lecture notes from Universite Libre de
  Bruxelles}}\BibitemShut {NoStop}%
\bibitem [{\citenamefont {Martin}(2016)}]{SUSY-Martin}%
  \BibitemOpen
  \bibfield  {author} {\bibinfo {author} {\bibfnamefont {S.}~\bibnamefont
  {Martin}},\ }\href@noop {} {\bibinfo {title} {{A Supersymmetry Primer}}}
  (\bibinfo {year} {2016}),\ \Eprint {https://arxiv.org/abs/hep-ph/9709356v7}
  {hep-ph/9709356v7} \BibitemShut {NoStop}%
\bibitem [{\citenamefont {Haber}\ and\ \citenamefont
  {Haskins}(2018)}]{SUSY-Haber}%
  \BibitemOpen
  \bibfield  {author} {\bibinfo {author} {\bibfnamefont {H.}~\bibnamefont
  {Haber}}\ and\ \bibinfo {author} {\bibfnamefont {L.}~\bibnamefont
  {Haskins}},\ }\href@noop {} {\bibinfo {title} {{Supersymmetric Theory and
  Models}}} (\bibinfo {year} {2018}),\ \Eprint
  {https://arxiv.org/abs/hep-ph/1712.05926v4} {hep-ph/1712.05926v4}
  \BibitemShut {NoStop}%
\bibitem [{\citenamefont {Bertolini}(2019)}]{SUSY-Bertolini}%
  \BibitemOpen
  \bibfield  {author} {\bibinfo {author} {\bibfnamefont {M.}~\bibnamefont
  {Bertolini}},\ }\href {https://people.sissa.it/~bertmat/susycourse.pdf}
  {\bibinfo {title} {{Lectures on Supersymmetry}}} (\bibinfo {year}
  {2019})\BibitemShut {NoStop}%
\bibitem [{\citenamefont {Binetruy}(2006)}]{Binetruy}%
  \BibitemOpen
  \bibfield  {author} {\bibinfo {author} {\bibfnamefont {P.}~\bibnamefont
  {Binetruy}},\ }\href@noop {} {\emph {\bibinfo {title} {Supersymmetry: Theory,
  Experiment, and Cosmology}}}\ (\bibinfo  {publisher} {Oxford University
  Press},\ \bibinfo {year} {2006})\BibitemShut {NoStop}%
\bibitem [{\citenamefont {Gates~Jr.}\ \emph {et~al.}(1983)\citenamefont
  {Gates~Jr.}, \citenamefont {Grisaru}, \citenamefont {Rocek},\ and\
  \citenamefont {Siegel}}]{SUSY-Gates}%
  \BibitemOpen
  \bibfield  {author} {\bibinfo {author} {\bibfnamefont {S.}~\bibnamefont
  {Gates~Jr.}}, \bibinfo {author} {\bibfnamefont {M.}~\bibnamefont {Grisaru}},
  \bibinfo {author} {\bibfnamefont {M.}~\bibnamefont {Rocek}},\ and\ \bibinfo
  {author} {\bibfnamefont {W.}~\bibnamefont {Siegel}},\ }\bibfield  {title}
  {\bibinfo {title} {{Superspace or One thousand and one lessons in
  supersymmetry}},\ }\href@noop {} {\bibfield  {journal} {\bibinfo  {journal}
  {Front. Phys.}\ }\textbf {\bibinfo {volume} {58}},\ \bibinfo {pages} {1}
  (\bibinfo {year} {1983})},\ \Eprint {https://arxiv.org/abs/hep-th/0108200}
  {hep-th/0108200} \BibitemShut {NoStop}%
\bibitem [{\citenamefont {Argyres}(2001)}]{SQCD-Argyres}%
  \BibitemOpen
  \bibfield  {author} {\bibinfo {author} {\bibfnamefont {P.}~\bibnamefont
  {Argyres}},\ }\href
  {http://homepages.uc.edu/~argyrepc/cu661-gr-SUSY/susy2001.pdf} {\bibinfo
  {title} {Intro to global supersymmetry}} (\bibinfo {year} {2001}),\ \bibinfo
  {note} {{Cornell University Course}}\BibitemShut {NoStop}%
\bibitem [{\citenamefont {Intriligator}\ and\ \citenamefont
  {Seiberg}(1996)}]{intriligator-seiberg}%
  \BibitemOpen
  \bibfield  {author} {\bibinfo {author} {\bibfnamefont {K.}~\bibnamefont
  {Intriligator}}\ and\ \bibinfo {author} {\bibfnamefont {N.}~\bibnamefont
  {Seiberg}},\ }\bibfield  {title} {\bibinfo {title} {Lectures on
  supersymmetric gauge theories and electric-magnetic duality},\ }\href@noop {}
  {\bibfield  {journal} {\bibinfo  {journal} {Nucl.Phys.Proc.Suppl.}\ }\textbf
  {\bibinfo {volume} {45BC}},\ \bibinfo {pages} {1} (\bibinfo {year} {1996})},\
  \Eprint {https://arxiv.org/abs/hep-th/9509066} {hep-th/9509066} \BibitemShut
  {NoStop}%
\bibitem [{\citenamefont {Bilal}(2001)}]{bilal}%
  \BibitemOpen
  \bibfield  {author} {\bibinfo {author} {\bibfnamefont {A.}~\bibnamefont
  {Bilal}},\ }\href {https://doi.org/10.48550/ARXIV.HEP-TH/0101055} {\bibinfo
  {title} {{Introduction to Supersymmetry}}} (\bibinfo {year} {2001}),\ \Eprint
  {https://arxiv.org/abs/hep-th/0101055} {hep-th/0101055} \BibitemShut
  {NoStop}%
\bibitem [{\citenamefont {Dreiner}\ \emph {et~al.}(2010)\citenamefont
  {Dreiner}, \citenamefont {Haber},\ and\ \citenamefont {Martin}}]{2component}%
  \BibitemOpen
  \bibfield  {author} {\bibinfo {author} {\bibfnamefont {H.~K.}\ \bibnamefont
  {Dreiner}}, \bibinfo {author} {\bibfnamefont {H.~E.}\ \bibnamefont {Haber}},\
  and\ \bibinfo {author} {\bibfnamefont {S.~P.}\ \bibnamefont {Martin}},\
  }\bibfield  {title} {\bibinfo {title} {{Two-component spinor techniques and
  Feynman rules for quantum field theory and supersymmetry}},\ }\href
  {https://doi.org/10.1016/j.physrep.2010.05.002} {\bibfield  {journal}
  {\bibinfo  {journal} {Physics Reports}\ }\textbf {\bibinfo {volume} {494}},\
  \bibinfo {pages} {1} (\bibinfo {year} {2010})},\ \Eprint
  {https://arxiv.org/abs/0812.1594} {0812.1594} \BibitemShut {NoStop}%
\bibitem [{\citenamefont {Seiberg}\ and\ \citenamefont
  {Witten}(1994)}]{seiberg-witten}%
  \BibitemOpen
  \bibfield  {author} {\bibinfo {author} {\bibfnamefont {N.}~\bibnamefont
  {Seiberg}}\ and\ \bibinfo {author} {\bibfnamefont {E.}~\bibnamefont
  {Witten}},\ }\bibfield  {title} {\bibinfo {title} {{Condensation, And
  Confinement In N=2 Supersymmetric Yang-Mills Theory}},\ }\href@noop {}
  {\bibfield  {journal} {\bibinfo  {journal} {Nuclear Physics B}\ }\textbf
  {\bibinfo {volume} {426}},\ \bibinfo {pages} {19} (\bibinfo {year} {1994})},\
  \Eprint {https://arxiv.org/abs/hep-th/9407087} {hep-th/9407087} \BibitemShut
  {NoStop}%
\bibitem [{\citenamefont {Seiberg}(1994{\natexlab{a}})}]{seiberg-duality}%
  \BibitemOpen
  \bibfield  {author} {\bibinfo {author} {\bibfnamefont {N.}~\bibnamefont
  {Seiberg}},\ }\bibfield  {title} {\bibinfo {title} {{Monopoles, Duality and
  Chiral Symmetry Breaking in N=2 Supersymmetric QCD}},\ }\href@noop {}
  {\bibfield  {journal} {\bibinfo  {journal} {Nuclear Physics B}\ }\textbf
  {\bibinfo {volume} {431}},\ \bibinfo {pages} {484} (\bibinfo {year}
  {1994}{\natexlab{a}})},\ \Eprint {https://arxiv.org/abs/hep-th/9408099}
  {hep-th/9408099} \BibitemShut {NoStop}%
\bibitem [{\citenamefont {Aad}\ and\ \citenamefont {{et
  al}}(2022{\natexlab{a}})}]{top-quark-polarization}%
  \BibitemOpen
  \bibfield  {author} {\bibinfo {author} {\bibfnamefont {G.}~\bibnamefont
  {Aad}}\ and\ \bibinfo {author} {\bibnamefont {{et al}}} (\bibinfo
  {collaboration} {{ATLAS}}),\ }\bibfield  {title} {\bibinfo {title}
  {Measurement of the polarisation of single top quarks and antiquarks produced
  in the t-channel at ${\ensuremath{\sqrt{s}}}$= 13 {TeV} and bounds on the
  {tWb} dipole operator from the {ATLAS} experiment},\ }\href
  {https://doi.org/10.1007/jhep11(2022)040} {\bibfield  {journal} {\bibinfo
  {journal} {Journal of High Energy Physics}\ }\textbf {\bibinfo {volume}
  {2022}},\ \bibinfo {pages} {40} (\bibinfo {year} {2022}{\natexlab{a}})},\
  \Eprint {https://arxiv.org/abs/2202.11382} {2202.11382} \BibitemShut
  {NoStop}%
\bibitem [{\citenamefont {Myhrer}\ and\ \citenamefont {Thomas}(2008)}]{pspin1}%
  \BibitemOpen
  \bibfield  {author} {\bibinfo {author} {\bibfnamefont {F.}~\bibnamefont
  {Myhrer}}\ and\ \bibinfo {author} {\bibfnamefont {A.}~\bibnamefont
  {Thomas}},\ }\bibfield  {title} {\bibinfo {title} {A possible resolution of
  the proton spin problem},\ }\href
  {https://doi.org/10.1016/j.physletb.2008.04.034} {\bibfield  {journal}
  {\bibinfo  {journal} {Physics Letters B}\ }\textbf {\bibinfo {volume}
  {663}},\ \bibinfo {pages} {302–305} (\bibinfo {year} {2008})},\ \Eprint
  {https://arxiv.org/abs/0709.4067} {0709.4067} \BibitemShut {NoStop}%
\bibitem [{\citenamefont {Zyla}\ \emph
  {et~al.}(2020{\natexlab{a}})\citenamefont {Zyla}, \citenamefont {{et al}},\
  and\ \citenamefont {{(Particle Data Group)}}}]{CKM}%
  \BibitemOpen
  \bibfield  {author} {\bibinfo {author} {\bibfnamefont {P.~A.}\ \bibnamefont
  {Zyla}}, \bibinfo {author} {\bibnamefont {{et al}}},\ and\ \bibinfo {author}
  {\bibnamefont {{(Particle Data Group)}}},\ }\href
  {https://pdg.lbl.gov/2021/reviews/rpp2020-rev-ckm-matrix.pdf} {\bibinfo
  {title} {{CKM Quark-Mixing Matrix}}} (\bibinfo {year} {2020}{\natexlab{a}}),\
  \bibinfo {note} {{Prog. Theor. Exp. Phys. 2020 and 2021 update,
  083C01}}\BibitemShut {NoStop}%
\bibitem [{\citenamefont {Rosner}\ \emph {et~al.}(2021)\citenamefont {Rosner},
  \citenamefont {Stone}, \citenamefont {Van~de Water},\ and\ \citenamefont
  {{(Particle Data Group)}}}]{CKM-Leptonic}%
  \BibitemOpen
  \bibfield  {author} {\bibinfo {author} {\bibfnamefont {J.}~\bibnamefont
  {Rosner}}, \bibinfo {author} {\bibfnamefont {S.}~\bibnamefont {Stone}},
  \bibinfo {author} {\bibfnamefont {R.}~\bibnamefont {Van~de Water}},\ and\
  \bibinfo {author} {\bibnamefont {{(Particle Data Group)}}},\ }\href
  {https://pdg.lbl.gov/2021/reviews/rpp2021-rev-pseudoscalar-meson-decay-cons.pdf}
  {\bibinfo {title} {Leptonic decays of charged pseudoscalar mesons}} (\bibinfo
  {year} {2021}),\ \bibinfo {note} {{Prog. Theor. Exp. Phys. 2020 and 2021
  update, 083C01}}\BibitemShut {NoStop}%
\bibitem [{\citenamefont {Athar}\ and\ \citenamefont {{et
  al}}(2022)}]{neutrino_review}%
  \BibitemOpen
  \bibfield  {author} {\bibinfo {author} {\bibfnamefont {M.~S.}\ \bibnamefont
  {Athar}}\ and\ \bibinfo {author} {\bibnamefont {{et al}}},\ }\bibfield
  {title} {\bibinfo {title} {Status and perspectives of neutrino physics},\
  }\href {https://doi.org/10.1016/j.ppnp.2022.103947} {\bibfield  {journal}
  {\bibinfo  {journal} {Prog. in Particle and Nucl. Physics}\ }\textbf
  {\bibinfo {volume} {124}},\ \bibinfo {pages} {103947} (\bibinfo {year}
  {2022})},\ \Eprint {https://arxiv.org/abs/2111.07586} {2111.07586}
  \BibitemShut {NoStop}%
\bibitem [{\citenamefont {Zyla}\ \emph {et~al.}(2021)\citenamefont {Zyla},
  \citenamefont {{et al}},\ and\ \citenamefont {{(Particle Data
  Group)}}}]{heavy-lepton}%
  \BibitemOpen
  \bibfield  {author} {\bibinfo {author} {\bibfnamefont {P.~A.}\ \bibnamefont
  {Zyla}}, \bibinfo {author} {\bibnamefont {{et al}}},\ and\ \bibinfo {author}
  {\bibnamefont {{(Particle Data Group)}}},\ }\href
  {https://pdg.lbl.gov/2021/listings/rpp2021-list-charged-lepton.pdf} {\bibinfo
  {title} {{Heavy Charged Lepton Searches}}} (\bibinfo {year} {2021}),\
  \bibinfo {note} {{Prog. Theor. Exp. Phys. 2020 and 2021 update,
  083C01}}\BibitemShut {NoStop}%
\bibitem [{\citenamefont {Sterling}\ and\ \citenamefont
  {Veltman}(1981)}]{sterling}%
  \BibitemOpen
  \bibfield  {author} {\bibinfo {author} {\bibfnamefont {T.}~\bibnamefont
  {Sterling}}\ and\ \bibinfo {author} {\bibfnamefont {M.}~\bibnamefont
  {Veltman}},\ }\bibfield  {title} {\bibinfo {title} {Decoupling in theories
  with anomalies},\ }\href
  {https://doi.org/https://doi.org/10.1016/0550-3213(81)90581-2} {\bibfield
  {journal} {\bibinfo  {journal} {Nuclear Physics B}\ }\textbf {\bibinfo
  {volume} {189}},\ \bibinfo {pages} {557} (\bibinfo {year}
  {1981})}\BibitemShut {NoStop}%
\bibitem [{\citenamefont {D'Hoker}\ and\ \citenamefont {Farhi}(1984)}]{dhoker}%
  \BibitemOpen
  \bibfield  {author} {\bibinfo {author} {\bibfnamefont {E.}~\bibnamefont
  {D'Hoker}}\ and\ \bibinfo {author} {\bibfnamefont {E.}~\bibnamefont
  {Farhi}},\ }\bibfield  {title} {\bibinfo {title} {Decoupling a fermion whose
  mass is generated by a yukawa coupling: The general case},\ }\href
  {https://doi.org/https://doi.org/10.1016/0550-3213(84)90586-8} {\bibfield
  {journal} {\bibinfo  {journal} {Nuclear Physics B}\ }\textbf {\bibinfo
  {volume} {248}},\ \bibinfo {pages} {59} (\bibinfo {year} {1984})}\BibitemShut
  {NoStop}%
\bibitem [{\citenamefont {Preskill}(1991)}]{anomaly-scale}%
  \BibitemOpen
  \bibfield  {author} {\bibinfo {author} {\bibfnamefont {J.}~\bibnamefont
  {Preskill}},\ }\bibfield  {title} {\bibinfo {title} {Gauge anomalies in an
  effective field theory},\ }\href
  {https://doi.org/https://doi.org/10.1016/0003-4916(91)90046-B} {\bibfield
  {journal} {\bibinfo  {journal} {Annals of Physics}\ }\textbf {\bibinfo
  {volume} {210}},\ \bibinfo {pages} {323} (\bibinfo {year} {1991})},\ \bibinfo
  {note} {preprint:
  https://www.preskill.caltech.edu/pubs/preskill-1991-anomalies.pdf}\BibitemShut
  {NoStop}%
\bibitem [{\citenamefont {Chapman}(2024{\natexlab{a}})}]{new-quark}%
  \BibitemOpen
  \bibfield  {author} {\bibinfo {author} {\bibfnamefont {S.}~\bibnamefont
  {Chapman}},\ }\href@noop {} {\bibinfo {title} {Fitting the exotic hadron
  spectrum with a seventh quark}} (\bibinfo {year} {2024}{\natexlab{a}}),\
  \Eprint {https://arxiv.org/abs/2203.03007} {2203.03007} \BibitemShut
  {NoStop}%
\bibitem [{\citenamefont {Chapman}(2025{\natexlab{b}})}]{HADRON2025}%
  \BibitemOpen
  \bibfield  {author} {\bibinfo {author} {\bibfnamefont {S.}~\bibnamefont
  {Chapman}},\ }\href {https://pos.sissa.it/500/089/pdf} {\bibinfo {title}
  {Fitting the exotic hadron spectrum with an additional quark}} (\bibinfo
  {year} {2025}{\natexlab{b}}),\ \bibinfo {note} {{Talk given at HADRON
  2025}}\BibitemShut {NoStop}%
\bibitem [{\citenamefont {Chapman}(2025{\natexlab{c}})}]{LHCbpres2025}%
  \BibitemOpen
  \bibfield  {author} {\bibinfo {author} {\bibfnamefont {S.}~\bibnamefont
  {Chapman}},\ }\href {https://indico.cern.ch/event/1543697} {\bibinfo {title}
  {Fitting the exotic hadron spectrum with an additional quark --with
  predictions for {LHCb}}} (\bibinfo {year} {2025}{\natexlab{c}}),\ \bibinfo
  {note} {{Recording of presentation made to LHCb (also available at YouTube
  {voN9o5y}{\textunderscore}{5lI})}}\BibitemShut {NoStop}%
\bibitem [{\citenamefont {Zyla}\ \emph
  {et~al.}(2020{\natexlab{b}})\citenamefont {Zyla}, \citenamefont {{et al}},\
  and\ \citenamefont {{(Particle Data Group)}}}]{CKM-Vcb}%
  \BibitemOpen
  \bibfield  {author} {\bibinfo {author} {\bibfnamefont {P.~A.}\ \bibnamefont
  {Zyla}}, \bibinfo {author} {\bibnamefont {{et al}}},\ and\ \bibinfo {author}
  {\bibnamefont {{(Particle Data Group)}}},\ }\href
  {https://pdg.lbl.gov/2020/reviews/rpp2020-rev-vcb-vub.pdf} {\bibinfo {title}
  {{Determination of $V_{cb}$ and $V_{ub}$}}} (\bibinfo {year}
  {2020}{\natexlab{b}}),\ \bibinfo {note} {{Prog. Theor. Exp. Phys. 2020 and
  2021 update, 083C01}}\BibitemShut {NoStop}%
\bibitem [{\citenamefont {Aaij}\ and\ \citenamefont {et~al}(2022)}]{Z-charm}%
  \BibitemOpen
  \bibfield  {author} {\bibinfo {author} {\bibfnamefont {R.}~\bibnamefont
  {Aaij}}\ and\ \bibinfo {author} {\bibnamefont {et~al}} (\bibinfo
  {collaboration} {LHCb}),\ }\bibfield  {title} {\bibinfo {title} {Study of
  $\mathit{Z}$ bosons produced in association with charm in the forward
  region},\ }\href {https://doi.org/10.1103/physrevlett.128.082001} {\bibfield
  {journal} {\bibinfo  {journal} {Phys. Rev. Lett.}\ }\textbf {\bibinfo
  {volume} {128}},\ \bibinfo {pages} {082001} (\bibinfo {year} {2022})},\
  \Eprint {https://arxiv.org/abs/2109.08084} {2109.08084} \BibitemShut
  {NoStop}%
\bibitem [{\citenamefont {Ball}\ \emph {et~al.}(2022)\citenamefont {Ball},
  \citenamefont {Candido}, \citenamefont {Cruz-Martinez}, \citenamefont
  {Forte}, \citenamefont {Giani}, \citenamefont {Hekhorn}, \citenamefont
  {Kudashkin}, \citenamefont {Magni},\ and\ \citenamefont
  {Rojo}}]{intrinsic_charm}%
  \BibitemOpen
  \bibfield  {author} {\bibinfo {author} {\bibfnamefont {R.~D.}\ \bibnamefont
  {Ball}}, \bibinfo {author} {\bibfnamefont {A.}~\bibnamefont {Candido}},
  \bibinfo {author} {\bibfnamefont {J.}~\bibnamefont {Cruz-Martinez}}, \bibinfo
  {author} {\bibfnamefont {S.}~\bibnamefont {Forte}}, \bibinfo {author}
  {\bibfnamefont {T.}~\bibnamefont {Giani}}, \bibinfo {author} {\bibfnamefont
  {F.}~\bibnamefont {Hekhorn}}, \bibinfo {author} {\bibfnamefont
  {K.}~\bibnamefont {Kudashkin}}, \bibinfo {author} {\bibfnamefont
  {G.}~\bibnamefont {Magni}},\ and\ \bibinfo {author} {\bibfnamefont
  {J.}~\bibnamefont {Rojo}} (\bibinfo {collaboration} {{NNPDF}}),\ }\bibfield
  {title} {\bibinfo {title} {Evidence for intrinsic charm quarks in the
  proton},\ }\href {https://doi.org/10.1038/s41586-022-04998-2} {\bibfield
  {journal} {\bibinfo  {journal} {Nature}\ }\textbf {\bibinfo {volume} {608}},\
  \bibinfo {pages} {483} (\bibinfo {year} {2022})},\ \Eprint
  {https://arxiv.org/abs/2208.08372} {2208.08372} \BibitemShut {NoStop}%
\bibitem [{\citenamefont {Zyla}\ and\ \citenamefont {{et
  al}}(2020)}]{PDG-Bp-2020}%
  \BibitemOpen
  \bibfield  {author} {\bibinfo {author} {\bibfnamefont {P.~A.}\ \bibnamefont
  {Zyla}}\ and\ \bibinfo {author} {\bibnamefont {{et al}}} (\bibinfo
  {collaboration} {Particle Data Group}),\ }\href
  {https://pdg.lbl.gov/2020/listings/rpp2020-list-b-prime-quark.pdf} {\bibinfo
  {title} {{b' (4th Generation) Quark, Searches for}}} (\bibinfo {year}
  {2020}),\ \bibinfo {note} {{Prog. Theor. Exp. Phys. 2020 and 2021 update,
  083C01}}\BibitemShut {NoStop}%
\bibitem [{\citenamefont {Olive}\ and\ \citenamefont {{et al}}(2022)}]{PDG_R}%
  \BibitemOpen
  \bibfield  {author} {\bibinfo {author} {\bibfnamefont {K.}~\bibnamefont
  {Olive}}\ and\ \bibinfo {author} {\bibnamefont {{et al}}} (\bibinfo
  {collaboration} {(Particle Data Group)}),\ }\href
  {https://pdg.lbl.gov/2022/hadronic-xsections/} {\bibinfo {title} {Data files
  and plots of cross-sections}} (\bibinfo {year} {2022}),\ \bibinfo {note}
  {{Prog. Theor. Exp. Phys. 2022, 083C01}}\BibitemShut {NoStop}%
\bibitem [{\citenamefont {Hayrapetyan}\ and\ \citenamefont
  {et~al}(2024)}]{CMS-6600}%
  \BibitemOpen
  \bibfield  {author} {\bibinfo {author} {\bibfnamefont {A.}~\bibnamefont
  {Hayrapetyan}}\ and\ \bibinfo {author} {\bibnamefont {et~al}} (\bibinfo
  {collaboration} {{CMS}}),\ }\bibfield  {title} {\bibinfo {title} {New
  structures in the $\ensuremath{J/\psi J/\psi}$ mass spectrum in proton-proton
  collisions at $\sqrt{s}$ = 13 {TeV}},\ }\href
  {https://doi.org/10.1103/physrevlett.132.111901} {\bibfield  {journal}
  {\bibinfo  {journal} {Physical Review Letters}\ }\textbf {\bibinfo {volume}
  {132}},\ \bibinfo {pages} {111901} (\bibinfo {year} {2024})},\ \Eprint
  {https://arxiv.org/abs/2306.07164} {2306.07164} \BibitemShut {NoStop}%
\bibitem [{\citenamefont {Aaboud}\ and\ \citenamefont {{et
  al}}(2023)}]{ATLAS-6600}%
  \BibitemOpen
  \bibfield  {author} {\bibinfo {author} {\bibfnamefont {M.}~\bibnamefont
  {Aaboud}}\ and\ \bibinfo {author} {\bibnamefont {{et al}}} (\bibinfo
  {collaboration} {ATLAS}),\ }\bibfield  {title} {\bibinfo {title} {Observation
  of an excess of di-charmonium events in the four-muon final state with
  {ATLAS} detector},\ }\href {https://doi.org/10.1103/PhysRevLett.131.151902}
  {\bibfield  {journal} {\bibinfo  {journal} {Phys.Rev. Lett.}\ }\textbf
  {\bibinfo {volume} {131}},\ \bibinfo {pages} {151902} (\bibinfo {year}
  {2023})},\ \Eprint {https://arxiv.org/abs/2304.08962} {2304.08962}
  \BibitemShut {NoStop}%
\bibitem [{\citenamefont {Aleev}\ and\ \citenamefont {{et.al.}}(1993)}]{3250}%
  \BibitemOpen
  \bibfield  {author} {\bibinfo {author} {\bibfnamefont {A.}~\bibnamefont
  {Aleev}}\ and\ \bibinfo {author} {\bibnamefont {{et.al.}}} (\bibinfo
  {collaboration} {EXCHARM}),\ }\bibfield  {title} {\bibinfo {title} {Narrow
  baryonia with open and hidden strangeness},\ }\href
  {https://lib-extopc.kek.jp/preprints/PDF/1993/9306/9306126.pdf} {\bibfield
  {journal} {\bibinfo  {journal} {Yadernaya Fizika}\ }\textbf {\bibinfo
  {volume} {56}},\ \bibinfo {pages} {100} (\bibinfo {year} {1993})}\BibitemShut
  {NoStop}%
\bibitem [{\citenamefont {Workman}\ \emph {et~al.}(2024)\citenamefont
  {Workman}, \citenamefont {{et al}},\ and\ \citenamefont {{(Particle Data
  Group)}}}]{PDG-mesons-2022}%
  \BibitemOpen
  \bibfield  {author} {\bibinfo {author} {\bibfnamefont {R.~L.}\ \bibnamefont
  {Workman}}, \bibinfo {author} {\bibnamefont {{et al}}},\ and\ \bibinfo
  {author} {\bibnamefont {{(Particle Data Group)}}},\ }\href
  {https://pdg.lbl.gov/} {\bibinfo {title} {{Particle Listings}}} (\bibinfo
  {year} {2024}),\ \bibinfo {note} {{Phys. Rev. D110, 030001}}\BibitemShut
  {NoStop}%
\bibitem [{\citenamefont {Besson}\ and\ \citenamefont {{et al}}(2007)}]{lowR2}%
  \BibitemOpen
  \bibfield  {author} {\bibinfo {author} {\bibfnamefont {D.}~\bibnamefont
  {Besson}}\ and\ \bibinfo {author} {\bibnamefont {{et al}}} (\bibinfo
  {collaboration} {CLEO}),\ }\bibfield  {title} {\bibinfo {title} {Measurement
  of total hadronic cross section in $\ensuremath{e^+e^-}$ annihilation below
  10.56 {GeV}},\ }\href {https://doi.org/10.1103/physrevd.76.072008} {\bibfield
   {journal} {\bibinfo  {journal} {Physical Review D}\ }\textbf {\bibinfo
  {volume} {76}},\ \bibinfo {pages} {072008} (\bibinfo {year} {2007})},\
  \Eprint {https://arxiv.org/abs/0706.2813} {0706.2813} \BibitemShut {NoStop}%
\bibitem [{\citenamefont {Chapman}(2024{\natexlab{b}})}]{exotics}%
  \BibitemOpen
  \bibfield  {author} {\bibinfo {author} {\bibfnamefont {S.}~\bibnamefont
  {Chapman}},\ }\href@noop {} {\bibinfo {title} {{Charmonium tetraquarks and
  pentaquarks or an additional quark?}}} (\bibinfo {year}
  {2024}{\natexlab{b}}),\ \Eprint {https://arxiv.org/abs/2204.00913}
  {2204.00913} \BibitemShut {NoStop}%
\bibitem [{\citenamefont {Apollinari}\ \emph {et~al.}(2005)\citenamefont
  {Apollinari}, \citenamefont {Barone}, \citenamefont {Carithers},
  \citenamefont {Dell'Orso}, \citenamefont {Dorigo}, \citenamefont {Fiori},
  \citenamefont {Franklin}, \citenamefont {Giannetti}, \citenamefont
  {Giromini}, \citenamefont {Happacher}, \citenamefont {Miscetti},
  \citenamefont {Parri}, \citenamefont {Ptohos},\ and\ \citenamefont
  {Velev}}]{narrow_search}%
  \BibitemOpen
  \bibfield  {author} {\bibinfo {author} {\bibfnamefont {G.}~\bibnamefont
  {Apollinari}}, \bibinfo {author} {\bibfnamefont {M.}~\bibnamefont {Barone}},
  \bibinfo {author} {\bibfnamefont {W.}~\bibnamefont {Carithers}}, \bibinfo
  {author} {\bibfnamefont {M.}~\bibnamefont {Dell'Orso}}, \bibinfo {author}
  {\bibfnamefont {T.}~\bibnamefont {Dorigo}}, \bibinfo {author} {\bibfnamefont
  {I.}~\bibnamefont {Fiori}}, \bibinfo {author} {\bibfnamefont
  {M.}~\bibnamefont {Franklin}}, \bibinfo {author} {\bibfnamefont
  {P.}~\bibnamefont {Giannetti}}, \bibinfo {author} {\bibfnamefont
  {P.}~\bibnamefont {Giromini}}, \bibinfo {author} {\bibfnamefont
  {F.}~\bibnamefont {Happacher}}, \bibinfo {author} {\bibfnamefont
  {S.}~\bibnamefont {Miscetti}}, \bibinfo {author} {\bibfnamefont
  {A.}~\bibnamefont {Parri}}, \bibinfo {author} {\bibfnamefont
  {F.}~\bibnamefont {Ptohos}},\ and\ \bibinfo {author} {\bibfnamefont
  {G.}~\bibnamefont {Velev}},\ }\bibfield  {title} {\bibinfo {title} {Search
  for narrow resonances below the upsilon mesons},\ }\href
  {https://doi.org/10.1103/physrevd.72.092003} {\bibfield  {journal} {\bibinfo
  {journal} {Physical Review D}\ }\textbf {\bibinfo {volume} {72}},\ \bibinfo
  {pages} {092003} (\bibinfo {year} {2005})},\ \Eprint
  {https://arxiv.org/abs/hep-ex/0507044} {hep-ex/0507044} \BibitemShut
  {NoStop}%
\bibitem [{\citenamefont {Aubert}\ and\ \citenamefont {{et
  al}}(2009)}]{Babar2008}%
  \BibitemOpen
  \bibfield  {author} {\bibinfo {author} {\bibfnamefont {B.}~\bibnamefont
  {Aubert}}\ and\ \bibinfo {author} {\bibnamefont {{et al}}} (\bibinfo
  {collaboration} {BABAR}),\ }\bibfield  {title} {\bibinfo {title} {Measurement
  of the $\ensuremath{e^+e^- \to b\bar{b}}$ cross section between
  $\ensuremath{\sqrt{s}}$ of 10.4 and 11.20 gev},\ }\bibfield  {journal}
  {\bibinfo  {journal} {Physical Review Letters}\ }\textbf {\bibinfo {volume}
  {102}},\ \href {https://doi.org/10.1103/physrevlett.102.012001}
  {10.1103/physrevlett.102.012001} (\bibinfo {year} {2009}),\ \Eprint
  {https://arxiv.org/abs/0809.4120} {0809.4120} \BibitemShut {NoStop}%
\bibitem [{\citenamefont {Dong}\ \emph {et~al.}(2020)\citenamefont {Dong},
  \citenamefont {Mo}, \citenamefont {Wang},\ and\ \citenamefont
  {Yuan}}]{bbar_China}%
  \BibitemOpen
  \bibfield  {author} {\bibinfo {author} {\bibfnamefont {X.-K.}\ \bibnamefont
  {Dong}}, \bibinfo {author} {\bibfnamefont {X.-H.}\ \bibnamefont {Mo}},
  \bibinfo {author} {\bibfnamefont {P.}~\bibnamefont {Wang}},\ and\ \bibinfo
  {author} {\bibfnamefont {C.-Z.}\ \bibnamefont {Yuan}},\ }\bibfield  {title}
  {\bibinfo {title} {Hadronic cross section of $\ensuremath{e^+e^-}$
  annihilation at bottomonium energy region},\ }\href
  {https://doi.org/10.1088/1674-1137/44/8/083001} {\bibfield  {journal}
  {\bibinfo  {journal} {Chinese Physics C}\ }\textbf {\bibinfo {volume} {44}},\
  \bibinfo {pages} {083001} (\bibinfo {year} {2020})}\BibitemShut {NoStop}%
\bibitem [{\citenamefont {Bodenstein}\ \emph {et~al.}(2012)\citenamefont
  {Bodenstein}, \citenamefont {Dominguez}, \citenamefont {Eidelman},
  \citenamefont {Spiesberger},\ and\ \citenamefont {Schilcher}}]{dim2con}%
  \BibitemOpen
  \bibfield  {author} {\bibinfo {author} {\bibfnamefont {S.}~\bibnamefont
  {Bodenstein}}, \bibinfo {author} {\bibfnamefont {C.~A.}\ \bibnamefont
  {Dominguez}}, \bibinfo {author} {\bibfnamefont {S.~I.}\ \bibnamefont
  {Eidelman}}, \bibinfo {author} {\bibfnamefont {H.}~\bibnamefont
  {Spiesberger}},\ and\ \bibinfo {author} {\bibfnamefont {K.}~\bibnamefont
  {Schilcher}},\ }\bibfield  {title} {\bibinfo {title} {Confronting
  electron-positron annihilation into hadrons with qcd: an operator product
  expansion analysis},\ }\bibfield  {journal} {\bibinfo  {journal} {Journal of
  High Energy Physics}\ }\textbf {\bibinfo {volume} {2012}},\ \href
  {https://doi.org/10.1007/jhep01(2012)039} {10.1007/jhep01(2012)039} (\bibinfo
  {year} {2012}),\ \Eprint {https://arxiv.org/abs/1110.2026} {1110.2026}
  \BibitemShut {NoStop}%
\bibitem [{\citenamefont {{ALEPH, DELPHI, L3, and OPAL
  Collaborations}}(2006)}]{Z-summary}%
  \BibitemOpen
  \bibfield  {author} {\bibinfo {author} {\bibnamefont {{ALEPH, DELPHI, L3, and
  OPAL Collaborations}}},\ }\bibfield  {title} {\bibinfo {title} {{Precision
  Electroweak Measurements on the Z Resonance}},\ }\href
  {https://doi.org/10.1016/j.physrep.2005.12.006} {\bibfield  {journal}
  {\bibinfo  {journal} {Physics Reports}\ }\textbf {\bibinfo {volume} {427}},\
  \bibinfo {pages} {257–454} (\bibinfo {year} {2006})},\ \Eprint
  {https://arxiv.org/abs/hep-ex/0509008} {hep-ex/0509008} \BibitemShut
  {NoStop}%
\bibitem [{\citenamefont {{\.{Z}}enczykowski}(2020)}]{WRHD}%
  \BibitemOpen
  \bibfield  {author} {\bibinfo {author} {\bibfnamefont {P.}~\bibnamefont
  {{\.{Z}}enczykowski}},\ }\bibfield  {title} {\bibinfo {title} {Revisiting
  weak radiative decays of hyperons},\ }\href
  {https://doi.org/10.5506/aphyspolb.51.2111} {\bibfield  {journal} {\bibinfo
  {journal} {Acta Physica Polonica B}\ }\textbf {\bibinfo {volume} {51}},\
  \bibinfo {pages} {2111} (\bibinfo {year} {2020})},\ \Eprint
  {https://arxiv.org/abs/2009.12552} {2009.12552} \BibitemShut {NoStop}%
\bibitem [{\citenamefont {Hara}(1964)}]{hara}%
  \BibitemOpen
  \bibfield  {author} {\bibinfo {author} {\bibfnamefont {Y.}~\bibnamefont
  {Hara}},\ }\bibfield  {title} {\bibinfo {title} {Nonleptonic decays of
  baryons and the eightfold way},\ }\href
  {https://doi.org/10.1103/PhysRevLett.12.378} {\bibfield  {journal} {\bibinfo
  {journal} {Phys. Rev. Lett.}\ }\textbf {\bibinfo {volume} {12}},\ \bibinfo
  {pages} {378} (\bibinfo {year} {1964})}\BibitemShut {NoStop}%
\bibitem [{\citenamefont {Aad}\ and\ \citenamefont {{et
  al}}(2022{\natexlab{b}})}]{ATLASttHb}%
  \BibitemOpen
  \bibfield  {author} {\bibinfo {author} {\bibfnamefont {G.}~\bibnamefont
  {Aad}}\ and\ \bibinfo {author} {\bibnamefont {{et al}}} (\bibinfo
  {collaboration} {{ATLAS}}),\ }\bibfield  {title} {\bibinfo {title}
  {Measurement of {Higgs} boson decay into b-quarks in associated production
  with a top-quark pair in pp collisions at sqrt s = 13 {TeV} with the {ATLAS}
  detector},\ }\bibfield  {journal} {\bibinfo  {journal} {Journal of High
  Energy Physics}\ }\textbf {\bibinfo {volume} {2022}},\ \href
  {https://doi.org/10.1007/jhep06(2022)097} {10.1007/jhep06(2022)097} (\bibinfo
  {year} {2022}{\natexlab{b}}),\ \Eprint {https://arxiv.org/abs/2111.06712}
  {2111.06712} \BibitemShut {NoStop}%
\bibitem [{\citenamefont {Hayrapetyan}\ \emph {et~al.}(2025)\citenamefont
  {Hayrapetyan}, \citenamefont {Tumasyan},\ and\ \citenamefont {{et
  al}}}]{CMSttHb}%
  \BibitemOpen
  \bibfield  {author} {\bibinfo {author} {\bibfnamefont {A.}~\bibnamefont
  {Hayrapetyan}}, \bibinfo {author} {\bibfnamefont {A.}~\bibnamefont
  {Tumasyan}},\ and\ \bibinfo {author} {\bibnamefont {{et al}}} (\bibinfo
  {collaboration} {{CMS}}),\ }\bibfield  {title} {\bibinfo {title} {Measurement
  of the {ttH} and {tH} production rates in the {H} to bb decay channel using
  proton-proton collision data at sqrt s = 13 {TeV}},\ }\bibfield  {journal}
  {\bibinfo  {journal} {Journal of High Energy Physics}\ }\textbf {\bibinfo
  {volume} {2025}},\ \href {https://doi.org/10.1007/jhep02(2025)097}
  {10.1007/jhep02(2025)097} (\bibinfo {year} {2025}),\ \Eprint
  {https://arxiv.org/abs/2407.10896} {2407.10896} \BibitemShut {NoStop}%
\bibitem [{\citenamefont {Misiak}\ and\ \citenamefont
  {Steinhauser}(2017)}]{Hpluslimit}%
  \BibitemOpen
  \bibfield  {author} {\bibinfo {author} {\bibfnamefont {M.}~\bibnamefont
  {Misiak}}\ and\ \bibinfo {author} {\bibfnamefont {M.}~\bibnamefont
  {Steinhauser}},\ }\bibfield  {title} {\bibinfo {title} {Weak radiative decays
  of the {B} meson and bounds on h+/- mass in the {Two-Higgs-Doublet Model}},\
  }\bibfield  {journal} {\bibinfo  {journal} {The European Physical Journal C}\
  }\textbf {\bibinfo {volume} {77}},\ \href
  {https://doi.org/10.1140/epjc/s10052-017-4776-y}
  {10.1140/epjc/s10052-017-4776-y} (\bibinfo {year} {2017})\BibitemShut
  {NoStop}%
\bibitem [{\citenamefont {Aad}\ and\ \citenamefont {{et
  al}}(2021)}]{ATLAShbmass}%
  \BibitemOpen
  \bibfield  {author} {\bibinfo {author} {\bibfnamefont {G.}~\bibnamefont
  {Aad}}\ and\ \bibinfo {author} {\bibnamefont {{et al}}} (\bibinfo
  {collaboration} {{ATLAS}}),\ }\bibfield  {title} {\bibinfo {title}
  {Measurements of {WH} and {ZH} production in the {H} to bb decay channel in
  pp collisions at 13 {TeV} with the {ATLAS} detector},\ }\bibfield  {journal}
  {\bibinfo  {journal} {The European Physical Journal C}\ }\textbf {\bibinfo
  {volume} {81}},\ \href {https://doi.org/10.1140/epjc/s10052-020-08677-2}
  {10.1140/epjc/s10052-020-08677-2} (\bibinfo {year} {2021}),\ \Eprint
  {https://arxiv.org/abs/2007.02873} {2007.02873} \BibitemShut {NoStop}%
\bibitem [{\citenamefont {Sirunyan}\ \emph {et~al.}(2018)\citenamefont
  {Sirunyan}, \citenamefont {Tumasyan},\ and\ \citenamefont {{et
  al}}}]{CMShbmass}%
  \BibitemOpen
  \bibfield  {author} {\bibinfo {author} {\bibfnamefont {A.}~\bibnamefont
  {Sirunyan}}, \bibinfo {author} {\bibfnamefont {A.}~\bibnamefont {Tumasyan}},\
  and\ \bibinfo {author} {\bibnamefont {{et al}}} (\bibinfo {collaboration}
  {{CMS}}),\ }\bibfield  {title} {\bibinfo {title} {Observation of {Higgs}
  boson decay to bottom quarks},\ }\bibfield  {journal} {\bibinfo  {journal}
  {Physical Review Letters}\ }\textbf {\bibinfo {volume} {121}},\ \href
  {https://doi.org/10.1103/physrevlett.121.121801}
  {10.1103/physrevlett.121.121801} (\bibinfo {year} {2018}),\ \Eprint
  {https://arxiv.org/abs/1808.08242} {1808.08242} \BibitemShut {NoStop}%
\bibitem [{\citenamefont {{The CMS Collaboration}}(2025)}]{CMSttHb2}%
  \BibitemOpen
  \bibfield  {author} {\bibinfo {author} {\bibnamefont {{The CMS
  Collaboration}}},\ }\href@noop {} {\bibinfo {title} {Simultaneous probe of
  the charm and bottom quark {Yukawa} couplings using {ttH} events}} (\bibinfo
  {year} {2025}),\ \Eprint {https://arxiv.org/abs/2509.22535} {2509.22535}
  \BibitemShut {NoStop}%
\bibitem [{\citenamefont {Chapman}(2025{\natexlab{d}})}]{VHcc}%
  \BibitemOpen
  \bibfield  {author} {\bibinfo {author} {\bibfnamefont {S.}~\bibnamefont
  {Chapman}},\ }\href
  {https://cms.cern/news/increased-confidence-higgs-boson-coupling-charm-quark}
  {\bibinfo {title} {Increased confidence for the {Higgs} boson coupling to the
  charm quark}} (\bibinfo {year} {2025}{\natexlab{d}})\BibitemShut {NoStop}%
\bibitem [{\citenamefont {Hayrapetyan}\ \emph {et~al.}(2024)\citenamefont
  {Hayrapetyan}, \citenamefont {Tumasyan},\ and\ \citenamefont {{et
  al}}}]{CMSVBF}%
  \BibitemOpen
  \bibfield  {author} {\bibinfo {author} {\bibfnamefont {A.}~\bibnamefont
  {Hayrapetyan}}, \bibinfo {author} {\bibfnamefont {A.}~\bibnamefont
  {Tumasyan}},\ and\ \bibinfo {author} {\bibnamefont {{et al}}} (\bibinfo
  {collaboration} {{CMS}}),\ }\bibfield  {title} {\bibinfo {title} {Measurement
  of the {Higgs} boson production via vector boson fusion and its decay into
  bottom quarks in proton-proton collisions at sqrt s = 13 {TeV}},\ }\bibfield
  {journal} {\bibinfo  {journal} {Journal of High Energy Physics}\ }\textbf
  {\bibinfo {volume} {2024}},\ \href {https://doi.org/10.1007/jhep01(2024)173}
  {10.1007/jhep01(2024)173} (\bibinfo {year} {2024}),\ \Eprint
  {https://arxiv.org/abs/2308.01253} {2308.01253} \BibitemShut {NoStop}%
\bibitem [{\citenamefont {Aad}\ and\ \citenamefont {{et
  al}}(2022{\natexlab{c}})}]{ATLASHiggs}%
  \BibitemOpen
  \bibfield  {author} {\bibinfo {author} {\bibfnamefont {G.}~\bibnamefont
  {Aad}}\ and\ \bibinfo {author} {\bibnamefont {{et al}}} (\bibinfo
  {collaboration} {{ATLAS}}),\ }\bibfield  {title} {\bibinfo {title} {A
  detailed map of {Higgs} boson interactions by the {ATLAS} experiment ten
  years after the discovery},\ }\href
  {https://doi.org/10.1038/s41586-022-04893-w} {\bibfield  {journal} {\bibinfo
  {journal} {Nature}\ }\textbf {\bibinfo {volume} {607}},\ \bibinfo {pages}
  {52–59} (\bibinfo {year} {2022}{\natexlab{c}})},\ \Eprint
  {https://arxiv.org/abs/2207.00092} {2207.00092} \BibitemShut {NoStop}%
\bibitem [{\citenamefont {Tumasyan}\ and\ \citenamefont {{et
  al}}(2023)}]{CMS95GeV}%
  \BibitemOpen
  \bibfield  {author} {\bibinfo {author} {\bibfnamefont {A.}~\bibnamefont
  {Tumasyan}}\ and\ \bibinfo {author} {\bibnamefont {{et al}}} (\bibinfo
  {collaboration} {{CMS}}),\ }\bibfield  {title} {\bibinfo {title} {Searches
  for additional {Higgs} bosons and for vector leptoquarks in di-tau final
  states in proton-proton collisions at sqrt s 13 {TeV}},\ }\bibfield
  {journal} {\bibinfo  {journal} {Journal of High Energy Physics}\ }\textbf
  {\bibinfo {volume} {2023}},\ \href {https://doi.org/10.1007/jhep07(2023)073}
  {10.1007/jhep07(2023)073} (\bibinfo {year} {2023}),\ \Eprint
  {https://arxiv.org/abs/2208.02717} {2208.02717} \BibitemShut {NoStop}%
\bibitem [{\citenamefont {Aad}\ and\ \citenamefont {{et
  al}}(2023)}]{ATLAStb2023}%
  \BibitemOpen
  \bibfield  {author} {\bibinfo {author} {\bibfnamefont {G.}~\bibnamefont
  {Aad}}\ and\ \bibinfo {author} {\bibnamefont {{et al}}} (\bibinfo
  {collaboration} {ATLAS}),\ }\bibfield  {title} {\bibinfo {title} {Search for
  vector-boson resonances decaying into a top quark and a bottom quark using pp
  collisions at sqrt s 13 tev with the {ATLAS} detector},\ }\bibfield
  {journal} {\bibinfo  {journal} {Journal of High Energy Physics}\ }\textbf
  {\bibinfo {volume} {2023}},\ \href {https://doi.org/10.1007/jhep12(2023)073}
  {10.1007/jhep12(2023)073} (\bibinfo {year} {2023}),\ \Eprint
  {https://arxiv.org/abs/2308.08521} {2308.08521} \BibitemShut {NoStop}%
\bibitem [{\citenamefont {Huang}(2025)}]{1710scalar}%
  \BibitemOpen
  \bibfield  {author} {\bibinfo {author} {\bibfnamefont {Y.}~\bibnamefont
  {Huang}},\ }\href {https://arxiv.org/abs/2503.13286} {\bibinfo {title}
  {Discovery of a glueball-like particle {X(2370) at BESIII}}} (\bibinfo {year}
  {2025}),\ \Eprint {https://arxiv.org/abs/2503.13286} {arXiv:2503.13286
  [hep-ex]} \BibitemShut {NoStop}%
\bibitem [{\citenamefont {Gorchtein}\ and\ \citenamefont
  {Seng}(2023)}]{Gorchtein_2023}%
  \BibitemOpen
  \bibfield  {author} {\bibinfo {author} {\bibfnamefont {M.}~\bibnamefont
  {Gorchtein}}\ and\ \bibinfo {author} {\bibfnamefont {C.-Y.}\ \bibnamefont
  {Seng}},\ }\bibfield  {title} {\bibinfo {title} {The standard model theory of
  neutron beta decay},\ }\href {https://doi.org/10.3390/universe9090422}
  {\bibfield  {journal} {\bibinfo  {journal} {Universe}\ }\textbf {\bibinfo
  {volume} {9}},\ \bibinfo {pages} {422} (\bibinfo {year} {2023})},\ \Eprint
  {https://arxiv.org/abs/2307.01145} {2307.01145} \BibitemShut {NoStop}%
\bibitem [{\citenamefont {Sasaki}\ \emph {et~al.}(2003)\citenamefont {Sasaki},
  \citenamefont {Orginos}, \citenamefont {Ohta},\ and\ \citenamefont
  {Blum}}]{Sasaki_2003}%
  \BibitemOpen
  \bibfield  {author} {\bibinfo {author} {\bibfnamefont {S.}~\bibnamefont
  {Sasaki}}, \bibinfo {author} {\bibfnamefont {K.}~\bibnamefont {Orginos}},
  \bibinfo {author} {\bibfnamefont {S.}~\bibnamefont {Ohta}},\ and\ \bibinfo
  {author} {\bibfnamefont {T.}~\bibnamefont {Blum}},\ }\bibfield  {title}
  {\bibinfo {title} {Nucleon axial charge from quenched lattice qcd with domain
  wall fermions},\ }\bibfield  {journal} {\bibinfo  {journal} {Physical Review
  D}\ }\textbf {\bibinfo {volume} {68}},\ \href
  {https://doi.org/10.1103/physrevd.68.054509} {10.1103/physrevd.68.054509}
  (\bibinfo {year} {2003}),\ \Eprint {https://arxiv.org/abs/hep-lat/0306007}
  {hep-lat/0306007} \BibitemShut {NoStop}%
\bibitem [{\citenamefont {Simonov}\ and\ \citenamefont
  {Trusov}(2009)}]{Simonov_2009}%
  \BibitemOpen
  \bibfield  {author} {\bibinfo {author} {\bibfnamefont {Y.~A.}\ \bibnamefont
  {Simonov}}\ and\ \bibinfo {author} {\bibfnamefont {M.~A.}\ \bibnamefont
  {Trusov}},\ }\bibfield  {title} {\bibinfo {title} {Nucleon matrix elements
  and baryon masses in the dirac orbital model},\ }\href
  {https://doi.org/10.1134/s1063778809060180} {\bibfield  {journal} {\bibinfo
  {journal} {Physics of Atomic Nuclei}\ }\textbf {\bibinfo {volume} {72}},\
  \bibinfo {pages} {1058–1062} (\bibinfo {year} {2009})},\ \Eprint
  {https://arxiv.org/abs/hep-ph/0607075} {hep-ph/0607075} \BibitemShut
  {NoStop}%
\bibitem [{\citenamefont {Ashman}\ and\ \citenamefont
  {et~al}(1988)}]{protonspincrisis}%
  \BibitemOpen
  \bibfield  {author} {\bibinfo {author} {\bibfnamefont {J.}~\bibnamefont
  {Ashman}}\ and\ \bibinfo {author} {\bibnamefont {et~al}} (\bibinfo
  {collaboration} {European Muon Collaboration}),\ }\bibfield  {title}
  {\bibinfo {title} {A measurement of the spin asymmetry and determination of
  the structure function g1 in deep inelastic muon-proton scattering},\ }\href
  {https://doi.org/https://doi.org/10.1016/0370-2693(88)91523-7} {\bibfield
  {journal} {\bibinfo  {journal} {Physics Letters B}\ }\textbf {\bibinfo
  {volume} {206}},\ \bibinfo {pages} {364} (\bibinfo {year}
  {1988})}\BibitemShut {NoStop}%
\bibitem [{\citenamefont {Jackiw}\ and\ \citenamefont
  {Rebbi}(1976)}]{solitonfermion}%
  \BibitemOpen
  \bibfield  {author} {\bibinfo {author} {\bibfnamefont {R.}~\bibnamefont
  {Jackiw}}\ and\ \bibinfo {author} {\bibfnamefont {C.}~\bibnamefont {Rebbi}},\
  }\bibfield  {title} {\bibinfo {title} {Solitons with fermion number
  \textonehalf{}},\ }\href {https://doi.org/10.1103/PhysRevD.13.3398}
  {\bibfield  {journal} {\bibinfo  {journal} {Phys. Rev. D}\ }\textbf {\bibinfo
  {volume} {13}},\ \bibinfo {pages} {3398} (\bibinfo {year}
  {1976})}\BibitemShut {NoStop}%
\bibitem [{\citenamefont {Krasznahorkay}\ and\ \citenamefont {{et
  al}}(2016)}]{X17-2016}%
  \BibitemOpen
  \bibfield  {author} {\bibinfo {author} {\bibfnamefont {A.~J.}\ \bibnamefont
  {Krasznahorkay}}\ and\ \bibinfo {author} {\bibnamefont {{et al}}},\
  }\bibfield  {title} {\bibinfo {title} {{Observation of Anomalous Internal
  Pair Creation in $^{8}\rm{Be}$: A Possible Indication of a Light, Neutral
  Boson}},\ }\href {https://doi.org/10.1103/PhysRevLett.116.042501} {\bibfield
  {journal} {\bibinfo  {journal} {Phys. Rev. Lett.}\ }\textbf {\bibinfo
  {volume} {116}},\ \bibinfo {pages} {042501} (\bibinfo {year}
  {2016})}\BibitemShut {NoStop}%
\bibitem [{\citenamefont {Krasznahorkay}\ and\ \citenamefont {{et
  al}}(2018)}]{X17-2018}%
  \BibitemOpen
  \bibfield  {author} {\bibinfo {author} {\bibfnamefont {A.~J.}\ \bibnamefont
  {Krasznahorkay}}\ and\ \bibinfo {author} {\bibnamefont {{et al}}},\
  }\bibfield  {title} {\bibinfo {title} {{New results on the $^{8}\rm{Be}$
  anomaly}},\ }\href {https://doi.org/10.1088/1742-6596/1056/1/012028}
  {\bibfield  {journal} {\bibinfo  {journal} {J. Phys.: Conf. Ser.}\ }\textbf
  {\bibinfo {volume} {1056}},\ \bibinfo {pages} {012028} (\bibinfo {year}
  {2018})}\BibitemShut {NoStop}%
\bibitem [{\citenamefont {Krasznahorkay}\ and\ \citenamefont {{et
  al}}(2019)}]{X17-2019}%
  \BibitemOpen
  \bibfield  {author} {\bibinfo {author} {\bibfnamefont {A.~J.}\ \bibnamefont
  {Krasznahorkay}}\ and\ \bibinfo {author} {\bibnamefont {{et al}}},\
  }\href@noop {} {\bibinfo {title} {{New evidence supporting the existence of
  the hypothetic X17 particle}}} (\bibinfo {year} {2019}),\ \Eprint
  {https://arxiv.org/abs/1910.10459} {1910.10459} \BibitemShut {NoStop}%
\bibitem [{\citenamefont {Krasznahorkay}\ and\ \citenamefont {{et
  al}}(2022)}]{X17-2022}%
  \BibitemOpen
  \bibfield  {author} {\bibinfo {author} {\bibfnamefont {A.~J.}\ \bibnamefont
  {Krasznahorkay}}\ and\ \bibinfo {author} {\bibnamefont {{et al}}},\
  }\bibfield  {title} {\bibinfo {title} {New anomaly observed in {C12} supports
  the existence and the vector character of the hypothetical {X17} boson},\
  }\bibfield  {journal} {\bibinfo  {journal} {Physical Review C}\ }\textbf
  {\bibinfo {volume} {106}},\ \href
  {https://doi.org/10.1103/physrevc.106.l061601} {10.1103/physrevc.106.l061601}
  (\bibinfo {year} {2022}),\ \Eprint {https://arxiv.org/abs/2209.10795}
  {2209.10795} \BibitemShut {NoStop}%
\bibitem [{\citenamefont {Krasznahorkay}\ and\ \citenamefont {{et
  al}}(2023)}]{X17-2023}%
  \BibitemOpen
  \bibfield  {author} {\bibinfo {author} {\bibfnamefont {A.~J.}\ \bibnamefont
  {Krasznahorkay}}\ and\ \bibinfo {author} {\bibnamefont {{et al}}},\
  }\href@noop {} {\bibinfo {title} {Observation of the {X17} anomaly in the
  decay of the giant dipole resonance of $^8$be}} (\bibinfo {year} {2023}),\
  \Eprint {https://arxiv.org/abs/2308.06473} {2308.06473} \BibitemShut
  {NoStop}%
\bibitem [{\citenamefont {Anh}\ and\ \citenamefont {{et
  al}}(2024)}]{X17-Hanoi}%
  \BibitemOpen
  \bibfield  {author} {\bibinfo {author} {\bibfnamefont {T.~T.}\ \bibnamefont
  {Anh}}\ and\ \bibinfo {author} {\bibnamefont {{et al}}},\ }\bibfield  {title}
  {\bibinfo {title} {Checking the 8be anomaly with a two-arm electron positron
  pair spectrometer},\ }\href {https://doi.org/10.3390/universe10040168}
  {\bibfield  {journal} {\bibinfo  {journal} {Universe}\ }\textbf {\bibinfo
  {volume} {10}},\ \bibinfo {pages} {168} (\bibinfo {year} {2024})},\ \Eprint
  {https://arxiv.org/abs/2401.11676} {2401.11676} \BibitemShut {NoStop}%
\bibitem [{\citenamefont {Barducci}\ and\ \citenamefont
  {Toni}(2023)}]{barducci}%
  \BibitemOpen
  \bibfield  {author} {\bibinfo {author} {\bibfnamefont {D.}~\bibnamefont
  {Barducci}}\ and\ \bibinfo {author} {\bibfnamefont {C.}~\bibnamefont
  {Toni}},\ }\bibfield  {title} {\bibinfo {title} {An updated view on the
  atomki nuclear anomalies},\ }\bibfield  {journal} {\bibinfo  {journal}
  {Journal of High Energy Physics}\ }\textbf {\bibinfo {volume} {2023}},\ \href
  {https://doi.org/10.1007/jhep02(2023)154} {10.1007/jhep02(2023)154} (\bibinfo
  {year} {2023}),\ \Eprint {https://arxiv.org/abs/2212.06453} {2212.06453}
  \BibitemShut {NoStop}%
\bibitem [{\citenamefont {Denton}\ and\ \citenamefont
  {Gehrlein}(2023)}]{X17-neutrino}%
  \BibitemOpen
  \bibfield  {author} {\bibinfo {author} {\bibfnamefont {P.~B.}\ \bibnamefont
  {Denton}}\ and\ \bibinfo {author} {\bibfnamefont {J.}~\bibnamefont
  {Gehrlein}},\ }\bibfield  {title} {\bibinfo {title} {Neutrino constraints and
  the atomki x17 anomaly},\ }\bibfield  {journal} {\bibinfo  {journal}
  {Physical Review D}\ }\textbf {\bibinfo {volume} {108}},\ \href
  {https://doi.org/10.1103/physrevd.108.015009} {10.1103/physrevd.108.015009}
  (\bibinfo {year} {2023}),\ \Eprint {https://arxiv.org/abs/2304.09877}
  {2304.09877} \BibitemShut {NoStop}%
\bibitem [{\citenamefont {Banerjee}\ and\ \citenamefont {{et
  al}}(2020)}]{NA64-2020}%
  \BibitemOpen
  \bibfield  {author} {\bibinfo {author} {\bibfnamefont {D.}~\bibnamefont
  {Banerjee}}\ and\ \bibinfo {author} {\bibnamefont {{et al}}} (\bibinfo
  {collaboration} {The NA64 Collaboration}),\ }\bibfield  {title} {\bibinfo
  {title} {{Improved limits on a hypothetical $X(16.7)$ boson and a dark photon
  decaying into ${e}^{+}{e}^{\ensuremath{-}}$ pairs}},\ }\href
  {https://doi.org/10.1103/PhysRevD.101.071101} {\bibfield  {journal} {\bibinfo
   {journal} {Phys. Rev. D}\ }\textbf {\bibinfo {volume} {101}},\ \bibinfo
  {pages} {071101} (\bibinfo {year} {2020})}\BibitemShut {NoStop}%
\bibitem [{\citenamefont {Affleck}\ \emph {et~al.}(1984)\citenamefont
  {Affleck}, \citenamefont {Dine},\ and\ \citenamefont
  {Seiberg}}]{Nonpert-SQCD-1}%
  \BibitemOpen
  \bibfield  {author} {\bibinfo {author} {\bibfnamefont {I.}~\bibnamefont
  {Affleck}}, \bibinfo {author} {\bibfnamefont {M.}~\bibnamefont {Dine}},\ and\
  \bibinfo {author} {\bibfnamefont {N.}~\bibnamefont {Seiberg}},\ }\href@noop
  {} {\bibfield  {journal} {\bibinfo  {journal} {Nucl. Phys. B}\ }\textbf
  {\bibinfo {volume} {241}},\ \bibinfo {pages} {493} (\bibinfo {year}
  {1984})}\BibitemShut {NoStop}%
\bibitem [{\citenamefont {Affleck}\ \emph {et~al.}(1985)\citenamefont
  {Affleck}, \citenamefont {Dine},\ and\ \citenamefont
  {Seiberg}}]{Nonpert-SQCD-2}%
  \BibitemOpen
  \bibfield  {author} {\bibinfo {author} {\bibfnamefont {I.}~\bibnamefont
  {Affleck}}, \bibinfo {author} {\bibfnamefont {M.}~\bibnamefont {Dine}},\ and\
  \bibinfo {author} {\bibfnamefont {N.}~\bibnamefont {Seiberg}},\ }\href@noop
  {} {\bibfield  {journal} {\bibinfo  {journal} {Nucl. Phys. B}\ }\textbf
  {\bibinfo {volume} {256}},\ \bibinfo {pages} {557} (\bibinfo {year}
  {1985})}\BibitemShut {NoStop}%
\bibitem [{\citenamefont {Seiberg}(1994{\natexlab{b}})}]{Nonpert-SQCD-Talk}%
  \BibitemOpen
  \bibfield  {author} {\bibinfo {author} {\bibfnamefont {N.}~\bibnamefont
  {Seiberg}},\ }\bibfield  {title} {\bibinfo {title} {{Exact Results on the
  Space of Vacua of Four Dimensional SUSY Gauge Theories}},\ }\href@noop {}
  {\bibfield  {journal} {\bibinfo  {journal} {Phys. Rev. D}\ }\textbf {\bibinfo
  {volume} {49}},\ \bibinfo {pages} {6857} (\bibinfo {year}
  {1994}{\natexlab{b}})},\ \Eprint {https://arxiv.org/abs/hep-th/9402044}
  {hep-th/9402044} \BibitemShut {NoStop}%
\end{thebibliography}
\end{document}